\documentclass[eqsecnum,showpacs]{revtex4}
\usepackage{epsf}
\usepackage{amsmath}

\begin{document}

\title{Non-analytic corrections to the Fermi-liquid behavior}
\author{Andrey V. Chubukov$^1$ and Dmitrii L. Maslov$^2$}
\affiliation{$^1$Department of Physics, University of Wisconsin-Madison,
1150 Univ. Ave., Madison, WI 53706-1390\\
$^2$Department of Physics, University of Florida, P. O. Box 118440,
Gainesville, FL 32611-8440}
\date{\today}

\begin{abstract}
The issue of non-analytic corrections to the Fermi-liquid behavior is
revisited. Previous studies have indicated that the corrections to the
Fermi-liquid forms of the specific heat and the static spin susceptibility scale
as $T^{D}$ and $T^{D-1}$, respectively (with extra logarithms for $D=1,3$).
In addition, the non-uniform spin susceptibility is expected to depend 
on the bosonic momentum $Q$ in a non-analytic way, i.e., as $Q^{D-1}$ (again
with extra logarithms for $D=1,3$). 
It is shown that these non-analytic corrections originate from the universal
singularities in the dynamical bosonic response functions of a generic Fermi
liquid. In contrast to the leading, Fermi-liquid forms which depend on the
interaction averaged over the Fermi surface, the non-analytic corrections
are parameterized by only two coupling constants, which are the components
of the interaction potential at momentum transfers $q=0$ and $q=2k_F$. For
3D systems, a recent result of Belitz, Kirkpatrick and Vojta for the spin
susceptibility is reproduced and the issue why a non-analytic momentum
dependence of the non-uniform spin susceptibility ($Q^{2}\ln |Q|$) is 
\emph{not }paralleled by a non-analyticity in the $T-$ dependence ($T^2$)
is clarified. For the case of a 2D system with a finite-range interaction,
explicit
forms of the corrections to the specific heat ($\propto T^2$), uniform
($\propto T$) and non-uniform ($\propto |Q|$) spin susceptibilities are obtained.
 It is shown that previous
calculations of the temperature dependences of these quantities in 2D were
 incomplete. Some of the results and conclusions of this paper have recently been
announced in a short communication
[A. V. Chubukov and D. L. Maslov, cond-mat/0304381].
\end{abstract}
\maketitle

\vspace{5mm}
\affiliation{$^1$Department of Physics, University of Wisconsin-Madison, 1150 Univ.
Ave., Madison WI 53706-1390\\
$^2$Department of Physics, University of Florida, P. O. Box 118440, Gainesville,
FL 32611-8440}

\vspace{5mm}
\affiliation{$^1$Department of Physics, University of Wisconsin-Madison, 1150 Univ.
Ave., Madison WI 53706-1390\\
$^2$Department of Physics, University of Florida, P. O. Box 118440, Gainesville,
FL 32611-8440}

\section{Introduction}

\label{sec:intro} The universal features of Fermi liquids and their physical
consequences continue to attract the attention of the condensed-matter
community for almost 50 years after the Fermi-liquid theory was developed by
Landau~\cite{landau}. A search for stability conditions of a Fermi liquid
and deviations from a Fermi-liquid behavior~\cite
{anderson,shankar,metzner,marston,nayak,chitov,kim}, particularly near quantum
critical points, intensified in recent years mostly due to the
non-Fermi-liquid features of the normal state of high $T_{c}$
superconductors~\cite{review_exp} and other materials.

In a generic Fermi liquid, the imaginary part of the retarded fermionic
self-energy $\Sigma _{R}(k,\omega )$ on the mass shell is determined solely
by fermions in a narrow $(\sim \omega $) energy range around the Fermi
surface and behaves as~\cite{agd,volumeIX,fetter,pines_noz}
\begin{equation}
\Sigma _{R}^{\prime \prime }=A(\omega ^{2}+(\pi T)^{2}).  \label{int1}
\end{equation}
Simultaneously, the real part of the self-energy scales as $\Sigma
_{R}^{\prime }=B\omega $, at small energies (Kramers-Kronig relations relate
constants $A$ and $B$ via an ultraviolet energy cutoff $W\sim E_{F}$). A
regular form of the self-energy has a profound effect on observable
quantities such as the specific heat and static spin and charge
susceptibilities, which have the same functional dependences as for free
fermions, e.g., specific heat $C(T)$ is linear in $T$, while the
susceptibilities $\chi _{s}(Q,T)$ and $\chi _{c}(Q,T)$ both approach
constant values at $Q=0$ and $T=0$. A regular behavior of the fermionic
self-energy is also in line with a general reasoning that turning on the
interaction in $D>1$ should not affect drastically the low-energy properties
of an electronic system, unless special circumstances, e.g., proximity to a
quantum phase transition, interfere~\cite{review_exp,qc}.

The subject of this paper is the analysis of the non-analytic, \emph{%
universal} corrections to the Fermi-liquid behavior that should be present
in a generic Fermi liquid. It has been known for some time that the
subleading terms in the $\omega -$ and $T-$ expansions of the fermionic
self-energy do not form regular, analytic series in $\omega ^{2}$ or $T^{2}$
(i.e., $\omega ^{3},\omega ^{5}$, etc. for $\Sigma ^{\prime }$ and $\omega
^{4},\omega ^{6}$, etc. for $\Sigma ^{\prime \prime }$)~\cite{galitskii}. In
particular, in $D=3$, the power counting shows the first subleading term in
the (retarded) on-shell self-energy at $T=0$ is~\cite{3dse}
\begin{eqnarray}
\delta \Sigma _{R}(\omega ) &=&\Sigma _{R}(\omega )-\Sigma _{R}^{FL}(\omega )
\notag \\
&=&B_{3D}\omega ^{3}\ln(-i\omega ).  \label{ya}
\end{eqnarray}
where $B_{3D}$ is real. For a generic $2<D<3$, this subleading term behaves
as $\omega ^{D}$. In 2D, it is again logarithmic~ \cite
{chaplik,hodges,bloom,fujimoto,quinn,bruch,randeria}
\begin{equation}
\delta \Sigma _{R}(\omega )=-iB_{2D}\omega ^{2}\ln(-i\omega )  \label{ya1}
\end{equation}
where $B_{2D}$ is real. From a formal perspective, the $\omega ^{2}\ln
\omega $ form of the correction term in 2D implies that at $\omega
\rightarrow 0$ it dominates over a Fermi-liquid, $\omega ^{2}$ -term, i.e.,
a conventional Fermi-liquid reasoning breaks down. This is true also for $%
D<2 $, as the correction term scales again as $\omega ^{D}$. However, as
long as $D>1$, ${\mbox{Re}}\Sigma ^{FL}(k_{F},\omega )\sim \omega $ is
asymptotically larger at low frequencies than ${\mbox{Im}}\Sigma
(k_{F},\omega )$, i.e., fermionic excitations remain well-defined. For a
complete set of references on this problem see Ref. \cite{aleinertauphi}.

The singularity in ${\mbox{Re}}\Sigma $ affects directly the subleading term
$\delta C(T)$ in the specific heat $C(T)=\gamma T+\delta C(T)$ via~\cite{agd}
\begin{equation}
\delta C(T)= 2 \pi V~\frac{\partial}{\partial T}
~\left[\frac{1}{T}~ \int \frac{d^D k}{(2\pi)^D} \int_{-\infty}^\infty
d\omega \omega \frac{\partial n (\omega)}{\partial \omega }
{\mbox{Re}}\Sigma (\omega,k) \delta (\omega - \epsilon_k) \right],
 \label{sp_he}
\end{equation}
where $V$ is the system volume. In 3D, the power counting yields $\delta
C(T)\propto T^{3}\ln T$~\cite{doniach}, while in 2D, $\mbox{Re}(\delta
\Sigma (\omega ))\propto \omega ^{2}$, and by power counting $\delta
C(T)\propto T^{2}$~\cite{previous_C(T),bedell}.

Belitz, Kirkpatrick and Vojta (BKV) argued~\cite{bkv} that the singularity
in the fermionic self-energy should also affect spin susceptibility and give
rise to a singular momentum expansion of the static $\chi _{s}(Q,T=0)$. A
similar idea was expressed by Misawa~\cite{misawa}. Indeed, the
susceptibility is a convolution of the two fermionic Green's functions (a
particle-hole bubble). For non-interacting fermions, $\chi _{s}(Q,0)$ is
given by the Lindhard function which is analytic in $Q$ for small $Q$ in all
$D$. The corrections to the Lindhard form are obtained by self-energy and
vertex-correction insertions into the particle-hole bubble (see Fig. \ref
{fig:diag}). The diagrams with self-energy insertions can be viewed as
convolutions of $G$ and $G_{0}$ where $G^{-1}=G_{0}^{-1}+\Sigma $.
Substituting the self-energy and expanding in $\Sigma $ and in $Q$, we
obtain
\begin{equation}
\delta \chi (Q,0)=\chi (Q,0)-\chi (0,0)\propto Q^{2}\int d\omega d\epsilon
_{q}\frac{\Sigma (\omega ,q)}{(i\omega -\epsilon _{q})^{5}}.  \label{i2}
\end{equation}
Substituting the singular part of $\Sigma (\omega )$ into (\ref{i2}) and
just counting powers, we find $\delta \chi (Q,0)\propto Q^{2}\ln |Q|$ for $%
D=3$, and $\delta \chi (Q,0)\propto Q^{D-1}$ for smaller $D$. (For $D=1$, a
more accurate estimate yields $\chi (Q,0)\propto \ln |Q|$).

To verify this reasoning, BKV explicitly computed $\delta \chi _{s}(Q,0)$ in
3D to second order in the interaction, and indeed demonstrated~\cite{bkv}
that $\delta \chi (Q,0)\propto Q^{2}\ln |Q|$, in agreement with power
counting. Based on this agreement, BKV conjectured that power counting
should be valid for all $D>1$, i.e., the fully renormalized spin
susceptibility should scale with momenta as $Q^{D-1}$.

Another non-analytic behavior was discovered in the analysis of the
temperature dependence of the uniform susceptibility in $2D$. Baranov, Kagan
and Marenko (BKM)~\cite{marenko} estimated $\chi _{s}(Q=0,T)$ using a
relation between the uniform susceptibility and the quasiparticle
interaction function~\cite{agd,volumeIX}, and argued that $\chi _{s}(0,T)$
is linear in $T$ in 2D. Chitov and Millis (CM) \cite{millis} later used the
same approach, but went beyond estimates and performed a detailed analysis
of the quasiparticle interaction function and the susceptibility. They also
found a linear-in-$T$ dependence.

Another example of non-analyticity in the leading corrections to a
Fermi-liquid behavior is linear-in-$T$ corrections to the impurity
scattering time in two dimensions
\cite{stern,gold,dassarma,dassarmahuang}. A general treatment of
this situation \cite{zna} shows that the correction to the
residual conductivity of a dirty Fermi liquid depends linearly on
the temperature in the ballistic regime, i.e., when $T$ is much
larger than the level width due to impurity scattering. Unlike the
familiar $\ln T$ -dependence of the conductivity in the diffusive
regime \cite{altshuler}, this linear $T-$ dependence originates
from the singular behavior of the response functions of a clean
Fermi liquid in 2D.

Our intension to pursue a further study on singular corrections to the
Fermi-liquid behavior is stimulated by several factors. First, we want to
clarify what actually causes the singularities in the fermionic self-energy,
specific heat and spin susceptibility. To illustrate the importance of
understanding this issue, we note that power counting arguments are not
rigorous and can lead to incorrect results. Indeed, let's apply power
counting to the susceptibility of noninteracting fermions, which, we know,
is a Lindhard function. Each Green's function of free fermions $%
G_{0}(p,\omega _{n})=[i\omega _{n}-v_{F}(k-k_{F})]^{-1}$ scales as one
inverse power of momentum and energy (the corresponding dynamical exponent $%
z_{F}=1$), so the convolution of the two Green's functions contributes two
powers of $k-k_{F}$ in the denominator of the integrand for $\chi (Q,0)$.
Expanding up to $Q^{2}$, one then adds two extra powers. The frequency
integration eliminates one, so there are three powers of momentum left in
the denominator. The prefactor for $Q^{2}$ should then scale as
\begin{equation}
\int \frac{d^{D}q}{q^{3}},
\end{equation}
where $q=p-k_{F}.$ The lower limit of the integration is of order $Q$, the
upper limit is of order $k_{F}$. The integral is infrared divergent for $%
D\leq 3$ , scales as $\ln |Q|$ for $D=3$, as $|Q|^{D-3}$ for $1<D<3$, and as
$|Q|^{-2}\ln |Q|$ for $D=1$. We see that a power counting predicts a
singular momentum dependence of the Lindhard function. The true Lindhard
function obviously does not obey this behavior -- it is analytic near $Q=0$
for all $D$. In 3D~\cite{ashcroft_mermin},
\begin{equation}
\chi _{0}(Q,T=0)=\chi _{0}^{3D}\left( 1-\frac{Q^{2}}{8k_{F}^{2}}\right)
\end{equation}
where $\chi _{0}^{3D}=mk_{F}/\pi ^{2}$. In 2D, it is just a constant for $%
|Q|<2k_{F}$~\cite{kagan,stern_1},
\begin{equation}
\chi _{0}(Q,0)=\chi _{0}^{2D},~~~~Q<2k_{F},
\end{equation}
where $\chi _{0}^{2D}=m/\pi $. In 1D
\begin{equation}
\chi _{0}(Q,0)=\chi _{0}^{1D}\left( 1+\frac{1}{12}\frac{Q^{2}}{k_{F}^{2}}%
\right) ,
\end{equation}
where $\chi _{0}^{2D}=2m/\pi k_{F}$. The failure of power counting arguments
to reproduce the behavior of the Lindhard function clearly calls for
understanding under which conditions they do work. The same problem holds
also for the self-energy, as the singular forms of Eqs. (\ref{ya}) and (\ref
{ya1}) are obtained by power counting, and there is no guarantee that the
coefficients are nonzero. In fact, CM computed the leading correction to the
real part of the self-energy in 2D and argued that it \emph{vanishes}.
This would imply that the coefficient $B_{2D}$ in (\ref{ya1}) vanishes, and
thus the $\omega ^{2}\ln \omega $ in ${\mbox{Im}}\Sigma _{R}$ is absent. Our
result will be different (see below) - we will find that $B_{2D}$ is
finite.

Another reason to look more deeply into the physics of singularities is the
discrepancy between momentum and temperature dependences of the
susceptibility. The fact that dynamical exponent $z_{F}=1$ would normally
imply that a non-analytic dependence $\delta \chi (Q,T=0)\propto Q^{D-1}$
should be paralleled by a non-analytic dependence of $\delta \chi
(0,T)\propto T^{D-1}$. In 3D, this analogy would mean that $\delta \chi
(0,T)\propto T^{2}\ln T$. Misawa \cite{misawa_2} did find a $T^{2}\ln T$
term in his calculations in early 70s. However, later Carneiro and Pethick ~
\cite{pethick}, and recently BKV~\cite{bkv} argued that the $T^{2}\ln T$
term is actually absent in 3D. Several explanations have been put forward to
explain this discrepancy. BKV suggested that the absence of the $T^{2}\ln T$
dependence in 3D is accidental and should not be regarded as a failure of
power counting arguments. They conjectured that for a generic $D<3$, the $%
T^{D-1}$ dependence of $\chi _{s}(0,T)$ should hold. This conjecture was
verified numerically by Hirashima and Takahashi~\cite{hirashima} for $D=2$,
but no definite conclusion has been drawn because of numerical difficulties.

As we already said, BKM~\cite{marenko} and CM \cite{millis} considered $\chi
(0,T)$ in 2D analytically and argued that the linear-in-$T$ term is present.
Both groups argued that
 $\delta \chi _{s}(0,T)\propto T$ comes from $2k_{F}$ effects 
(our results are in full agreement with this). 
 BKM also argued that a $T-$ dependence is caused by the singular
behavior of the quasiparticle interaction function for fermions away from
the Fermi surface (in equivalent diagrammatic language - by the singular
frequency dependence of the particle hole bubble near $2k_{F}$). CM found
that the linear-in-$T$ behavior is caused not only by this effect, but also 
 by the non-analytic temperature
behavior of the quasiparticle interaction function for fermions at the Fermi
surface (in diagrammatic language, by the singular $T$ dependence of the
static particle-hole near $2k_{F}$). The relation between the singularity in
the particle hole bubble and non-analyticity of $\chi _{s}(0,T)$ follows
from the fact that a generic diagram for for the correction to a
Fermi-liquid susceptibility, e.g., diagram 1 in Fig. \ref{fig:diag},
contains a combination
\begin{equation}
\delta \chi (0,T)\sim T\sum_{\omega _{n}}\int d^{2}kG^{3}\left( k,\omega
_{n}\right) T\sum_{\Omega _{m}}\int d^{2}qG\left( \mathbf{k+q},\omega
_{n}+\Omega _{m}\right) \Pi \left( q,\Omega _{m},T\right),
\end{equation}
where $G\left( k,\omega _{n}\right) =\left( i\omega _{n}-\epsilon
_{k}\right) ^{-1}$ is the fermionic propagator. At $T=0$, a static
particle-hole polarization bubble $\Pi (q,\omega =0,T=0)$ in $D=2$ has an
asymmetric square root singularity at $q\rightarrow 2k_{F}+0$~\cite
{kagan,stern_1,tremblay,chubukov}. A finite $T$ or finite $\omega $ soften the
singularity and yield $\Pi (q,\omega ,T)-\Pi (q,0,0)\propto \sqrt{\text{max\{%
}T,\omega \}}$ in the momentum range $|q-2k_{F}|\sim T/v_{F}$~\cite
{stern_1,fukuyama,millis}. A simple calculation shows that fermions which
contribute to $\delta \chi _{s}(0,T)$ have energies of order $\sim T$ and
are located in a narrow angular range where the angle $\theta $ between
vectors $\mathbf{k}$ and $\mathbf{q}$ is almost $\pi :$ $\pi -\theta \sim
\left( T/E_{F}\right) ^{1/2}$. Using this and assembling the powers, one
obtains that $\delta \chi (0,T)\propto T.$

In 3D, an analogous reasoning yields the $T^{2}\ln T$ behavior. CM
suggested~ \cite{millis} that previous computations in 3D might have missed
the crucial $2k_{F}$ effects and hinted that Misawa may be right in that the
$T^{2}\ln T$ term may actually be present in 3D.

In the present communication, we analyze in detail the physical origin of
the non-analytic corrections to the Fermi liquid and clarify the discrepancy
between earlier papers. We obtain explicit results in $D=2$ for the
fermionic self-energy, the effective mass, the specific heat, and for spin
and charge susceptibilities at finite $Q$ and $T=0$, and at finite $T$ and $%
Q=0$. We also verify earlier results for $D=3$.

We argue that a proper treatment of non-analyticities in the fermionic
self-energy and in $\chi _{s}(Q,0)$ requires the knowledge of the \emph{%
dynamical} particle-hole response function. We show explicitly that the
non-analyticity in the static Lindhard function near $2k_{F}$ does not give
rise to a non-analytic behavior of the self-energy due to extra
cancelations. For the spin susceptibility, the computation with the static
Lindhard function does yield linear in $|Q|$ and $T$-terms, due to $2k_{F}$
effects, but with incorrect prefactors. We also demonstrate that
non-analytic terms in the self-energy and the spin susceptibility can be
viewed equivalently as coming either from the non-analyticity in the
dynamical particle-hole bubble near $q=0$, \emph{or} $q=2k_{F}$, \emph{or}
from the non-analyticity in the dynamical particle-particle bubble near zero
total momentum. Our results do agree with that of BKV who formally
considered only $q=0$ contributions. However, we show explicitly that they
indeed computed all possible non-analytic contributions to the static
susceptibility, including $2k_{F}$ effects, but just used an unconventional
labeling of internal momenta in the diagrams. As an essential step beyond
the BKV work, we show explicitly that the non-analytic terms in all diagrams
for $\chi _{s}(Q,0)$ come exclusively from the vertices in which the
transferred momentum is either $0$ or $2k_{F}$, and \emph{simultaneously}
the total momentum is $0.$ There are only such vertices. They can be viewed
as two parts of the scattering amplitude with zero momentum transfer \emph{%
and} zero total momentum:
\begin{equation}
\Gamma _{\alpha ,\beta ;\gamma ,\delta }(k,-k;k,-k)=U(0)\delta _{\alpha
\gamma }\delta _{\beta \delta }-U(2k_{F})\delta _{\alpha \delta }\delta
_{\beta \gamma}.
\end{equation}
This restriction to just one scattering amplitude is rather non-trivial, as
it implies that non-analytic terms in all diagrams for the susceptibility
depend only on $U(0)$ and $U(2k_{F})$ but not on averaged interactions over
the Fermi surface, as in the BKV analysis. 
 A similar result has been obtained recently for the
conductivity in the ballistic regime in 2D\cite{zna}: for a short-range
interaction, the conductivity has a non-analytic $T-$ dependent piece, whose
prefactor depends only on $U(0)$ and $U(2k_{F})$ rather than on the
interaction averaged over the Fermi surface.

Some of the results and conclusions of this paper have recently been
announced in a short communication \cite{short}.

The paper is organized as follows. In Sec. II we briefly review three known
non-analyticities in the response functions of a Fermi liquid. In the next
four sessions we consider a fermionic system with a contact, i.e., $q-$%
independent interaction. In Sec. III we discuss the leading corrections to
the self-energy for interacting fermions in 2D. We show that the on-shell
self-energy has the form of Eq. (\ref{ya1}) with a nonzero $B_{2D}$, and
this gives rise to a linear-in-$T$ correction to the effective mass, and to $%
T^{2}$ correction to the specific heat. We show that a correction to the
effective mass is not observable in a magneto-oscillations experiment due to
a peculiar cancelation between two $T-$ dependent terms in the self-energy.
We also briefly discuss self-energy corrections for $D=3$.

In Sec. IV-VI we consider in detail a non-analytic perturbation theory for
the charge- and spin-susceptibilities. We use the self-energy calculated in
Sec. II along with the dynamical Lindhard functions near $q=0$ and $q=2k_{F}$
and the dynamical particle-hole bubble near the zero total momentum as
building blocks, and obtain analytic expressions for charge- and
spin-susceptibilities. More specifically, in Sec. IV we present, for
completeness, the expressions for the spin susceptibility of noninteracting
fermions (the Lindhard function) for various $D$. In Sec. V we consider the
susceptibility at $T=0$ and finite $Q$. We present the first analytic
calculation of $\chi _{s}(Q,0)$ in 2D. We explicitly show that it scales as $%
|Q|$ and compute the prefactor. These 2D calculations require substantially
more effort than in 3D since the internal momenta in the diagrams are all of
order $Q$, and one cannot simply expand in $Q^{2}$ and then cut the infrared
divergence of the prefactor by $Q,$ because in $2D$ the divergence is power
law rather than logarithmic. We then discuss the 3D case for which we
reproduce in a novel way the result of BKV that $\delta \chi
_{s}(Q,0)\propto Q^{2}\ln |Q|$. We explicitly verify that non-analytic ($%
|Q|) $ terms obtained either via a ``conventional'' approach to treat $%
2k_{F} $ contributions and the technique invented by BKV are the same. We
also discuss briefly the $1D$ case.

In Sec. VI we consider the static susceptibility at finite $T$. We show that
in 2D, $\chi _{s}(0,T)$ scales as $T$ with a universal prefactor. We also
show that the linear-in-$T$ dependence come from two effects: from the
thermal smearing of the \emph{static} Lindhard function for particles at the
Fermi surface, and from the frequency dependence of the dynamical Lindhard
function (i.e., from particles outside the Fermi surface). In particular, we
show that the linear-in-$T$ piece is present in all diagrams for $\chi (0,T)$%
, including the ones for which the momentum transfer in the Lindhard
function is near $q=0$ (the linear-in-$T$ terms coming from near $q=0$ and $%
q=2k_{F}$ are equal). Near $q=0$, the static Lindhard function is analytic,
and a linear-in-$T$ susceptibility comes entirely from the non-analyticity
in the dynamical part of the Lindhard function. BKM considered only the
second source of the $O(T)$ behavior, CM included both effects. Our result
differs by a factor of 2 compared to that of CM -- we could not detect the
reason for the discrepancy. We further analyze in detail the physical origin
for the linear-in-$T$ term in $2D$ (and $T^{D-1}$ for a general $D\leq 3$),
and discuss to which extent it is related to $|Q|^{D-1}$ term in $\chi
_{s}(Q,0)$. We show that the physics behind $T^{D-1}$ term in $\chi
_{s}(0,T) $ and $|Q|^{D-1}$ term in $\chi _{s}(Q,0)$ is, in fact, different.
We discuss how the non-analytic term in $\chi (0,T)$ $T$ evolves with $D$,
and show that for $D=3$, $\chi _{s}(0,T)\propto T^{2}$ without an extra
logarithmic factor. This agrees with Carneiro and Pethick~\cite{pethick} and
BKV results that $\chi _{s}(0,T)$ in 3D is free from non-analyticities to
order $T^{2}$. We also show that although $\chi _{s}(0,T)$ goes smoothly
through $D=2$, the $2D$ case is still somewhat special. Finally, we analyzed
charge susceptibility and found that non-analytic terms in $\chi _{c}(Q,T)$
are all cancelled out, i.e., the first corrections to the Fermi-liquid form
for the charge susceptibility are all analytic. For a $2D$ case, this result
fully agrees with CM.

In Sec. VII we consider the case of a finite-range interaction with $q-$%
dependent $U(q)$. We demonstrate that non-analytic terms appear in a way
similar to anomalies in quantum field theory, and depend \emph{only} on $%
U(0) $ and $U(2k_{F})$, not on the momentum-averaged interaction. We show
that at both $T=0$ and finite $T$, the non-analytic correction to the
self-energy depends on $U^{2}(0)+U^{2}(2k_{F})-U(0)U(2k_{F})$, while the
total non-analytic correction to $\chi _{s}$ depends only on $U^{2}(2k_{F})$%
. We show that the charge susceptibility does not have a non-trivial $Q$
dependence--all non-analytic terms from individual diagrams cancel out even
when $U=U(q)$.

In Sec. VIII we present our conclusions. Appendices A-D show details of some
calculations.

\section{non-analyticities in the bosonic response functions}

\label{sec:singbosresp} We will demonstrate in this paper that the
non-analytic corrections to the Fermi-liquid theory are universally related
to the \emph{Fermi-liquid} non-analyticities in the dynamical bosonic
response functions. To set the stage, we review briefly these
non-analyticities.

There are three physically distinct bosonic non-analyticities in a generic
Fermi liquid at $T=0$~\cite{agd,volumeIX,fetter}. The first is the
non-analyticity in the particle-hole response function,
\begin{equation}
\Pi _{ph}(Q,\Omega _{m})=-\int \int \frac{d^{D}pd\omega _{n}}{(2\pi )^{D+1}}%
G(p,\omega _{n})G(\mathbf{p}+\mathbf{Q},\omega _{n}+\Omega _{m})
\end{equation}
at small momentum and frequency transfers. For $D=2$,
\begin{equation}
\Pi _{ph}^{Q\rightarrow 0}(Q,\Omega _{m})=\frac{m}{2\pi }~\left( 1-\frac{%
|\Omega _{m}|}{\sqrt{(v_{F}Q)^{2}+\Omega _{m}^{2}}}\right) .  \label{2.01}
\end{equation}
For $D=3$,
\begin{equation}
\Pi _{ph}^{Q=0}(q,\Omega _{m})=\frac{mk_{F}}{2\pi ^{2}}~\left( 1-\frac{%
|\Omega _{m}|}{v_{F}q}\tan ^{-1}\frac{v_{F}Q}{|\Omega _{m}|}\right) .
\label{2.011}
\end{equation}
The zero frequency results: $\Pi _{ph}(0,0)=m/2\pi $ in 2D and $\Pi
_{ph}(0,0)=mk_{F}/2\pi ^{2}$ in 3D, are the densities of states of free
fermions per one spin orientation.

The non-analyticity in the particle-hole bubble at small momenta introduces
the dependence of $\Pi _{ph}(Q\rightarrow 0,\omega \rightarrow 0)$ on the
ratio $\Omega /v_{F}Q$, and eventually gives rise to the emergence of a
zero-sound collective mode in a Fermi liquid~\cite{agd,volumeIX}.

The second is the non-analyticity in the particle-hole response function at
momentum transfer near $2k_{F}$. For $D=2$
\begin{equation}
\Pi _{ph}^{2k_{F}}(Q,\Omega _{m})=\frac{m}{2\pi }\left( 1-\sqrt{\frac{\tilde{%
Q}}{2k_{F}}+\left[ \left( \frac{\Omega _{m}}{2v_{F}k_{F}}\right) ^{2}+\left(
\frac{\tilde{Q}}{2k_{F}}\right) ^{2}\right] ^{1/2}}\right) ,  \label{2.02}
\end{equation}
where $\tilde{Q}\equiv Q-2k_{F}$ and $\left| \tilde{Q}\right| \ll 2k_{F}$.
In the static limit, the non-analyticity is one-sided~\cite
{kagan,stern_1,tremblay,chubukov}:
\begin{eqnarray}
\Pi _{ph}^{2k_{F}}(Q,0) &=&\frac{m}{2\pi },~~{\text{f}}\text{{or}}~~Q<2k_{F};
\notag \\
\Pi _{ph}^{2k_{F}}(Q,0) &=&\frac{m}{2\pi }\left( 1-\left( \frac{Q-2k_{F}}{%
k_{F}}\right) ^{1/2})\right) ,~~\text{for}~~Q>2k_{F}.  \label{2.022}
\end{eqnarray}
In $D=3$, this non-analyticity is logarithmic and odd with respect to $%
\tilde{Q}$~\cite{ashcroft_mermin}. In the static limit
\begin{equation}
\Pi _{ph}^{2k_{F}}(q,0)=\frac{mk_{F}}{4\pi ^{2}}\left( 1-\frac{\tilde{Q}}{%
2k_{F}}\ln {\frac{4k_{F}}{|\tilde{Q}|}}\right).  \label{2.0222}
\end{equation}
The dynamical expression is rather complex in 3D, and we refrain from
presenting it.

The $2k_{F}$ non-analyticity gives rise to long-range Friedel oscillations
of electron density in a Fermi liquid~\cite{kl} and eventually accounts for $%
p-$wave pairing in electron systems with short-range repulsive interaction~
\cite{ch_kag}.

The third is the logarithmic singularity in the particle-particle response
function
\begin{equation}
\Pi _{pp}(Q,\Omega _{m})=-\int \int \frac{d^{D}pd\omega _{n}}{(2\pi )^{D+1}}%
G(p,\omega _{n})G(-\mathbf{p}+\mathbf{Q},-\omega _{n}+\Omega _{m})
\end{equation}
at small total momentum $q$ and frequency $\omega $. In 2D,
\begin{equation}
\Pi _{pp}(Q,\Omega _{m})=\frac{m}{2\pi }\ln {\frac{|\Omega _{m}|+\sqrt{%
\Omega _{m}^{2}+(v_{F}Q)^{2}}}{W}},  \label{2.03}
\end{equation}
where $W\sim E_{F}$. In 3D, the functional form is similar. If the full
irreducible interaction between electrons is attractive for at least one
value of the angular momentum, this singularity gives rise to
superconductivity at $T=0$~\cite{agd}. In the weak-coupling regime that we
will be focusing on, the instability occurs at only exponentially small
frequencies, and we will neglect it, assuming that the system remains normal
down to $T=0$. Still, as we will see, a non-analytic dependence on the ratio
$\Omega _{m}/v_{F}q$ in $\Pi _{pp}(Q,\Omega _{m})$ will give rise to a
non-analyticity in the self-energy and susceptibility.

In the rest of the paper we show that these non-analyticities give rise to
universal subleading terms in the fermionic self-energy, effective mass,
specific heat, and static spin susceptibility.

\section{fermionic self-energy. effective mass, specific heat, and the
amplitude of magneto-oscillations}

\label{sec:selfenergy}

In this Section we obtain non-analytic corrections to the fermionic
self-energy and consider how they affect observable quantities such as the
effective mass and the specific heat. We will mostly focus on $D=2$, but for
the sake of completeness will also discuss the situation in $D=3$ and $D=1$.
We also assume for simplicity that the interaction is a contact one, i.e.,
its Fourier transform is independent of momentum. We will restore the
momentum dependence of $U\left( q\right) $ in Sec. VII.

\subsection{Self-energy of a generic Fermi liquid}

\label{sec:sigmageneric} 
\begin{figure}[tbp]
\centerline{\epsfxsize=6in
 \epsfbox{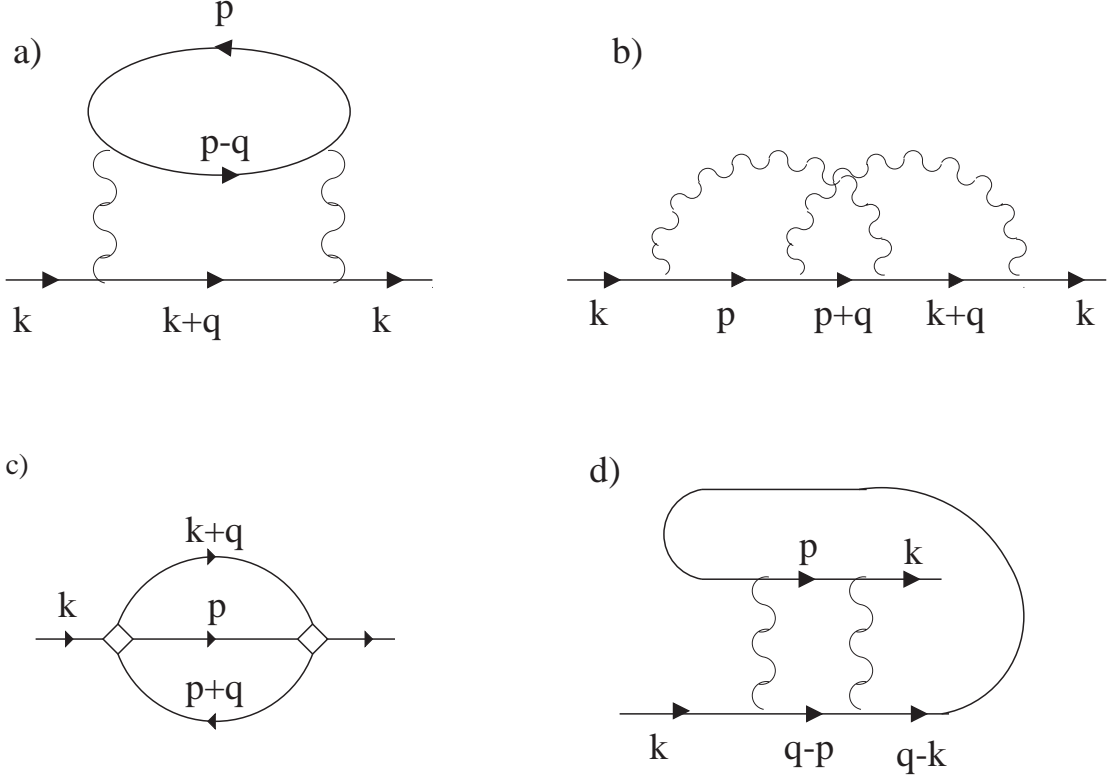}}
\caption{a) and b) the two non-trivial second-order diagrams for the
self-energy; c) an equivalent form of diagrams a) and b) (``sunrise''
diagram; d) diagram b) in an explicit particle-particle form. }
\label{fig1}
\end{figure}

The (Matsubara) fermionic self-energy is related to
the Green's function via
\begin{equation}
G^{-1}(k,\omega _{n})=G_{0}^{-1}(k,\omega _{n})+\Sigma (k,\omega _{n}),
\end{equation}
where $G_{0}^{-1}(k,\omega _{n})=i\omega _{n}-\epsilon _{k}$ and $\epsilon
_{k}=\left( k^{2}-k_{F}^{2}\right) /2m.$ The two nontrivial second-order
diagrams for $\Sigma (k,\omega _{n})$ are presented in Fig. \ref{fig1}.

For a contact interaction with a coupling constant $U$, the diagrams a) and
b) in Fig. \ref{fig1} yield equal functional forms of the self-energy, and
only differ in the combinatorial factor resulting from the spin summation
and the number of closed loops. This factor is equal to (-2) for diagram (a)
in Fig.~\ref{fig1} and to (1) for diagram (b). The result for $\Sigma
(k,\omega _{n})$ can then be re-expressed as a single diagram Fig.\ref{fig1}%
c in which the diamond stands for the interaction vertex $iU.$ %\[
%\Gamma _{a,\beta ;\gamma ,\delta }=U\left( \delta _{\alpha \beta }\delta
%_{\gamma \delta }-\delta _{\alpha \delta }\delta _{\beta ,\gamma }\right)
%\label{vert}
%\]
In the analytic form, we have
\begin{equation}
\Sigma (k,\omega
_{n})=U^{2}\sum_{k_{1},k_{2},k_{3}}G_{0}(k_{1})G_{0}(k_{2})G_{0}(k_{3})%
\delta (k_{1}+k_{2}-k_{3}-k).  \label{2.101}
\end{equation}
For brevity, we introduced temporarily a ``relativistic'' notation $k=(%
\mathbf{k},\omega _{n})$. The diagram in Fig.\ref{fig1}c can be equally
re-expressed either via particle-hole polarization operator $\Pi
_{ph}(Q,\Omega _{m})$, as
\begin{equation}
\Sigma (k,\omega _{n})=-TU^{2}\sum_{\Omega _{m}}\int \frac{d^{D}Q}{(2\pi
)^{D}}~G_{0}(\mathbf{k}+\mathbf{Q},\omega _{n}+\Omega _{m})\Pi
_{ph}(Q,\Omega _{m}),  \label{2.1}
\end{equation}
or via the particle-particle polarization operator, as
\begin{equation}
\Sigma (k,\omega _{n})=-TU^{2}\sum_{\Omega _{m}}\int \frac{d^{D}Q}{(2\pi
)^{D}}~G_{0}(\mathbf{Q}-\mathbf{k},\Omega _{m}-\omega _{n})\Pi
_{pp}(Q,\Omega _{m}).  \label{2.102}
\end{equation}
We illustrate the last representation in Fig.\ref{fig1}d. Here and
thereafter $\omega _{n}=\pi \left( 2n+1\right) T$ and $\Omega _{m}=2\pi mT.$

For definiteness, we will proceed with the form of Eq. (\ref{2.1}) and
discuss how the non-analyticity in the particle-hole bubble gives rise to
the non-analyticity in the fermionic self-energy. To shorten the notations,
we will use $\Pi _{ph}(Q,\Omega _{m})=\Pi (Q,\Omega _{m})$ until otherwise
specified. We then show that a non-analytic part of the self-energy can be
viewed equivalently as coming from the non-analyticity in the
particle-particle bubble.

For the analysis of the specific heat, effective mass and fermionic damping,
we will need the retarded fermionic self-energy $\Sigma _{R}(k,\omega )$ in
real frequencies and at finite temperatures. In some cases, it can be
obtained directly from $\Sigma (k,\omega _{n})$ via a replacement $i\omega
_{n}\rightarrow \omega +i\delta $. In general though it is rather difficult
to deal with discrete Matsubara sums. The approach we adopt here will be to
find \ the imaginary part of the retarded self-energy $\Sigma _{R}^{\prime
\prime }(\mathbf{k},\omega ).$ The real part of the self-energy, $\Sigma
_{R}^{\prime }(k,\omega )$ is then obtained via the Kramers-Kronig relation.

Applying the spectral representation
\begin{equation}
f(i\omega _{n})=\frac{1}{\pi }\int dz\frac{f_{R}^{\prime \prime }(z)}{%
z-i\omega _{n}},  \label{spectral}
\end{equation}
to (\ref{2.1}), and using $\mbox{Im}G_{0}^{R}(\mathbf{k}+\mathbf{Q},\omega
)=-\pi \delta (\omega -\epsilon _{\mathbf{k}+\mathbf{Q}})$, we find
\begin{equation}
\Sigma _{R}^{\prime \prime }(\mathbf{k},\omega )=\frac{1}{2}U^{2}~\int
d\Omega \int \frac{d^{D}Q}{\left( 2\pi \right) ^{D}}\delta (\Omega +\omega
-\epsilon _{\mathbf{k}+\mathbf{Q}})\Pi _{R}^{\prime \prime }(Q,\Omega )\left[
\coth \frac{\Omega }{2T}-\tanh \frac{\omega +\Omega }{2T}\right] .
\label{2.2}
\end{equation}

We first remind a reader how the Fermi-liquid form of $\Sigma _{R}^{\prime
\prime }(\mathbf{k},\omega )$ is obtained. Suppose that $\omega \ll \epsilon
_{F}$. A simple analysis of (\ref{2.2}) shows that because of the last term
in (\ref{2.2}), typical $\Omega $ are of order of $\omega $, i.e., they are
also small compared to $\epsilon _{F}$. The imaginary part of the retarded $%
\Pi _{R}^{\prime \prime }(Q,\Omega )$ is an odd function of frequency, and
hence for small frequencies $\Pi ^{\prime \prime }(Q,\Omega )=\Omega
F(Q,\Omega ).$ Let's now \emph{assume} that typical $v_{F}Q$ are much larger
than typical $\Omega $. Then $F(Q,\Omega )\approx F(Q,0)$. Substituting this
into (\ref{2.2}), we obtain
\begin{equation}
\Sigma _{R}^{\prime \prime }(\mathbf{k},\omega )=\frac{1}{2}U^{2}~\int \int
\frac{d^{D}Q}{\left( 2\pi \right) ^{D}}\delta (\epsilon _{\mathbf{k}+\mathbf{%
Q}})F(Q,0)~\int d\Omega \Omega \left[ \coth \frac{\Omega }{2T}-\tanh \frac{%
\omega +\Omega }{2T}\right] .  \label{2.21}
\end{equation}
We see that as long as the momentum integral is infrared convergent, it is
dominated by large $Q\simeq k_{F}$. The momentum integral is then fully
separated from the frequency integral and yields a constant prefactor. That
typical $Q\simeq k_{F}$ also justifies \emph{a posteriori} the assumption
that $F(Q,\Omega )\approx F(Q,0)$. The easiest way to do the remaining
frequency integration is to integrate in a finite range $-W<\Omega <W$.
Shifting the variable in the second term as $\Omega +\omega \rightarrow
\Omega $, and then setting $W=\infty $ we find
\begin{equation}
\Sigma _{R}^{\prime \prime }(\mathbf{k},\omega )=C\left[ \omega ^{2}+(\pi
T)^{2}\right] ,  \label{2.40}
\end{equation}
where $C$ is a constant. This is a well-known result in the Fermi-liquid
theory \cite{agd}.

The form of $\Sigma _{R}^{\prime \prime }(\mathbf{k},\omega )$ given by Eq.(%
\ref{2.40}) is generic to any Fermi liquid provided that the momentum
integral is dominated by large momenta $Q\gg \Omega /v_{F}$. Higher order
terms in $\Pi _{R}^{\prime \prime }(Q,\Omega )$ form a series in $\Omega
^{2n+1}$. If we assume that the prefactors depend on $Q$ in a regular way,
we obtain higher powers of $\omega ^{2}$ and $T^{2}$ in $\Sigma _{R}^{\prime
\prime }$ . As we already mentioned, this form of $\Sigma _{R}^{\prime
\prime }(k\omega )$ yields, upon Kramers-Kronig transformation, a regular
frequency expansion of the real part $\Sigma _{R}^{\prime }(k,\omega
)=A\omega +B\omega ^{3}+...$, where the prefactors are regular functions of $%
T^{2}$ . Of particular importance here is the absence of $\omega T$ term
that would result in a linear-in-$T$ renormalization of the effective mass.
It then follows that non-analytic corrections to $\Sigma _{R}^{\prime }$ can
only emerge if the regular expansion of $\Pi _{R}^{\prime \prime }(Q,\Omega
) $ breaks down for typical momenta that contribute to $\Sigma _{R}^{\prime
\prime }$. This is only possible if $\Pi _{R}^{\prime \prime }(Q,\Omega )$
contains non-analytic terms that break a regular expansion in odd powers of $%
\Omega $, at least for some momenta. The momentum integration should then
show at which order of the expansion in $\Omega $ the prefactor will be
divergent enough to make the momentum integral in (\ref{2.21}) infrared
non-analytic.

We now show that such non-analytic terms do exist and give rise to
non-analytic corrections to the Fermi-liquid behavior. One of non-analytic
corrections comes from the non-analyticity in $\Pi (Q,\Omega )$ at small $Q$%
, another comes from the non-analyticity in $\Pi (Q,\Omega )$ at $Q=2k_{F}$.
We focus on the 2D case and analyze how these two non-analyticities affect
the self-energy. We then show that the non-analytic correction to $\Sigma $
can be viewed equivalently as coming from the non-analyticity in the
particle-particle response function.

\subsection{A non-analytic contribution to the self-energy from $Q=0$}

\label{sec:sigmaQ=0} We begin with the non-analyticity in $\Pi ^{\prime
\prime }(Q,\Omega )$ at small $Q$. Converting (\ref{2.01}) to real
frequencies, we find
\begin{equation}
\Pi _{R}^{\prime \prime }(Q,\Omega )=\left\{
\begin{array}{l}
\frac{m}{2\pi }\frac{\Omega }{((v_{F}Q)^{2}-\Omega ^{2})^{1/2}}\;\mathrm{for}%
\;|\Omega |<v_{F}Q;\nonumber \\
0,\;\mathrm{otherwise.}
\end{array}
\right.  \label{2.2R}
\end{equation}
For notational simplicity we suppress in this subsection the superindex $Q=0$%
. We see that the expansion of $\Pi _{R}^{\prime \prime }$ holds in powers
of $\Omega /v_{F}Q$. Obviously, at some order of the expansion, the momentum
integral becomes infrared non-analytic, which violates the assumption that
momentum and frequency integrals in the diagram for the self-energy are
decoupled.

In $D=2$, this happens already at the leading order in $\Omega $. Indeed,
substituting (\ref{2.2R}) into (\ref{2.2}), linearizing the quasiparticle
dispersion as $\epsilon _{\mathbf{k}+\mathbf{Q}}=\epsilon _{k}+v_{F}Q\cos
\theta $ and integrating first over $\theta $ and then over $Q$ with
logarithmic accuracy, we obtain
\begin{equation}
\Sigma _{R}^{\prime \prime }(\mathbf{k},\omega )=\frac{mU^{2}}{16\pi
^{3}v_{F}^{2}}\int_{-\infty }^{\infty }d\Omega ~\Omega ~\ln {\frac{W^{2}}{%
|\omega -\epsilon _{k}||2\Omega +(\omega -\epsilon _{k})|}}~\left[ \coth
\frac{\Omega }{2T}-\tanh \frac{\omega +\Omega }{2T}\right] .  \label{2.5}
\end{equation}
where $W\sim E_{F}$ is the upper cutoff in the integration over $v_{F}q$. We
see that the momentum integral is infrared-singular and introduces an extra
logarithmic dependence on frequency.

The calculation of $\Sigma _{R}^{\prime \prime }(\mathbf{k},\omega )$ in $%
D=2 $ requires certain care as $\Sigma _{R}^{\prime \prime }(\mathbf{k}%
,\omega ), $ given by Eq.(\ref{2.5}), diverges logarithmically on the mass
shell $\left( \omega =\epsilon _{k}\right) $. However, we will see that this
divergence does not affect the real part of the self-energy at the mass
shell and hence does not affect the specific heat. We therefore proceed in
this subsection with the self-energy (\ref{2.5}) obtained with the
linearized spectrum. In Appendix \ref{app_F}, we consider the mass-shell
singularity in more detail and show that it is in artifact cured by taking
into account either a finite curvature of the electron spectrum or higher
orders in the expansion in $U$.

The frequency integral in (\ref{2.5}) can be evaluated analytically at $T=0$%
. and in the limiting cases at a finite $T$. At $T=0$, Eq.(\ref{2.2})
reduces to (at $\omega >0$)
\begin{equation}
\Sigma _{R}^{\prime \prime }(\mathbf{k},\omega )=\frac{mU^{2}}{8\pi
^{3}v_{F}^{2}}\int_{0}^{\omega }d\Omega ~\Omega ~\ln \frac{W^{2}}{|\omega
-\epsilon _{k}||2\Omega -(\omega -\epsilon _{k})|}~.  \label{sigmams2}
\end{equation}
The integration over $\Omega $ is now straightforward and yields
\begin{equation}
\Sigma _{R}^{\prime \prime }(\mathbf{k},\omega )=\frac{mU^{2}}{16\pi
^{3}v_{F}^{2}}\left[ \left\{ \omega ^{2}-\frac{1}{4}\left( \omega -\epsilon
_{k}\right) ^{2}\right\} \ln \frac{W}{\left| \omega +\epsilon _{k}\right| }%
+\left\{ \omega ^{2}+\frac{1}{4}\left( \omega -\epsilon _{k}\right)
^{2}\right\} \ln \frac{W}{\left| \omega -\epsilon _{k}\right| }\right] +....
\label{sigmams1}
\end{equation}
where the $\dots $ represent the regular $\omega ^{2}$ term. Away from a
near vicinity of $\omega =-\epsilon _{k}$, the term with $(\omega -\epsilon
_{k})^{2}$ is irrelevant (to logarithmic accuracy) and $\Sigma ^{\prime
\prime }(k,\omega )$ can be written as
\begin{subequations}
\begin{eqnarray}
\Sigma _{R}^{\prime \prime }(\mathbf{k},\omega ) &=&\Sigma _{1}^{\prime
\prime }(\mathbf{k},\omega )+\Sigma _{2}^{\prime \prime }(\mathbf{k},\omega
),  \label{sigma12} \\
\Sigma _{1}^{\prime \prime }(\mathbf{k},\omega ) &=&\frac{mU^{2}}{16\pi
^{3}v_{F}^{2}}\omega ^{2}~\ln \frac{W}{|\omega +\epsilon _{k}|},
\label{sigma1} \\
\Sigma _{2}^{\prime \prime }(\mathbf{k},\omega ) &=&\frac{mU^{2}}{16\pi
^{3}v_{F}^{2}}\omega ^{2}\ln \frac{W}{|\omega -\epsilon _{k}|}.
\label{sigma2}
\end{eqnarray}
We see from (\ref{sigma12} that for $\epsilon _{k}\sim \omega $, both terms
scale as $\omega ^{2}\ln \omega $. In particular, at $\epsilon _{k}=0$,
\end{subequations}
\begin{equation}
\Sigma _{R}^{\prime \prime }(\mathbf{k},\omega )=\frac{mU^{2}}{16\pi
^{3}v_{F}^{2}}\omega ^{2}~\ln \frac{W^{2}}{\omega ^{2}}
\end{equation}
Tracing Eq. (\ref{sigma12}) back to (\ref{2.5}), we observe that the first
term $\Sigma _{1}^{\prime \prime }(\mathbf{k},\omega )$ comes from the $%
\Omega $-dependent part of the logarithm in (\ref{2.5}), and the second term
$\Sigma _{2}^{\prime \prime }(\mathbf{k},\omega )$ comes from the $\Omega $%
-independent part of the logarithm. We see that for $\Sigma _{2}^{\prime
\prime }(\mathbf{k},\omega )$, the factorization of the momentum and
frequency integrations still holds, and as in a Fermi liquid, the momentum
integral just adds an overall factor that logarithmically depends on the
external $\omega $ and $\epsilon _{k}$. On the contrary, for $\Sigma
_{1}^{\prime \prime }(\mathbf{k},\omega )$, the momentum and frequency
integrals are coupled.

The zero-temperature result for the self-energy can be also obtained
directly in Matsubara frequencies, without doing the analytic continuation
first. Expanding in small momentum transfer $Q,$ we have for the Matsubara
self-energy at $T=0$,
\begin{eqnarray}
\Sigma (k,\omega _{n})|_{T=0} &=&-\frac{mU^{2}}{8\pi ^{4}}%
\int_{0}^{W/v_{F}}QdQ\int_{-\infty }^{\infty }d\Omega _{m}\int_{0}^{\pi
}d\theta \frac{1}{v_{F}Q\cos \theta +\epsilon _{k}-i(\omega _{n}+\Omega _{m})%
} \\
\times &&~\frac{|\Omega _{m}|}{\sqrt{(v_{F}Q)^{2}+\Omega _{m}^{2}}}.
\end{eqnarray}
The integration over $\theta $ is elementary and yields
\begin{eqnarray}
\Sigma (k,\omega _{n})|_{T=0} &=&-i\frac{mU^{2}}{8\pi ^{3}v_{F}^{2}}%
\int_{-\infty }^{\infty }d\Omega _{m}|\Omega _{m}|\mathrm{sgn}(\omega
_{n}+\Omega _{m}) \\
\times &&~\int_{0}^{W}dx\frac{x}{\sqrt{x^{2}+\Omega _{m}^{2}}\sqrt{%
x^{2}+(\omega _{n}+\Omega _{m}+i\epsilon _{k})^{2}}},
\end{eqnarray}
where we introduce $x=v_{F}Q$. Performing finally the integration over $x$,
we obtain with logarithmic accuracy, for $\omega _{n}>0$,
\begin{eqnarray}
&&\Sigma (k,\omega _{n})|_{T=0} =-i\frac{mU^{2}}{8\pi ^{3}v_{F}^{2}}%
\int_{0}^{\omega _{n}}d\Omega _{m}\Omega _{m}\left( \ln {\frac{W}{\omega
_{n}+i\epsilon _{k}}}+\ln {\frac{W}{2\Omega _{m}+\omega _{n}+i\epsilon _{k}}}%
\right)  \notag \\
&& = -i~\frac{mU^{2}}{16\pi ^{3}v_{F}^{2}}\left[ \left( \omega _{n}^{2}+\frac{1%
}{4}~(\omega _{n}+i\epsilon _{k})^{2}\right) \ln \frac{W}{\omega
_{n}+i\epsilon _{k}}+\left( \omega _{n}^{2}-\frac{1}{4}(\omega +i\epsilon
_{k})^{2}\right) \ln \frac{W}{\omega _{n}-i\epsilon _{k}}\right].
\label{2.50111}
\end{eqnarray}
Continuing to real frequencies, ($i\omega _{n}\rightarrow \omega +i0),$ we
indeed obtain (\ref{sigmams1}) for $\Sigma _{R}^{\prime \prime }$. The
Matsubara self-energy can also be partitioned into $\Sigma _{1}(k,\omega
_{n})$ and $\Sigma _{2}(k,\omega _{n})$. The first term is singular near the
mass surface, while for the second we have (to logarithmic accuracy) for a
generic $\epsilon _{k}/\omega _{n}$,
\begin{equation}
\Sigma _{2}(k,\omega _{n})|_{T=0}=-i\frac{mU^{2}}{16\pi ^{3}v_{F}^{2}}%
~\omega _{n}^{2}\ln \frac{W}{\omega _{n}}
\end{equation}
Continuing to real frequencies, we obtain
\begin{equation}
\Sigma _{2}(k,\omega )|_{T=0}=\frac{mU^{2}}{16\pi ^{3}v_{F}^{2}}\omega ^{2}(-%
\frac{\pi }{2}\text{sgn}\omega +i\ln \frac{W}{|\omega |}).  \label{3.29}
\end{equation}

At finite $T,$ instead of (\ref{sigmams2}) we have
\begin{equation}
\Sigma _{R}^{\prime \prime }(\mathbf{k},\omega )=\frac{mU^{2}}{16\pi
^{3}v_{F}^{2}}\int_{-\infty }^{\infty }d\Omega ~\Omega ~\ln {\frac{W^{2}}{%
|\omega -\epsilon _{k}||2\Omega +(\omega -\epsilon _{k})|}}~\left[ \coth {%
\frac{\Omega }{2T}}-\tanh {\frac{\omega +\Omega }{2T}}\right] .
\label{2.5_1}
\end{equation}
It is again convenient to split the self-energy into two parts, $\Sigma
_{1}^{\prime \prime }(\omega )$ and $\Sigma _{2}^{\prime \prime }(\omega ),$
coming from $\Omega $-dependent and $\Omega $-independent pieces of the
logarithm in (\ref{2.5_1}). For the $\Omega $-independent part of the
logarithm, the frequency integration is the same as in a Fermi liquid, hence
\begin{equation}
\Sigma _{1}^{\prime \prime }(\mathbf{k},\omega )=\frac{mU^{2}}{16\pi
^{3}v_{F}^{2}}~\left[ \omega ^{2}+(\pi T)^{2}\right] ~\ln \frac{W}{|\omega
-\epsilon _{k}|}.  \label{2.51}
\end{equation}
For the second term, we have
\begin{equation}
\Sigma _{2}^{\prime \prime }(\mathbf{k},\omega )=\frac{mU^{2}}{16\pi
^{3}v_{F}^{2}}\int d\Omega ~\Omega ~\ln \frac{W}{|2\Omega +(\omega -\epsilon
_{k})|}~\left[ \coth \frac{\Omega }{2T}-\tanh \frac{\omega +\Omega }{2T}%
\right] .  \label{2.52}
\end{equation}
In this last term, the dependence on the ratio $\omega /\epsilon _{k}$ is
not singular and can be neglected, to logarithmic accuracy. Using series
representations for the hyperbolic functions we can then re-express the
r.h.s. of (\ref{2.52}) as
\begin{equation}
\Sigma _{2}^{\prime \prime }(\omega )=-\frac{mU^{2}}{16\pi ^{3}v_{F}^{2}}%
\left( [(\pi T)^{2}+\omega ^{2}]\ln (T/{\bar{A}})+\omega ^{2}~f\left( \frac{%
\omega }{\pi T}\right) \right) ,  \label{2.6}
\end{equation}
where ${\bar{A}}$ is a constant, and
\begin{equation}
f(x)=0.79+\mathcal{P}\int dy~\tanh \frac{\pi xy}{2}~\left( y\ln \frac{y^{2}}{%
|y^{2}-1|}+\frac{1}{y}-\ln \frac{y+1}{|y-1|}\right) .  \label{2.7}
\end{equation}
One can easily make sure that the expansion of $f(x)$ holds in even powers
of $x$. %The plot of $f(x)$ is presented in Fig.~\ref{fig2}.
At large $x$, $f(x)\approx \ln x$, i.e., at $\omega \gg T$, this expression
reproduces $\Sigma ^{\prime \prime }(\omega )\propto \omega ^{2}\ln \omega $%
. At small $x$, i.e., at $\omega \ll T$, $f(x)\approx 0.79+0.35x^{2}$.

\begin{figure}[tbp]
\centerline{\epsfxsize=5in \epsfbox{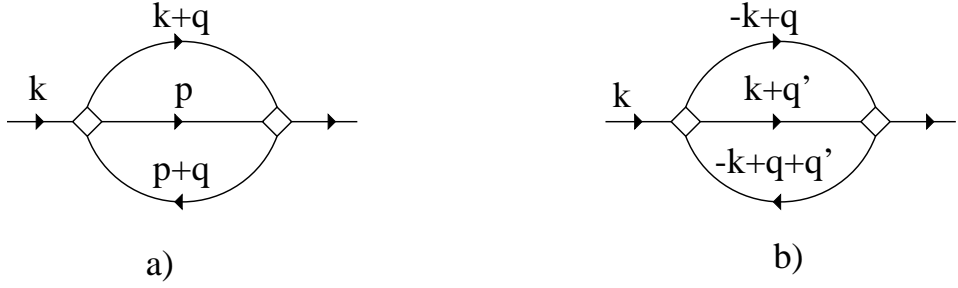}}
\caption{a) $q=0$ contribution to the self-energy; b) $q=2k_F$ contribution
to the self-energy.}
\label{fig3}
\end{figure}

\subsection{A non-analytic contribution to the self-energy from $q\approx
2k_{F}$}

\label{sec:sigmatwokf}

We next consider a singular contribution to $\Sigma _{R}^{\prime \prime }(%
\mathbf{k},\omega )$ from momentum transfers close to $2k_{F}$. To perform
computations along the same lines as for Q near $0$, we would need to know
the form of $\Pi (Q,\Omega )$ at finite $\Omega $ and $T$, which is rather
involved. However, we actually would not need this form at all, as we
demonstrate that the contribution to the self-energy from $Q\approx 2k_{F}$
is \textit{exactly the same} as $\Sigma _{2}(\mathbf{k},\omega )$ defined in
(\ref{sigma12}). The most straightforward way to see this is to go back to a
diagram representation of the self-energy in terms of three fermionic
propagators (Fig.\ref{fig1}c). In analytical form, the $^{\prime \prime
}q=0^{\prime \prime }$ contribution to the self-energy is
\begin{equation}
\Sigma ^{q=0}(k)=U^{2}\int \frac{d^{D+1}q}{\left( 2\pi \right) ^{D+1}}\int
\frac{d^{D+1}p}{\left( 2\pi \right) ^{D+1}}G_{k+q}G_{p}G_{p+q}.
\end{equation}
where $q$ is assumed to be small. We again use ``relativistic'' notation $%
k\equiv \left( \mathbf{k,}\omega \right) $ and $q\equiv \left( \mathbf{Q,}%
\Omega \right) $. Integrating over $p$ first, we obtain
\begin{equation}
\Sigma ^{q=0}=-U^{2}\int \frac{d^{D+1}q}{\left( 2\pi \right) ^{D+1}}%
G_{k+q}\Pi (q),
\end{equation}
where $\Pi (q)$ is a particle-hole bubble at small momentum and frequency.
This expression we used in the previous subsection. We found that two
singular contributions to $\Sigma ^{q=0}$, $\Sigma _{1}(\mathbf{k},\omega )$
and $\Sigma _{2}(\mathbf{k},\omega )$, and that $\Sigma _{2}(\mathbf{k}%
,\omega )$ comes from the momentum region where two of the internal momenta
are close to $-\mathbf{k}$ and the third one is close to $\mathbf{k}$, i.e.,
from the range of $p$ which are nearly antiparallel to $k$. Since $p+k$ are
small (of order of external momenta), we can relabel the momenta as shown in
Fig. \ref{fig3}b and re-express $\Sigma _{2}(\mathbf{k},\omega )$ as
\begin{equation}
\Sigma _{2}(k)=U^{2}\int \frac{d^{D+1}q}{\left( 2\pi \right) ^{D+1}}\int
\frac{d^{D+1}q^{\prime }}{\left( 2\pi \right) ^{D+1}}G_{k+q}G_{-k+q^{\prime
}}G_{-k+q+q^{\prime }}.
\end{equation}
where now \textit{both} $q$ and $q^{\prime }$ are small. Integrating over $%
q^{\prime }$ first, we obtain a conventional expression for $\Sigma _{2}(k)$
in terms of the polarization operator with small momentum transfer. On the
other hand, changing the order of integration and integrating over $q$
first, we obtain
\begin{equation}
\Sigma _{2}(k)=-U^{2}\int \frac{d^{D+1}q^{\prime }}{\left( 2\pi \right)
^{D+1}}~G_{-k+q^{\prime }}{\tilde{\Pi}}({2k-q^{\prime }})  \label{2.0001}
\end{equation}
where
\begin{equation}
{\tilde{\Pi}}(2k-q^{\prime })=-\int \frac{d^{D+1}q}{\left( 2\pi \right)
^{D+1}}G_{k+q}G_{-k+q+q^{\prime }}.
\end{equation}
In general, ${\tilde{\Pi}}(2k-q)$ is not equivalent to the polarization
bubble $\Pi (q)$ with momentum near $2k_{F}$, as our re-writing is only
valid if internal $q$ are small. However, the singular parts of the two
bubbles coincide because the singular part in $\Pi (Q\approx 2k_{F},\Omega )$
(proportional to $\sqrt{|Q-2k_{F}|}\theta (|Q-2k_{F}|)$ in the static case)
comes from the momentum range where the two internal momenta in the
particle-hole bubble are close to $\pm \mathbf{k}$, i.e., from exactly the
same range that is covered in ${\tilde{\Pi}}(2k-q^{\prime })$. We show this
explicitly in the Appendix \ref{app_A}. This equivalence implies that the
r.h.s. of (\ref{2.0001}) is just the singular part of the ``$2k_{F}$''
contribution to the self-energy. We see therefore that $\Sigma
^{q=2k_{F}}(k)=\Sigma _{2}^{q=0}(k)$. The total self-energy is then
\begin{equation}
\Sigma (k)=\Sigma ^{q=0}(k)+\Sigma ^{q=2k_{F}}=\Sigma _{1}(k)+2\Sigma
_{2}(k).
\end{equation}
For momentum-dependent interaction $U=U(q)$, the computation of the $2k_{F}-$%
contribution requires more care and we present it in Sec. VII.

That the $2k_{F}$-singularity comes from nearly antiparallel internal
fermionic momenta has been implicitly used in the Kohn-Luttinger analysis of
superconducting instability with large angular momenta~of Cooper pairs\cite
{kl}. In the context of corrections to the Fermi-liquid theory, Belitz et
al.~\cite{bkv} argued that all singular contributions to the spin
susceptibility can be described as small $q$ effects, although they did not
emphasize that some of their small $q$ effects are in fact equivalent to $%
2k_{F}$ contributions in conventional notations.

That both $q=0$ and $q=2k_{F}$ singularities in the polarization bubble
contribute to the self-energy was first emphasized by CM \cite{millis}.
However, the relative sign of the two terms is different in their and our
calculations. We found that the singular terms add, while they argued that
singular contributions from $q=0$ and $q=2k_{F}$ cancel each other. Since
the interplay between $q=0$ and $q=2k_{F}$ contributions to the self-energy
is crucial to the issue of whether or not there is a $T^{2}$-term in the
specific heat and linear-in-$T$ term in the effective mass (CM argued that
both are absent due to cancelation between $q=0$ and $q=2k_{F}$ terms), we
present in Appendix B \ref{app_B} an explicit computation of the $2k_{F}$
contribution to the second-order self-energy at $T=0$. This calculation
confirms that $\Sigma ^{q=2k_{F}}=\Sigma ^{q=0}$.

\subsection{An alternative analysis, in terms of $\Pi_{pp} (Q, \Omega)$}

\label{sec:sigmacooper} We next demonstrate that the backscattering
non-analyticity in the fermionic self-energy can be viewed equivalently as
coming from the non-analyticity in the particle-particle bubble at small
total momentum and frequency. This readily follows from our consideration of
the ``$2k_{F}$'' diagram. Indeed, since both $q$ and $q^{\prime }$ are
small, the full self-energy can be re-expressed as
\begin{equation}
\Sigma (k)=-U^{2}\int \frac{d^{D+1}q}{\left( 2\pi \right) ^{D+1}}\int \frac{%
d^{D+1}q^{\prime }}{\left( 2\pi \right) ^{D+1}}G_{-k+q+q^{\prime }}\Pi
_{pp}(q+q^{\prime }).
\end{equation}
Performing the same analysis as in the previous section, we observe that the
deviation from the Fermi-liquid form of $\Sigma $ is only possible if the
expansion of $\Pi _{pp}^{\prime \prime }(Q,\Omega )$ in odd powers of $%
\Omega $ breaks down due to infrared divergences of momentum dependent
prefactors. This is precisely what happens in $\Pi _{pp}(Q,\Omega )$ given
by (\ref{2.03}) as the frequency expansion holds in $\Omega /v_{F}Q$, i.e.,
the prefactors are non-analytic at vanishing $Q$. We emphasize that the
logarithmic divergence of $\Pi _{pp}(\mathbf{Q},\Omega )$ at vanishing $Q$
and $\Omega $ is by itself not essential; what matters is a non-analytic
dependence on the ratio $\Omega /v_{F}Q$.

We see, therefore, that the non-analytic piece in the self-energy can be
viewed equivalently as coming from a non-analyticity in the particle-hole
bubble, \textit{or} from a non-analyticity in the particle-particle bubble.
To further verify this, we explicitly compute in Appendix \ref{app_B} the
non-analytic part of $\Sigma (k)$ at $T=0$ using the ``particle-particle
formalism'' and indeed find it to be equal to the non-analytic $2\Sigma
_{2}(k,\omega )$ that we obtained in the ``particle-hole formalism,'' i.e.,
\begin{equation}
\Sigma _{pp}^{(Q=0)}(k)=2\Sigma _{2}(k).
\end{equation}
The term $\Sigma _{1}(k,\omega )$ can be also reproduced in the
particle-particle formalism, but this contribution comes from large $%
q+q^{\prime }\approx 2k$, and we refrain from re-deriving this piece.

Our results on this issue again disagree with those by CM~ \cite{millis}.
They performed a complimentary analysis of the self-energy based on the
evaluation of an effective vertex function to second order in $U$, and
argued that there is a cancelation between non-analytic contributions
coming from the $2k_{F}$ non-analyticity in the particle-hole channel and
the non-analyticity in the particle-particle channel. We, on the contrary,
find that the contribution from the particle-particle non-analyticity is
twice the ``$2k_{F}$'' contribution from the particle hole channel.

Summarizing the results of the last two subsections, we see that the
non-analytic part of the fermionic self-energy in 2D consists of two parts.
The first part, $\Sigma _{1}^{\prime \prime }(k)$, comes from forward
scattering. It has the same functional form, $\omega ^{2}+(\pi T)^{2}$, as
in a Fermi liquid, but the prefactor logarithmically depends on $\omega
-\epsilon _{k}$. The second part, $\Sigma _{2}^{\prime \prime }(k)$, comes
from the processes which involve the scattering amplitude with near-zero
total and transferred momentum. This $\Sigma _{2}^{\prime \prime }(k)$ has a
non-Fermi-liquid form, and can be equally attributed to the $Q=0$
non-analyticity in the particle-hole polarization bubble, or to the $2k_{F}$
non-analyticity in the same bubble, or to the $Q=0$ singularity in the
particle-particle bubble. In the next section we show that only $\Sigma
_{2}(k)$ actually contributes to the thermodynamics.

\subsection{Effective Mass and Specific Heat}

\label{sec:mstarCofT} We first use the result for $\Sigma ^{\prime \prime }$
obtained in Sec.\ref{sec:sigmaQ=0} and compute the real part of the
self-energy on the mass shell. We then use $\Sigma ^{\prime }(\omega
=\epsilon _{k})$ to find the effective mass and specific heat.

The Kramers-Kronig relation on the mass shell is
\begin{equation}
\Sigma _{R}^{\prime }(\omega )=\frac{1}{\pi }\mathcal{P}\int dE\frac{\Sigma
^{\prime \prime }(E,\epsilon _{k}=\omega )}{E-\omega }.  \label{2.71}
\end{equation}
We begin with $\Sigma _{1}(k)$. Substituting $\Sigma _{1}^{\prime \prime }(%
\mathbf{k},\omega )$ from (\ref{sigma1})into (\ref{2.71}), we find that on
the mass shell
\begin{equation}
\Sigma _{1}^{\prime }(k,\omega )|_{\omega =\epsilon _{k}}=\frac{mU^{2}}{%
16\pi ^{4}v_{F}^{2}}\mathcal{P}\int_{-\infty }^{\infty }dz\frac{z^{2}+(\pi
T)^{2}}{z-\omega }~\ln \frac{W}{|z-\omega |}.  \label{KK}
\end{equation}
By dimensional analysis, the integral in (\ref{KK}) is of order $\omega ^{2}$%
. However, the prefactor in front of $\omega ^{2}$ turns out to be zero. The
easiest way to see this is to evaluate the integral in finite limits $-W<z<W$
and to search for the universal term that would be independent of $W$.
Performing elementary manipulations, we find that $\Sigma _{1}^{\prime
}(\omega )$ does not contain such a term. Foreshadowing, we note that the
same result holds for the static spin susceptibility which we discuss in
detail in Sections \ref{sec:chiQ} and \ref{sec:chiT}. We will see there that
the inclusion of the $\Sigma _{2}(k,\omega )$ into a particle-hole bubble
with external momentum $Q$ yields a non-analytic $|Q|$ term in $\chi _{s}(Q)$%
. On the contrary, the susceptibility diagram with an extra $\Sigma _{1}(%
\mathbf{k},\omega )$ scales, in Matsubara frequencies, as
\begin{equation}
\delta \chi \propto \int d\omega _{n}\omega _{n}\int d\epsilon _{k}\frac{\ln %
\left[ W/\left( \epsilon {_{k}-}i\omega _{n}\right) \right] }{(\epsilon
_{k}-i\omega _{n})^{2}\left[ (\epsilon _{k}-i\omega _{n})^{2}-(v_{F}Q)^{2}%
\right]}.
\end{equation}
By power counting, the leading $Q$ dependence of the integral should be $|Q|$%
. However, a straightforward computation shows that the prefactor again
vanishes. The outcome of this analysis is that the divergence of $\Sigma
_{1}^{\prime \prime }(\mathbf{k},\omega )$ on the mass shell does not give
rise to non-analytic corrections to Fermi-liquid form of the thermodynamic
observables.

We next consider $\Sigma _{2}^{\prime }(\mathbf{k},\omega )$. Substituting $%
\Sigma _{2}^{\prime \prime }(\mathbf{k},\omega )$ from Eq.(\ref{2.5_1}) into
Eq.(\ref{2.71}), we obtain after simple manipulations
\begin{eqnarray}
\Sigma _{2}^{\prime }(\omega ) &=&-\frac{mU^{2}}{16\pi ^{4}v_{F}^{2}}~\omega
~\int_{-\infty }^{\infty }d\Omega \Omega \text{ }\mathcal{P}\int_{0}^{\infty
}\frac{dE}{E^{2}-\omega ^{2}}~\left( \coth \frac{\Omega }{2T}-\tanh \frac{%
\Omega +E}{2T}\right)  \notag \\
&&~\left( \frac{E}{\omega }\ln \left| \frac{2\Omega +E-\omega }{2\Omega
+E+\omega }\right| +\ln \frac{|(2\Omega +E)^{2}-\omega ^{2}|}{W^{2}}\right).
\label{2.712}
\end{eqnarray}
Integrations over $\Omega $ and $E$ can be performed exactly. We give the
details of this calculation in Appendix~\ref{app_se} and present just the
results here. At $T=0$, we obtain
\begin{equation}
\Sigma _{2}^{\prime }(\omega )=-\frac{mU^{2}}{32\pi ^{2}v_{F}^{2}}\omega
|\omega |.  \label{2.713}
\end{equation}
This coincides with Eq.(\ref{3.29}) obtained via analytic continuation of
the Matsubara self-energy.

In the opposite limit of small $\omega /T$, we have
\begin{equation}
\Sigma _{2}^{\prime }(\omega )=~-~\frac{mU^{2}\ln 2}{8\pi ^{2}v_{F}^{2}}%
~~\omega T.  \label{2.714}
\end{equation}
As the self-energy in this region is linear in $\omega ,$ Eq. (\ref{2.714})
implies that the effective mass of subthermal quasiparticles, i.e., with $%
\omega \ll T,$ scales linearly with $T$. Using the fact that the full $%
\Sigma (k,\omega )=\Sigma _{1}(k,\omega )+2\Sigma _{2}(k,\omega )$ and that $%
\Sigma _{1}(k,\omega )$ does not contribute to thermodynamics, we obtain
\begin{equation}
m^{\ast }(T)=m^{\ast }(T=0)\left( 1-2\ln 2\left( \frac{mU}{4\pi }\right) ^{2}%
\frac{T}{E_{F}}\right) .  \label{2.11}
\end{equation}
This result disagrees with CM--they argued that the linear-in-$T$
term in the mass renormalization is absent. 

In a very  recent
study Das Sarma, Galitski, and Zhang \cite{dassarma_mass} did
find a linear-in-T correction to the effective mass for 
 the Coulomb interaction in $D=2$. Although the sign of
their linear-in-$T$ term is opposite to
that in Eq.(\ref{2.11}), we believe that there is no contradiction
here as there are no general restrictions on the
 sign of the prefactor.  It is therefore quite possible  that the
 sign of the $O(T)$ term  is
different for short- and long-range interactions. Note in this regard
 that the effect of the interaction on the effective mass is
different for these two cases even at $T=0$ : a short-range
interaction increases $m^{*}$, while the Coulomb interaction decreases
$m^*$ in the limit $r_s\ll 1$ \cite{agd}.

For generic $\omega /T$, the non-analytic part of the full $\Sigma ^{\prime
}(\omega )$ can be cast into the following scaling form
\begin{equation}
\Sigma ^{\prime }(\omega )=-\frac{mU^{2}}{16\pi ^{2}v_{F}^{2}}\omega |\omega
|g\left( \frac{|\omega |}{T}\right) ,  \label{2.715}
\end{equation}
where
\begin{equation}
g(x)=1+\frac{4}{x^{2}}\left[ \frac{\pi ^{2}}{12}+\text{Li}_{2}\left(
-e^{-x}\right) \right],  \label{2.111}
\end{equation}
and Li$_{2}(x)$ is a polylogarithmic function.

Note that $g(\infty )=1$ and $g(x\ll 1)\approx 4\ln 2/x$. Substituting these
limiting expressions into Eq.(\ref{2.715}) we indeed reproduce Eq.(\ref
{2.713}) and Eq.(\ref{2.714}).

The full functional form of $g(x)$ is required for the computation of the
specific heat, as the frequency integral for $C(T)$ given by Eq.(\ref{sp_he}%
) is confined to $\omega \sim T$. Previous work ~\cite{previous_C(T)} on $%
C(T)$ used only the $T=0$ form of the self-energy and hence yielded
incorrect prefactors. Substituting our result for $\Sigma ^{\prime }$ into (%
\ref{sp_he}) we obtain in 2D
\begin{equation}
\delta C(T)= C_{FL} \frac{48 K}{\pi }~\left( \frac{mU}{4\pi }\right) ^{2}%
\frac{T}{E_{F}},  \label{2.10}
\end{equation}
where $C_{FL}(T)=m \pi T /3$ is the Fermi gas result for the specific heat
and
\begin{equation}
K=\int_{0}^{\infty }\frac{dxx}{\cosh ^{2}x}~\left[ x^{2}+\frac{\pi ^{2}}{12}+%
\text{Li}_{2}\left( -e^{-2x}\right) \right] =1.803.
\end{equation}

As it was to be anticipated, the non-analytic correction to the fermionic
self-energy gives rise to the $T^{2}$-term in the specific heat. It is
essential that this non-analytic term comes only from fermions in a near
vicinity of the Fermi surface and is thus model-independent. The same is
true for the linear-in-$T$ correction to the effective mass. In other words,
the leading corrections to the Fermi-liquid forms of $m$ and $C(T)$ are
fully universal.

The $T^{2}$-dependence of the correction to the specific heat agrees with
the results by Coffey and Bedell~\cite{bedell} and Misawa~\cite{misawa}.
However, Coffey and Bedell did not explicitly compute the prefactor and
apparently only included small momentum transfers (i.e., no $2k_{F}$
effects). Misawa did compute the prefactor, but he neglected the temperature
dependence of the fermionic self-energy. We found above that this $T$
dependence cannot be neglected, and our prefactor disagrees with that by
Misawa.

\subsection{Amplitude of quantum magneto-oscillations}

\label{sec:dhva} In previous sections, we found the general form of
non-analytic corrections to the real and imaginary parts of the self-energy.
We now discuss whether these corrections can be observed experimentally via
magneto-oscillations. Naively speaking, one might have expected the finite
quasiparticle relaxation rate, $T^{2}\ln T,$ to damp the amplitude of the
oscillations as a contribution to the ``Dingle temperature'', whereas the $%
T- $ dependent effective mass might affect the thermal smearing factor.
However, we argue below that quadratic and quadratic-times-log terms in the
self-energy are not detectable by measuring the amplitude of
magneto-oscillations in $D=2$.

In the Luttinger formalism \cite{luttinger}, the amplitude of the $k^{th}-$
harmonic of magneto-oscillations is given by
\begin{equation}
A_{k}=\frac{4\pi ^{2}kT}{\Omega _{c}}\sum_{\omega _{n}>0}\exp \left( -\frac{%
2\pi k\left[ \omega _{n}-i\Sigma \left( \omega _{n},T\right) \right] }{%
\Omega _{c}}\right) ,  \label{amplitude}
\end{equation}
where $\Omega _{c}$ is the cyclotron frequency. It is essential for our
consideration that the amplitude is determined by the self-energy in the
Matsubara representation rather than by the real and imaginary parts of the
retarded self-energy \cite{engelsberg}. By itself, $\Sigma _{R}^{\prime }$
and $\Sigma _{R}^{\prime \prime }$ determine the fermion dispersion and
lifetime, respectively; however in (\ref{amplitude}) this distinction is
lost.

The assumption made in deriving (\ref{amplitude}) is that the dependence of
the self-energy on the magnetic field can be neglected. In 3D, this
assumption is well justified as the effect of the magnetic field on the
self-energy yields corrections to $A_{k}$ which are small in $1/\sqrt{N},$
where $N=\epsilon _{F}/\Omega _{c}\gg 1$ is the total number of Landau
levels. In 2D, however, the effect of the magnetic field is
non-perturbative, and at $T=0$ and in the absence of disorder, the
field-induced oscillations of the self-energy are as important as the
oscillations of the thermodynamic potential itself \cite{curnoe}. Eq.(\ref
{amplitude}) is then only applicable as long as oscillations of the
thermodynamic potential are exponentially small due to either finite
temperature and/or disorder. In this paper we disregard effects of disorder
(considered recently in \cite{maslov}), thus the amplitude is only
controlled by the finite temperature. In this case, the restriction of the
small amplitude in its turn implies that the sum over Matsubara frequencies
in (\ref{amplitude}) can be truncated to only the $n=0$ term. Notice that
this restriction is mandatory in $D=2$ within the Luttinger formalism but
depends on the choice of experimental conditions in $D=3.$ The amplitude of
the first (largest) harmonic then simplifies to
\begin{equation}
A_{1}=\frac{4\pi ^{2}T}{\Omega _{c}}\exp \left( -\frac{2\pi \left[ \pi
T-i\Sigma \left( \pi T,T\right) \right] }{\Omega _{c}}\right) .  \label{bbbb}
\end{equation}
The temperature enters the Matsubara self-energy $\Sigma \left( \omega
_{n},T\right) $ in two ways: first, as the Matsubara frequency, and second,
as the physical temperature determining the thermal distribution of the
degrees of freedom. For the lowest frequency, $\omega _{0}=\pi T$, the
interplay between the two effects leads to a peculiar cancelation.

Indeed, consider for a moment a generic Fermi liquid, for which
\begin{equation}
\Sigma \left( \omega _{n},T\right) =\left( \frac{m^{\ast }}{m}-1\right)
i\omega _{n}+iC\left[ \left( \pi T\right) ^{2}-\omega _{n}^{2}\right] +\dots
,  \label{se3D}
\end{equation}
where $C$ is a constant, $\dots $ stand for the higher order terms $\left[
\mathcal{O}\left( \epsilon _{n}^{3},T^{3}\right) \right] $, and $m^{\ast }/m$
has a regular expansion in powers of $T^{2}$. The analytic continuation of (%
\ref{se3D}) to real frequencies yields the correct retarded self-energy (\ref
{int1}). We see that the second term $\Sigma \left( \omega _{n},T\right) $
vanishes for $\omega _{n}=\pm \pi T$, i.e., the self-energy that enters into
the formula for $A_{k}$ contains terms only of order $T^{3}$ and higher. In
other words, the quadratic in $T$ piece present in the imaginary part of the
retarded self-energy and associated observables, does not affect the
amplitude of magneto-oscillations, which to order $T^{3}$ is given by
\begin{equation}
A_{1}=\frac{4\pi ^{2}T}{\Omega _{c}}\exp \left( -\frac{2\pi ^{2}T}{\Omega
_{c}^{\ast }}\right) ,\text{ }\Omega _{c}^{\ast }\equiv \frac{m}{m^{\ast }}%
\Omega _{c},
\end{equation}
where $m^{\ast }/m$ is a regular mass renormalization which comes from
fermions far away from the Fermi surface. This rather remarkable result was
previously obtained specifically for electron-phonon interaction and is
known as a ``Fowler-Prange theorem'' \cite{prange}.

We found that a similar cancelation occurs also for our self-energy in $D=2$%
. To logarithmic accuracy, the second term in Eq.(\ref{se3D}) is replaced by
\begin{equation}
\tilde{\Sigma}\left( \omega _{n},T\right) =-i{\tilde{C}}~T\sum_{\Omega _{m}}%
\text{sgn}\left( \omega _{n}-\Omega _{m}\right) \left| \Omega _{m}\right|
\ln \frac{\left| \Omega _{m}\right| }{W},
\end{equation}
where ${\tilde{C}}$ is a real constant, and the factor of sgn$\left( \omega
_{n}-\Omega _{m}\right) $ resulted from the angular integration of the
Green's function. A simple transformation of the Matsubara sum reduces $%
\tilde{\Sigma}\left( \omega _{n},T\right) $ to
\begin{equation}
\tilde{\Sigma}\left( \omega _{n},T\right) =-2iT{\tilde{C}}~\sum_{\Omega
_{m}=0}^{\omega _{n}-\pi T}\Omega _{m}\ln \frac{\Omega _{m}}{W}.
\label{3.50}\end{equation}
Expression (\ref{3.50}) obviously vanishes for $\omega _{n}=\pi T$, i.e., therefore
$\Sigma \left( \pi T,T\right) $ in (\ref{bbbb}) does not contain a
contribution from $\tilde{\Sigma}.$ Due to this cancelation, the
exponential factor in $A_{1}$ does not contain terms of order $T^{2}\ln T$.
A more detailed analysis \cite{maslov}, shows that $T^{2}$ terms are also
absent, i.e., both quadratic terms and quadratic-times-log terms in the
self-energy (and thus the linear-in-$T$ effective mass [Eq.(\ref{2.11}) ])
are not observable in a magneto-oscillation experiments.

\section{spin and charge susceptibilities}

\label{sec:chi} We next proceed to the analysis of the corrections to the
Fermi-liquid forms of spin and charge susceptibilities.

The charge and spin operators are bilinear combinations of fermions:
\begin{equation}
C(q)=\sum_{\mathbf{k},\alpha }c_{\mathbf{k}+\mathbf{q},\alpha }^{\dagger }c_{%
\mathbf{k},\alpha }
\end{equation}
for charge, and
\begin{equation}
{\vec{S}}(q)=\sum_{\mathbf{k},\alpha ,\beta }{\vec{\sigma}}_{\alpha \beta
}c_{\mathbf{k}+\mathbf{q},\alpha }^{\dagger }c_{\mathbf{k},\beta }
\end{equation}
for spin. The corresponding susceptibilities for a system of interacting
fermions are given by fully renormalized particle-hole bubbles with side
vertices
\begin{equation}
\Gamma _{c}=\delta _{\alpha ,\beta };~~\Gamma _{s}^{i}=\sigma _{\alpha
,\beta }^{i},
\end{equation}
where $c$ and $s$ refer to charge and spin, respectively.

For non-interacting fermions, the spin and charge susceptibilities are equal
and given by the Lindhard function that coincides, up to an overall factor,
with the polarization operator $\Pi (Q,\Omega _{m})$:
\begin{equation}
\chi _{0}^{c}(Q,\Omega _{m})=\chi _{0}^{s}(Q,\Omega _{m})=2\Pi (Q,\Omega
_{m}),
\end{equation}
where $\chi _{0}^{s}(Q,\Omega _{m})\equiv \left[ \chi _{0}^{s}(Q,\Omega _{m})%
\right] _{ii}$ and $i=1,2,3,$ and
\begin{eqnarray}
\Pi (Q,\Omega _{m}) &=&-T\sum_{m}\int \frac{d^{D}k}{(2\pi )^{d}}~  \notag \\
&&G_{0}(\mathbf{k},\omega _{n})~G_{0}(\mathbf{k}+\mathbf{Q},\omega
_{n}+\Omega _{m}).
\end{eqnarray}

At $T=0,$the charge and spin susceptibilities can be evaluated exactly for
any $Q$ and $\Omega _{m}.$ In the static limit, $\Omega _{m}=0,$ they
acquire particularly simple forms. For $D=3$, we have~\cite{ashcroft_mermin}
\begin{equation}
\chi _{0}^{c}(Q,0)=\chi _{0}^{s}(Q,0)=\chi _{0}^{3D}\left[ \frac{1}{2}+\frac{%
4k_{F}-Q^{2}}{8Qk_{F}}\ln {\frac{Q+2k_{F}}{|Q-2k_{F}|}}\right] ,
\label{lind3D}
\end{equation}
where $\chi _{0}^{3D}=mk_{F}/\pi ^{2}$. In $D=2$, the corresponding
expression is~\cite{kagan,stern_1}
\begin{eqnarray}
\chi _{0}^{c}(Q,0) &=&\chi _{0}^{s}(Q,0)=\chi _{0}^{2D},~~Q<2k_{F};  \notag
\\
\chi _{0}^{c}(Q,0) &=&\chi _{0}^{s}(Q,0)=\chi _{0}^{2D}\left[ 1-\left( 1-%
\frac{4k_{F}^{2}}{Q^{2}}\right) ^{1/2}\right] ,~~Q>2k_{F},  \label{lind2D}
\end{eqnarray}
where $\chi _{0}^{2D}=m/\pi $. In 1D, we have~\cite{DL_1}
\begin{equation}
\chi _{0}^{c}(Q,0)=\chi _{0}^{s}(Q,0)=\chi _{0}^{1D}\frac{k_{F}}{Q}~\ln
\left| \frac{k_{F}+\frac{Q}{2}}{k_{F}-\frac{Q}{2}}\right| ,  \label{lind1D}
\end{equation}
where $\chi _{0}^{1D}=2/(\pi v_{F}).$ As it was mentioned in the
Introduction, $\chi _{0}^{c,s}(Q,0)$ is analytic in $Q$ for small $Q$ in all
dimensions.

The first nontrivial corrections to $\chi _{0}^{c,s}(Q,0)$ come from the
diagrams presented in Fig.\ref{fig:diag}. These diagrams represent
self-energy and vertex-correction insertions into the bare particle- hole
bubble~\cite{bkv}. Diagrams $1-5$ are nonzero for both $\chi _{s}$ and $\chi
_{c}$. Diagrams $6$ and $7$ are finite for $\chi _{c}$, but vanish for $\chi
_{s}$ upon the spin summation ($\sum_{\alpha }\sigma _{\alpha \alpha }^{i}=0$%
). The internal parts of all diagrams contain fermionic bubbles:
particle-hole bubbles for diagrams $1,2,3,5$ and particle-particle bubble
for the diagram $4$.

In the next two sections we analyze the form of the static susceptibility
first at a finite $Q$ and zero temperature, and then at finite $T$ and $Q=0$.

\subsection{Spin and charge susceptibilities at finite $Q$ and $T=0$}

\label{sec:chiQ}

\begin{figure}[tbp]
\centerline{\epsfxsize=6in \epsfbox{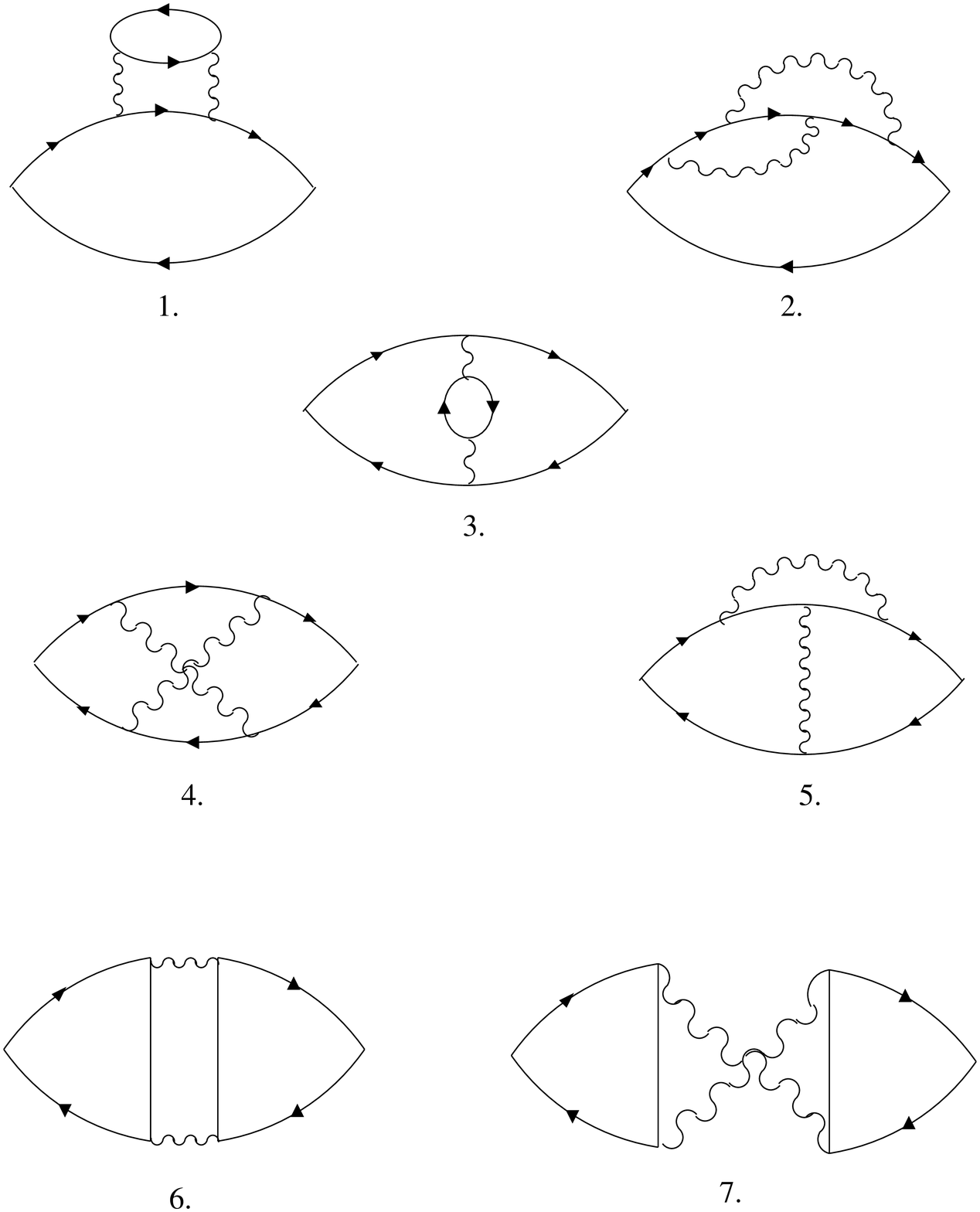}}
\caption{Each of \ the seven diagrams in this figure give singular
corrections to spin and charge susceptibilities. }
\label{fig:diag}
\end{figure}

As in Sec. III, we assume that the interaction is independent of momentum.
We explicitly computed all $7$ diagrams Fig. \ref{fig:diag} and found that
each of the diagrams (except for diagrams 6 and 7 which vanish identically
for the spin channel) contributes a correction $\delta \chi (Q,0)\propto |Q|$%
, and that this non-analyticity is a direct consequence of the dynamical
singularities in the particle-hole and particle-particle bubbles.

\subsubsection{$D=2$}

\label{sec:chiQ2D} As we mentioned in the Introduction, the calculation in $%
D=2$ is more difficult to perform than in $D=3$ because all typical internal
momenta and energies are of the same order as the external ones ($Q$ and $%
v_{F}Q$, respectively); thus no expansion is possible. In 3D, where $\delta
\chi (Q,0)\propto Q^{2}\ln Q$, typical internal momenta are larger than
external $Q$, and one could expand the integrand in $Q^{2}$ and evaluate the
prefactor to logarithmic accuracy.

We begin with diagram 1 which represents the self-energy insertion into the
particle-hole bubble. This diagram yields the same contribution for spin and
charge channels, so we will drop the subscript and denote $\chi _{1}\equiv
\chi _{1s}=\chi _{1c}$.

An analytic form of diagram 1 in the Matsubara representation is given by
\begin{equation}
\delta \chi _{1}(Q,0)=-8U^{2}\int \frac{d^{2}k~d^{2}q~d\omega d\Omega }{%
(2\pi )^{6}}G_{0}^{2}(\mathbf{k},\omega )G_{0}(\mathbf{k}+\mathbf{Q},\omega
)G_{0}(\mathbf{k}+\mathbf{q},\omega +\Omega )\Pi (q,\Omega ).  \label{c1}
\end{equation}
The combinatorial factor of $8$ includes two factors of $2$ due to spin
summation and an extra factor of $2$ associated with the fact that the
self-energy can be added to any of the two fermionic lines in the bubble.
Non-analytic contributions to $\delta \chi _{1}(Q,0)$ come from two regions
of momentum transfers: $q$ near zero and $q$ near $2k_{F}$. Since we have
already shown in Sec.~\ref{sec:selfenergy} that the contributions to the
self-energy from these two regions are equal for a contact interaction (up
to a forward scattering piece in $\Sigma ^{q=0}$ that, as we demonstrated,
does not contribute to $|Q|$ term in the susceptibility), we do not have to
calculate the $q=0$ and $q=2k_{F}$ contributions to $\chi _{1}(Q,0)$
separately--the two are just equal:
\begin{equation}
\delta \chi _{1}^{q=0}(Q,0)=\delta \chi _{1}^{q=2k_{F}}(Q,0).  \label{cccc}
\end{equation}
This implies that we only have to compute $\delta \chi _{1}^{q=0}(Q,0)$, the
full $\delta \chi _{1}(Q,0)$ will be twice that value. To be on a safe side,
we verified this reasoning by explicitly computing $\delta \chi
_{1}^{q=2k_{F}}(Q,0)$. We present the calculations in Appendix \ref{app_C}.
We indeed found it to be equal to $\delta \chi _{1}^{q=0}(Q,0)$.

We now compute $\delta \chi _{1}^{0}(Q,0)$ Since the non-analyticity in $%
\chi _{1}(Q,0)$ is expected to come from the vicinity of the Fermi surface,
the fermionic spectra $\epsilon _{k}$, $\epsilon _{k+q}$ and $\epsilon
_{k+Q} $ can be expanded to first order in $k-k_{F}$:
\begin{equation}
\epsilon _{\mathbf{k}}=v_{F}(k-k_{F}),~~\epsilon _{\mathbf{k}+\mathbf{Q}%
}=\epsilon _{k}+v_{F}Q\cos {\theta _{1}},~~\epsilon _{\mathbf{k}+\mathbf{q}%
}=\epsilon _{k}+v_{F}q\cos {\theta _{2}}.  \label{linear}
\end{equation}
Substituting this expansion into Eq.(\ref{c1}) and performing elementary
integrations over $k$, $\omega $, and $\theta _{1}$, we obtain
\begin{eqnarray}
\delta \chi _{1}^{q=0}(Q,0) &=&-\frac{2mU^{2}}{\pi ^{4}}~\int_{0}^{\infty
}qdq\int_{0}^{\infty }{\Omega }d{\Omega }\Pi (q,\Omega ) \\
&&\times \int_{0}^{\pi }d\theta _{2}~\frac{1}{(i{\Omega }-v_{F}q\cos \theta
_{2})^{2}}~\frac{1}{\sqrt{(v_{F}Q)^{2}+({\Omega }+iv_{F}q\cos \theta {_{2}}%
)^{2}}},  \label{c2}
\end{eqnarray}
where $\Pi (q,\omega )$ at small $q$ and $\omega $ is given by Eq.(\ref{2.01}%
). Rescaling the remaining variables as ${\tilde{q}}=q/Q,~{\tilde{\omega}}%
=\Omega /(v_{F}Q)$ and introducing polar coordinates as ${\tilde{q}}=r\cos
\phi ,{\tilde{\Omega}}=r\sin \phi $, we obtain from (\ref{c2})
\begin{eqnarray}
\delta \chi _{1}^{q=0}(Q,0) &=&-\frac{2mU^{2}|Q|}{\pi ^{4}v_{F}}%
~\int_{0}^{\pi /2}d\phi \sin \phi \cos \phi \Pi (\phi ) \\
&&\int_{0}^{\pi }d\theta _{2}~\int rdr~\frac{1}{(\cos \phi ~\cos \theta
_{2}-i\sin \phi )^{2}}~\frac{1}{\sqrt{1+r^{2}(\sin \phi +i\cos \phi ~\cos {%
\theta _{2}})^{2}}},  \notag
\end{eqnarray}
where $\Pi (\phi )=(m/2\pi )(1-\sin \phi )$. The upper limit of the integral
over $r$ is $r_{max}=\mathcal{O}(k_{F}/Q)\gg 1$. The integration over $r$ is
straightforward and yields
\begin{eqnarray}
\delta \chi _{1}^{q=0}(Q,0) &=&\frac{2mU^{2}}{\pi ^{4}v_{F}}~\int_{0}^{\pi
/2}d\phi \sin \phi \cos \phi \Pi (\phi )\int_{0}^{\pi }d\theta _{2}~\frac{1}{%
(\cos \phi ~\cos \theta _{2}-i\sin \phi )^{4}} \\
&&\times \left[ \sqrt{Q^{2}+(Q~r_{max})^{2}(\sin \phi +i\cos \phi ~\cos {%
\theta _{2}})^{2}}-|Q|\right] .  \notag  \label{c4}
\end{eqnarray}
As $Q~r_{max}\sim k_{F}$, the dominant piece in $\delta \chi _{1}^{q=0}(Q,0)$
comes from high energies and accounts for the non-universal correction to
the uniform susceptibility $\chi (0,0)$. We, however, are interested in the
first subleading term which scales as $|Q|$ and does not depend on $r_{max}$%
. Performing the integration over $\theta _{2}$, we obtain for this
universal contribution
\begin{equation}
\delta \chi _{1}^{q=0}(Q,0)=-\frac{mU^{2}|Q|}{\pi ^{3}v_{F}}~\int_{0}^{\pi
/2}d\phi \sin ^{2}{\phi }\cos {\phi }~(5\sin ^{2}\phi -3)~\Pi (\phi ).
\label{c5}
\end{equation}
Finally, introducing $z=\cos \phi $ [so that $\Pi (z)=(m/2\pi )(1-z)]$, we
obtain
\begin{equation}
\delta \chi _{1}^{q=0}(Q,0)=-\frac{m^{2}U^{2}|Q|}{2\pi ^{4}v_{F}}%
~\int_{0}^{1}dz~(5z^{4}-3z^{2})~(1-z).  \label{c1_1}
\end{equation}
The relevance of the non-analyticity in the polarization bubble is now
transparent: if $\Pi (z)$ was $z$-independent, the integral over $z$ would
vanish. However, because of the non-analyticity, $\Pi (z)$ varies linearly
with $z$. The integral over $z$ then does not vanish, and performing the
integration we obtain
\begin{equation}
\delta \chi _{1}^{q=0}(Q,0)=\chi _{0}~\frac{2}{3\pi }~\left( \frac{mU}{4\pi }%
\right) ^{2}~\frac{|Q|}{k_{F}},  \label{c5_5}
\end{equation}
where $\chi _{0}=2\Pi (0,0)=m/\pi $ is the static susceptibility of
noninteracting fermions.

Using (\ref{cccc}), we then obtain the total contribution of diagram 1:
\begin{equation}
\delta \chi _{1}(Q,0)=2\delta \chi _{1}^{q=0}(Q,0)=\chi _{0}~\frac{4}{3\pi }%
~\left( \frac{mU}{4\pi }\right) ^{2}~\frac{|Q|}{k_{F}}.
\end{equation}

Diagram 2 is another self-energy insertion into the particle-hole bubble.
For a contact interaction, $\delta \chi _{2}$ is exactly $(-1/2)$ of $\delta
\chi _{1}$, the rescaling factor $-1/2$ comes from the fact that compared to
diagram 1, diagram 2, has one lass fermionic loop with more than one vertex,
and lacks the factor of two due to spin summation. Therefore
\begin{equation}
\delta \chi _{2}(Q,0)=-\chi _{0}~\frac{2}{3\pi }~\left( \frac{mU}{4\pi }%
\right) ^{2}~\frac{|Q|}{k_{F}}.  \label{xa-xa}
\end{equation}

The next diagram, diagram 3, represents a vertex correction to the
particle-hole bubble. The $q=0$ contribution to this diagram can be shown to
be of the same magnitude but opposite sign as the $q=0$ part of diagram 1.
To see this, we write the $q=0$ contribution to diagram 3 as
\begin{eqnarray}
\delta \chi _{3}^{q=0}(Q,0) &=&-4U^{2}\int \int \int \int \frac{%
d^{2}k~d^{2}q~d\omega d\Omega }{(2\pi )^{6}}G_{0}(\mathbf{k},\omega )G_{0}(%
\mathbf{k}+\mathbf{q},\omega +\Omega ) \\
&&\times G_{0}(\mathbf{k}+\mathbf{Q}+\mathbf{q},\omega +\Omega )G_{0}(%
\mathbf{k}+\mathbf{Q},\omega )\Pi (q,\Omega )
\end{eqnarray}
and consider a combination
\begin{equation}
C=\frac{1}{2}\delta \chi _{1}^{q=0}+\delta \chi _{3}^{q=0}.  \label{defc}
\end{equation}
Linearizing the fermionic spectra according to Eq.(\ref{linear}), we
re-write $C$ as

\begin{equation}
C=-4U^{2}\nu _{1}\int \int \int \frac{d^{2}qd\omega d\Omega }{\left( 2\pi
\right) ^{4}}\int d\theta _{1}\Pi \left( q,\Omega \right) \left[ S_{1}+S_{3}%
\right] ,  \label{c}
\end{equation}
where
\begin{equation}
S_{1}=\int d\epsilon _{k}G_{0}^{2}(\mathbf{k},\omega )G_{0}(\mathbf{k}+%
\mathbf{Q},\omega )G_{0}(\mathbf{k}+\mathbf{q},\omega +\Omega )  \label{s1}
\end{equation}
and
\begin{equation}
S_{3}=\int d\epsilon _{k}G_{0}(\mathbf{k},\omega )G_{0}(\mathbf{k}+\mathbf{q}%
,\omega +\Omega )G_{0}(\mathbf{k}+\mathbf{Q}+\mathbf{q},\omega +\Omega
)G_{0}(\mathbf{k}+Q,\omega ).  \label{s3}
\end{equation}
Integrating over $\epsilon _{k}$ in Eqs.(\ref{s1},\ref{s3}) yields
\begin{eqnarray*}
S_{1} &=&-2\pi i\text{sgn}\left( \Omega \right) \theta \left( \omega \left(
\Omega -\omega \right) \right) \frac{1}{\left( i\Omega +v_{F}\mathbf{\hat{k}}%
\cdot \mathbf{q}\right) ^{2}}\frac{1}{i\Omega +v_{F}\mathbf{\hat{k}}\cdot
\mathbf{q-}v_{F}\mathbf{\hat{k}}\cdot \mathbf{Q}}, \\
S_{3} &=&2\pi i\text{sgn}\left( \Omega \right) \theta \left( \omega \left(
\Omega -\omega \right) \right) \frac{1}{v_{F}\mathbf{\hat{k}}\cdot \mathbf{Q}%
}\frac{1}{i\Omega +v_{F}\mathbf{\hat{k}}\cdot \mathbf{q}} \\
&&\times \left( \frac{1}{i\Omega +v_{F}\mathbf{\hat{k}}\cdot \mathbf{q-}v_{F}%
\mathbf{\hat{k}}\cdot \mathbf{Q}}-\frac{1}{i\Omega +v_{F}\mathbf{\hat{k}}%
\cdot \mathbf{Q}}\right) .
\end{eqnarray*}
Adding $S_{1}$ and $S_{3}$ and performing some elementary transformations,
we obtain
\begin{equation*}
S_{1}+S_{3}=2\pi i\text{sgn}\left( \Omega \right) \theta \left( \omega
\left( \Omega -\omega \right) \right) \frac{1}{\left( i\Omega +v_{F}\mathbf{%
\hat{k}}\cdot \mathbf{q}\right) ^{2}}\frac{1}{i\Omega +v_{F}\mathbf{\hat{k}}%
\cdot \mathbf{q+}v_{F}\mathbf{\hat{k}}\cdot \mathbf{Q}}.
\end{equation*}
Substituting the last expression back into Eq.(\ref{c}) and making the
change of variables $\mathbf{k\rightarrow -k,}$ $\mathbf{q\rightarrow -q}$
results in
\begin{equation}
C=-\frac{1}{2}\delta \chi _{1}^{q=0},
\end{equation}
Together with (\ref{defc}), this proves that
\begin{equation}
\delta \chi _{3}^{q=0}(Q,0)=-\delta \chi _{1}^{q=0}(Q,0)=-\chi _{0}~\frac{2}{%
3\pi }~\left( \frac{mU}{4\pi }\right) ^{2}~\frac{|Q|}{k_{F}}.  \label{c11}
\end{equation}

The $2k_{F}$-contribution from diagram 3 must be computed independently. The
computations are performed along the same lines as for diagram 1. We present
them in the Appendix \ref{app_C}. We obtain
\begin{equation}
\delta \chi _{3}^{2k_{F}}(Q,0)=\chi _{0}~\frac{2}{3\pi }~\left( \frac{mU}{%
4\pi }\right) ^{2}~\frac{|Q|}{k_{F}}.
\end{equation}
Comparing this with Eq.(\ref{xa-xa}), we see that, for a constant
interaction, the $\mathcal{O}(|Q|)$ contributions to diagram 3 from the
singularities at $q=0$ and $q=2k_{F}$ cancel each other. This result appears
to be quite general (the same is true for $D=3$ and $D=1$ also (see below),
but we do not know how to prove it other than to explicitly compute the
diagrams.

\begin{figure}[tbp]
\centerline{\epsfxsize=6in \epsfbox{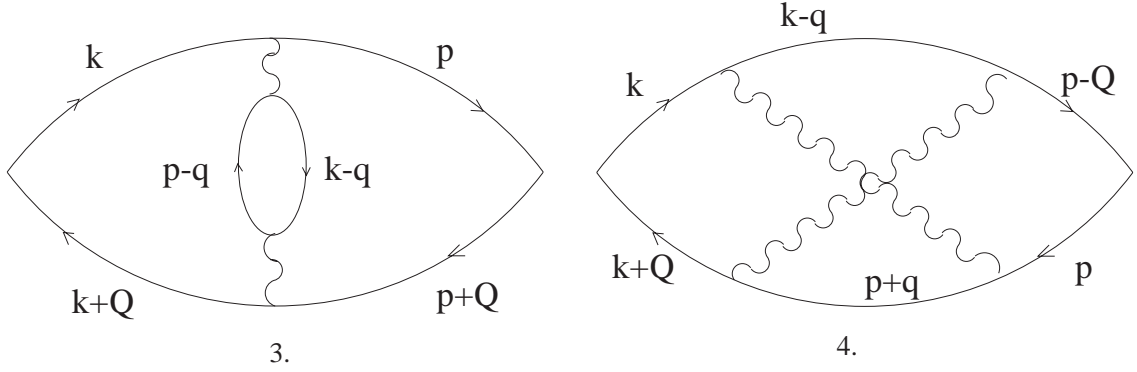}}
\caption{A reduction of $\protect\delta \protect\chi _{4}$ to the $2k_{F}$
part of $\protect\delta \protect\chi _{3}.$}
\label{fig:dia34}
\end{figure}

Next we consider diagram 4, which is obtained by inserting the
particle-particle bubble into the original particle-hole bubble. Expressing $%
\delta \chi _{4}$ via the product of four Greens's functions and the
particle-particle bubble, we obtain
\begin{eqnarray}
\delta \chi _{4}(Q,0) &=&-2U^{2}\int \int \int \int \frac{%
d^{2}k~d^{2}q~d\omega d\Omega }{(2\pi )^{6}}G_{0}(\mathbf{k},\omega )G_{0}(%
\mathbf{k}+\mathbf{Q},\omega )  \notag \\
&&\times G_{0}(\mathbf{q}-\mathbf{k},\Omega -\omega )G_{0}(\mathbf{q}-%
\mathbf{k}-\mathbf{Q},\Omega -\omega )\Pi _{pp}(q,\Omega ),  \label{chi4}
\end{eqnarray}
where $\Pi _{pp}(q,\Omega )$ is given by (\ref{2.03}).

In principle the result for $\delta \chi _{4}$ can be found by substituting
the particle-particle propagator into (\ref{chi4}). However, a
straightforward approach is very cumbersome in this case. There is a more
elegant way to compute $\delta \chi _{4}$ as the non-analytic part of this
diagram is related to the non-analytic $2k_{F}$ contribution from diagram 3,
which we have already found. Indeed, it is easy to make sure that a
non-analytic $(\propto |Q|)$ contribution from diagram 4 comes from internal
momenta for which one of the internal 3-momentum transfers is small. We can
then label the internal momenta in diagram 4 as shown in Fig. \ref{fig:dia34}
and set 3-momentum $q$ to be small (there is a combinatorial factor of $2$
associated with this choice). We can then represent diagram 3 as an
integral-over-$q$ of a product of two terms (``triads'') each containing a
product of three Green's functions:
\begin{equation}
\delta \chi _{4}=-2\times 2U^{2}\int \int \frac{d^{2}qd\Omega }{\left( 2\pi
\right) ^{3}}I\left( \mathbf{q},\Omega ;\mathbf{Q}\right) I\left( -\mathbf{q}%
,-\Omega ;-\mathbf{Q}\right) ,  \label{cooper}
\end{equation}
where a ``triad'' is defined as
\begin{equation}
I\left( q,\Omega ;Q\right) =\int \int \frac{d^{2}kd\omega }{\left( 2\pi
\right) ^{3}}G\left( \mathbf{k},\omega \right) G\left( \mathbf{k-q,}\omega
-\Omega \right) G\left( \mathbf{k+Q,}\omega \right) .  \label{triad}
\end{equation}
An extra overall factor of $-2$ in (\ref{chi3new}) is due to spin summation
and the presence of one closed fermionic loop. At the same time, we can use
the fact that in the $2k_{F}$ part of diagram 3, one of the two momenta in
the internal particle-hole bubble is close to incoming ones. Using the
labeling as in Fig. \ref{fig:dia34}, we can express the $2k_{F}$ part of
diagram 3 as
\begin{equation}
\delta \chi _{3}^{2k_{F}}=4U^{2}\int \int \frac{d^{2}qd\Omega }{\left( 2\pi
\right) ^{3}}\left[ I\left( \mathbf{q},\Omega ;\mathbf{Q}\right) \right]
^{2},  \label{chi3new}
\end{equation}
Carrying out integrations over $\epsilon _{k}$ and $\omega $ in Eq.(\ref
{triad}), we find that
\begin{equation}
I\left( -\mathbf{q},-\Omega ;\mathbf{Q}\right) =-I\left( \mathbf{q},\Omega ;%
\mathbf{Q}\right) ,
\end{equation}
and hence
\begin{equation}
\delta \chi _{4}(Q,0)=\delta \chi _{3}^{2k_{F}}(Q,0)=\chi _{0}~\frac{2}{3\pi
}~\left( \frac{mU}{4\pi }\right) ^{2}~\frac{|Q|}{k_{F}}.  \label{pp}
\end{equation}

Similarly, diagram 5 differs by a factor of $-1$ from diagram 3 (the lack of
the spin factor of two, compared to diagram 3, is compensated by an extra
combinatorial factor of two). For a contact interaction, the non-analytic
part of this diagram vanishes in the same way as it does for diagram 3.

Finally, for the charge susceptibility, diagram 6 just differs by $-1$ from
diagram 3, and diagram 7 differs by an extra $-2$ from diagram 4. For
diagram 6, the extra $-1$ is due to the fact that, compared to diagram 3, $%
q=0$ and $q=2k_{F}$ contributions are interchanged. For diagram 7, the extra
factor is due to the spin summation and reflects the presence of two closed
fermionic loops in diagram 7, as opposed to one loop in diagram 4.

Collecting all terms, we obtain
\begin{eqnarray}
&&\delta \chi _{1}(Q,0)=\chi _{0}~\frac{4}{3\pi }~\left( \frac{mU}{4\pi }%
\right) ^{2}~\frac{|Q|}{k_{F}},  \notag \\
\delta \chi _{2}(Q,0) &=&-\frac{1}{2}\delta \chi _{1}(Q,0),~~\delta \chi
_{3}(Q,0)=0,~~\delta \chi _{4}(Q,0)=\frac{1}{2}\delta \chi _{1}(Q,0),~~ \\
\delta \chi _{5}(Q,0) &=&0,~~\delta \chi _{6}(Q,0)=0, ~~\delta \chi
_{7}(Q,0)=-\delta \chi _{1}(Q,0).
\end{eqnarray}
As a result,
\begin{eqnarray}
\delta \chi _{s}^{2D}(Q,0) &=&\chi _{0}^{2D}~\frac{4}{3\pi }~\left( \frac{mU%
}{4\pi }\right) ^{2}~\frac{|Q|}{k_{F}};  \notag \\
\delta \chi _{c}(Q,0) &=&0.  \label{c15}
\end{eqnarray}
This result is consistent with the conjecture by BKV, who found that the
spin susceptibility has a $Q^{2}\ln |Q|$-dispersion in 3D, and conjectured
that $\chi _{s}(Q,0)$ should scale as $|Q|$ in 2D. We emphasize, however,
that we present for the first time an explicit calculation of $\chi
_{s}(Q,0) $ in 2D. BKV did not explicitly consider the charge
susceptibility, but the absence of the non-analytic momentum dependence of $%
\chi _{c}$ can be readily extracted from their analysis.

\subsubsection{$D=3$ and $D=1$}

\label{sec:chi3D1D} For completeness, we also performed full calculations in
$D=3$ and $D=1$. In both cases, the results, $\delta \chi
_{s}^{3D}(Q,0)\propto Q^{2}\ln Q$, $\delta \chi _{s}^{1D}(Q,0)\propto \ln Q$%
, have logarithmic non-analyticities in $Q$ , which allows one to expand in $%
Q$ from the very beginning. Doing so, we reproduced the results by BKV.

In 3D, we obtained for the spin susceptibility
\begin{eqnarray}
&&\delta \chi _{3}(Q,0)=\delta \chi _{5}(Q,0)=0;\delta \chi _{2}(Q,0)=-\frac{%
1}{2}\delta \chi _{1}(Q,0);\delta \chi _{4}(Q,0)=\frac{1}{2}\delta \chi
_{1}(Q,0);  \notag \\
&&\delta \chi _{1}(Q,0)=2\delta \chi _{1}^{q=0}(Q,0)=\frac{1}{18}~\chi
_{0}^{3D}\left( \frac{ak_{F}}{\pi }\right) ^{2}~\left[ \left( \frac{Q}{k_{F}}%
\right) ^{2}~\ln \frac{k_{F}}{Q}\right] ,  \label{3d_1}
\end{eqnarray}
where $\chi _{0}^{3D}=mk_{F}/\pi ^{2}$ is the static spin susceptibility and
$a=mU/4\pi $ is the scattering length. Combining all contributions we obtain
\begin{equation}
\delta \chi _{s}^{3D}(Q,0)=\frac{1}{18}~\chi _{0}^{3D}\left( \frac{ak_{F}}{%
\pi }\right) ^{2}~\left[ \left( \frac{Q}{k_{F}}\right) ^{2}~\ln \frac{k_{F}}{%
Q}\right] .  \label{3D}
\end{equation}
Eqs. (\ref{3d_1}) and (\ref{3D}) precisely coincide with the earlier results
by BKV~\cite{bkv}. We also considered the charge susceptibility and found
that, as in 2D, it does not possess a non-analytic dependence on $Q$.

In 1D, the relations between various components of $\delta \chi
_{s}^{1D}(Q,0)$ are the same as in 3D, and
\begin{equation}
\delta \chi _{s}^{1D}(Q,0)=\delta \chi _{1}(Q,0)=2\delta \chi
_{1}^{q=0}(Q,0)=-2\chi _{0}^{1D}\left( \frac{U}{2\pi v_{F}}\right) ^{2}~~\ln
\frac{k_{F}}{Q}.  \label{1D}
\end{equation}
This form $\delta \chi _{s}^{1D}(Q,0)$ agrees with the well-known result by
Dzyaloshinskii and Larkin~\cite{DL_1}.

\subsection{Spin and charge susceptibilities at finite $T$ and $Q=0$}

\label{sec:chiT}

In this Section, we consider the uniform ($Q=0$) spin and charge
susceptibilities at finite $T$. Of particular interest here is the question
whether a non-analytic momentum dependence of the static susceptibility at $%
T=0$ is accompanied by that of the static susceptibility.\textbf{\ } We
remind that in $D=3$, according to Carneiro and Pethick~ \cite{pethick} and
BKV, $\chi (Q,0)-\chi \left( 0,0\right) $ behaves as $Q^{2}\ln |Q|$, but $%
\chi (0,T)-\chi \left( 0,0\right) $ is analytic and behaves as $T^{2}$.
Misawa~\cite{misawa_2}, on the contrary, did find a $T^{2}\ln T$-behavior.
BKV conjectured that for a generic $D$, the momentum and temperature
dependences of $\chi _{s}$ should have the same exponents.

As it was pointed out in the Introduction, there were two microscopic
calculations of $\chi (0,T)$ in 2D: by BKM \cite{marenko} and CM \cite
{millis}. Both groups found $\chi _{s}(0,T)\propto T$ and associated this
non-analytical $T$ dependence with the square-root singularity in the
quasiparticle interaction function $f\left( \mathbf{k},\mathbf{k}^{\prime
}\right) $ caused by $2k_{F}$ scattering. We recall that the quasiparticle
interaction function, $f\left( \mathbf{k},\mathbf{k}^{\prime }\right),$ is
obtained by computing the vertex $\Gamma \left( \mathbf{k,}\omega ;\mathbf{k}%
^{\prime },\omega ^{\prime };\mathbf{q,}\Omega \right) $ to the second order
in the interaction and using the relation \cite{agd},\cite{volumeIX}
\begin{equation}
f\left( \mathbf{k},\mathbf{k}^{\prime }\right) =A\Gamma \left( \mathbf{k,}%
\epsilon _{k};\mathbf{k}^{\prime }\epsilon _{k^{\prime }};q/\Omega
\rightarrow 0\right),  \label{Landau}
\end{equation}
where $A$ is a normalization factor, BKM \cite{marenko} explored the
singularity $f\left( \mathbf{k},\mathbf{k}^{\prime
}\right)$ at $T=0$ and for small but finite quasiparticle energies, $\epsilon _{k}$ and $\epsilon
_{k^{\prime }}$. In
their approach, the $T-$dependence comes from the Fermi functions. In the
diagrammatic language, the approximation made by BKM accounts to evaluating
the particle-hole polarization bubble near $2k_{F}$ at $T=0$ but at a finite
frequency. CM included this effect into their consideration, but they also
took into account a $\sqrt{T}-$singularity associated with the thermal smearing of
the $2k_{F}$-feature in the susceptibility.

We compute $\chi _{s}(Q=0,T)$ in a straightforward diagrammatic approach
(the same we employed for the case of $Q\neq 0,$ $T=0),$ in which all possible
sources of $T$-dependence are taken into account automatically.
Our result differs by a factor of 2 compared to that of CM. We could not
establish the reason for the discrepancy.

We first report our results for $D=2$ first and then analyze the case of
arbitrary $D$.

\subsubsection{$D=2$}

The analysis of $\chi (0,T)$ proceeds in the same way as in Sec.\ref
{sec:chiQ}. We found that the interplay between the non-analytic terms in
various diagrams for the susceptibility at $T\neq 0$ is \emph{exactly} the
same as at $T=0$. Namely, the non-analytic pieces originate from the $q=0$
and $2k_{F}$ non-analyticities in the particle-hole susceptibility, or
alternatively, from the $q=0$ non-analyticity in the particle-particle
susceptibility. We explicitly verified that the relative coefficients
between non-analytic terms are the same as at $T=0$. This implies that (i)
just as at $T=0$, there is no non-analytic $T$ dependence in the charge
susceptibility, and (ii) to obtain the full correction the spin
susceptibility, it is sufficient to evaluate just one non-analytic
contribution, e.g., $\delta \chi _{1}^{q=0}(0,T)$. The full $\delta \chi
_{s}(0,T)$ is then given by
\begin{equation}
\delta \chi _{s}(0,T)=2\delta \chi _{1}^{q=0}(0,T).  \label{c16}
\end{equation}
At finite $T$ and $Q=0$, a general form of $\delta \chi _{1}^{q=0}(0,T)$ is
\begin{equation}
\delta \chi _{1}^{q=0}(0,T)=-8U^{2}T^{2}\sum_{\omega _{n},\Omega _{m}}~\int
\int \frac{d^{2}k~d^{2}q}{(2\pi )^{4}}G_{0}^{3}(k,\omega )G_{0}(\mathbf{k+q}%
,\omega _{n}+\Omega _{m})\Pi (q,\Omega ).  \label{c17}
\end{equation}
Expanding the quasiparticle spectra near the Fermi surface, integrating over
$\epsilon _{k}$ and then evaluating the sum over $\omega _{n}$, we obtain,
after simple algebra,
\begin{equation}
\delta \chi _{1}^{q=0}(0,T)=-4\chi _{0}^{2D}~\left( \frac{mU}{4\pi }\right)
^{2}~I(T),  \label{c18}
\end{equation}
where $\chi _{0}^{2D}=m/\pi ,$
\begin{equation}
I(T)=\frac{T}{E_{F}}\sum_{m}\int dxx~\frac{\Omega _{m}^{2}(2\Omega
_{m}^{2}-x^{2})}{(\Omega _{m}^{2}+x^{2})^{3}},  \label{c181}
\end{equation}
and $x\equiv v_{F}q$. Expression (\ref{c181}) is rather tricky, because $%
I(T) $ is formally ultraviolet-divergent. The most straightforward way to
get rid of the ultraviolet divergence is to introduce a short-range
(lattice) cutoff in the momentum integral so that $x\leq X_{0}\sim W.$
Evaluating the integral over $x$ first we obtain
\begin{equation}
I(T)=\frac{T}{4E_{F}}\sum_{m}S\left( m\right) ,  \label{c181_1}
\end{equation}
where
\begin{equation}
S\left( m\right) =1+2\frac{\Omega _{m}^{2}}{\Omega _{m}^{2}+X_{0}^{2}}-3%
\frac{\Omega _{m}^{4}}{\left( \Omega _{m}^{2}+X_{0}^{2}\right) ^{2}}.
\end{equation}
For $\Omega _{m}\ll X_{0},$ $S(m)$ is close to $1$, i.e., $S\left( m\right)
=1+\mathcal{O}((\Omega _{m}^{2}/X_{0})^{2})$, whereas for $\Omega _{m}\gg
X_{0}$ it falls off rapidly $[$as ($X_{0}/\Omega _{m})^{2}]$. The vanishing
of $S(m)$ at large $m$ ensures the convergence of the sum in \textbf{\ }Eq%
\textbf{.(}\ref{c181_1}\textbf{).} and allows one to use Euler-Maclaurin
formula \cite{arfken}. Applying it, the sum reduces to
\begin{equation}
\frac{T}{4E_{F}}\sum_{m=-\infty }^{\infty }S\left( m\right) =\frac{T}{2E_{F}}%
\int_{0}^{\infty }dmS\left( m\right) -\frac{T}{24E_{F}}S^{\prime }\left(
0\right) +\dots ,  \label{c181_2}
\end{equation}
where $\dots $ stands for higher-order derivatives of $S$. All derivatives
of $S(m)$ obviously vanish in the continuum limit $W\rightarrow \infty .$
The remaining integral term in \textbf{(}\ref{c181_2}) gives
\begin{equation}
\frac{T}{2E_{F}}\int_{0}^{\infty }dmS\left( m\right) =\frac{5}{16}\frac{X_{0}%
}{E_{F}},
\end{equation}
which is a $T$-independent contribution. As a result, the above computation
does not yield a linear-in-$T$ piece in $\delta \chi _{1}^{q=0}(0,T).$

A more careful inspection of the steps we took to arrive at this result
reveals a problem. Namely, it is obvious from (\ref{c181}) that the term
with $m=0,$ i.e., with $\Omega _{m}=0,$ vanishes for any finite $q$.
However, in the sum in (\ref{c181_1}) the $m=0$ term is present and
contributes $T/4E_{F}$. As the static susceptibility is properly defined as
the limit of $\chi (Q,T)$ at $Q\rightarrow 0$, one should always keep $q$
finite at the intermediate steps of the computations. Alternatively, one can
perform calculations for a finite system and then extend the system size to
infinity. In both cases, there exists a lower cutoff in the integral over $q$%
. This cutoff plays no role for all terms with $m\neq 0$ but it eliminates
the term with $m=0$. Subtracting off this term from \textbf{(}\ref{c181_1}%
\textbf{), } and using our previous results we obtain a \emph{universal},
linear-in-$T$ piece in $I(T)$
\begin{equation}
I\left( T\right) =-\frac{T}{4E_{F}}.  \label{c181_c}
\end{equation}

An alternative way to arrive at Eq.(\ref{c181_c}) is to perform the
summation over $\Omega _{m}$ in (\ref{c181}) first, keeping $q$ finite, and
then integrate over $q$. Performing the summation, we obtain
\begin{equation}
I(T)=\frac{1}{4E_{F}}\int_{0}^{\infty }dy\left( y\frac{\partial ^{2}}{%
\partial y^{2}}\left[ y^{1/2}\left\{ n_{B}(y^{1/2})+\frac{1}{2}\right\} %
\right] +2\frac{\partial ^{2}}{\partial y^{2}}\left[ y^{3/2}\left\{
n_{B}(y^{1/2})+\frac{1}{2}\right\} \right] \right) ,  \label{c21}
\end{equation}
where $n_{B}\left( z\right) =\left( \exp \left( z/T\right) -1\right) ^{-1}$
is the Bose distribution function and $y=v_{F}^{2}q^{2}$. Integrating by
parts, we obtain from (\ref{c21})
\begin{equation}
I(T)=-\frac{1}{2E_{F}}\left( 1+\frac{1}{2T}\int_{0}^{\infty }dy\frac{%
\partial }{\partial y}\left\{ y^{1/2}n_{B}(y^{1/2})\right\} \right) =-\frac{T%
}{4E_{F}},  \label{c22}
\end{equation}
in agreement with (\ref{c181_c}).

The above analysis shows that $\chi _{s}(0,T)$ does indeed contain a
linear-in-$T$ term in $D=2,$ as earlier studies conjectured. However, the
physics behind this term is very different from the one that leads to the $%
|Q|$ piece in $\chi _{s}(Q,0)$.

Substituting (\ref{c181_c}) into (\ref{c18}) and then using (\ref{c16}), we
obtain
\begin{equation}
\delta \chi _{s}(0,T)=2\chi _{0}^{2D}~\left( \frac{mU}{4\pi }\right) ^{2}~%
\frac{T}{E_{F}}.  \label{c20}
\end{equation}
This is the central result of this subsection.

We remind the reader that the full $\chi _{s}(0,T)$, given by (\ref{c20})
comes from {the dynamical} particle-hole bubble. To emphasize this point, in
Appendix~\ref{app_d} we compute $\delta \chi _{1}^{2k_{F}}$ neglecting the
frequency dependence of the polarization bubble, and show that this yields
an incorrect prefactor in the linear-in-$T$ piece.

We did not attempt to verify our $\delta \chi_{1}^{2k_F}$ by explicitly computing a linear in $T$ contribution from $2k_{F}$ polarization bubble at a finite $T$
 (as we did for $|Q|$ term at $T=0$). This calculation would 
require, as an input,
 the analytical expression for the dynamical polarization bubble near $2k_F$ at a finite $T$. We couldn't obtain this expression in a manageable form, nor we could find it in the literature. It would be interesting, however, to
 verify our  $\delta \chi_{1}^{2k_F}$  numerically by using the
 numerical results for  $\Pi (q, \omega, T)$~\cite{zheng}.

\subsubsection{other dimensions}

\label{sec:chiTanyD} For arbitrary $D$, the consideration analogous to the
one for $D=2$ yields, instead of Eq. (\ref{c18}),
\begin{equation}
\delta \chi _{D}(0,T)=-CU^{2}I_{D}(T),  \label{y1}
\end{equation}
where $C$ is a positive constant,
\begin{equation}
I_{D}(T)=TE_{F}^{1-D}\sum_{m}\int dxx^{D-1}~\frac{\Omega _{m}^{4}}{(\Omega
_{m}^{2}+x^{2})^{3}}  \label{y1_1}
\end{equation}
and $x=v_{F}q.$ For $D=2$, (\ref{y1_1}) coincides with (\ref{c181}) modulo a
piece [$T\sum_{m}\int dxx\left( \Omega _{m}^{2}-x^{2}\right) /(\Omega
_{m}^{2}+x^{2})^{3}]$ that vanishes upon integration over $x$. The ambiguity
with the order of summation and integration was resolved in the previous
section; now we know that it is safe to sum over $\Omega _{m}$ first and the
integrate over $x.$ Performing the summation with the help of the well-known
formula
\begin{equation}
\sum_{m}\frac{1}{\Omega _{m}^{2}+x^{2}}=\frac{1+2n_{B}(x)}{2x},
\end{equation}
we find
\begin{equation}
I_{D}(T)=\frac{1}{2E_{F}^{D-1}}\int dxx^{D-1}\left[ \frac{1+2n_{B}(x)}{2x}%
+2x^{2}\frac{\partial }{\partial x^{2}}\frac{1+2n_{B}(x)}{2x}+\frac{x^{4}}{2}%
\frac{\partial ^{2}}{\partial x^{2}}\frac{1+2n_{B}(x)}{2x}\right] .
\label{y2}
\end{equation}
Evaluating the integral over $x$ and introducing an infinitesimally-small $%
\delta $ to eliminate infrared divergences at intermediate steps, we find
the $T-$independent part of $I_{D}\left( T\right) $ for $D\geq 2$ to be
given by
\begin{eqnarray}
I_{D}(T) &=&-\frac{(D-2)(4-D)}{8}\left( \frac{T}{E_{F}}\right)
^{D-1}~\int_{0}^{\infty }\frac{dzz^{D-2}}{e^{z}-1}  \label{y3} \\
&=&-\frac{(D-2)(4-D)}{8}\Gamma \left( D-1\right) \zeta \left( D-1\right) ,
\end{eqnarray}
where $\Gamma \left( x\right) $ and $\zeta \left( x\right) $ are the Euler
and Riemann functions, respectively. For $D\rightarrow 2$, the pole of the $%
\zeta $ -function, $\zeta (D-1)\rightarrow 1/(D-2),$ is canceled by the
prefactor $D-2,$ so that $I_{2}(T)$ is finite and equal to $-T/4E_{F},$ in
agreement with (\ref{c22}).

For $D<2$, care has to be taken to ensure the cancelation of the divergent
terms. The final result for this case is
\begin{equation}
I(T)=-\frac{(2-D)(4-D)}{8}\left( \frac{T}{E_{F}}\right)
^{D-1}~\int_{0}^{\infty }\frac{dz}{z^{2-D}}\left( \frac{1}{z}-\frac{1}{%
e^{z}-1}\right) .  \label{y4}
\end{equation}
We see that for arbitrary $D$, the function $I_{D}(T)$ (and thus the spin
susceptibility) scales as $T^{D-2}$. In an explicit form,
\begin{equation}
\delta \chi (0,T)=-CU^{2}\left( \frac{T}{E_{F}}\right) ^{D-1}~f(D).
\label{y1_d}
\end{equation}
Function $f\left( D\right) $ diverges logarithmically for $D=1$ (and at $D=1$%
, $\delta \chi \propto \ln T$). Near $D=3$ function $f(D)$ is perfectly
regular and equal to
\begin{equation}
f(3)=-\frac{\pi ^{2}}{48}.  \label{y1_e}
\end{equation}
As we see from Eqs.(\ref{y1_d}) and (\ref{y1_e}), this last result implies
that in 3D, the leading temperature correction to the susceptibility scales
as $T^{2}$, and there is no logarithmic prefactor. This agrees with the
results of Carneiro and Pethick~\cite{pethick} and BKV.

Obviously, the absence of the $T^{2}\ln T$-behavior of $\chi (0,T)$ in 3D,
and $Q^{2}\ln Q$-behavior of $\chi (Q,0)$ implies that there is no
one-to-one correspondence between thermal corrections and quantum
corrections at finite $T$. Our consideration indeed shows that thermal and
quantum corrections are not equivalent.

We also see that although $f(D)$ goes smoothly through $D=2$, the functional
form of $f(D)$ changes between $D>2$ and $D<2$. The consequences of this
fact are, however, unclear to us.

\section{finite-range interaction}

\label{sec:Uofq} In the previous sections we considered the model case of a
contact interaction, characterized by a single coupling constant $U$ which
is independent of the momentum transfer. Now we analyze the more realistic
case of a finite-range interaction when the coupling is a function of the
momentum transfer $U\rightarrow U\left( q\right) ,$ where $U\left( q\right) $
is such that $U(0)$ and $U(2k_{F})$ are finite. \ Our key result is that
only these two parameters are important.

\subsection{Self-energy}

\label{sec:sigmaUofq} 
\begin{figure}[tbp]
\centerline{\epsfxsize=3in
\epsfbox{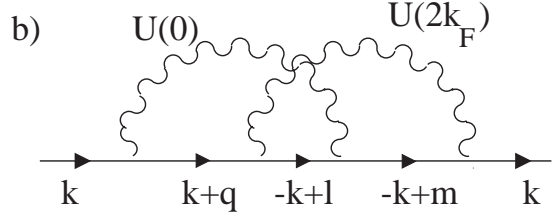}}
\caption{One of the self-energy diagrams for $U(0)\neq U(2k_F)$. Momenta $%
q,l $ and $m$ are small compared to $k$.}
\label{fig:crossed}
\end{figure}

We begin with the self-energy. For momentum-dependent $%
U(q)$ the two self-energy diagrams in Fig. \ref{fig1} have to be considered
separately. For diagram shown in Fig. \ref{fig1}a, the extension to $U=U(q)$
is straightforward--the factor $2U^{2}$ for that part of the self-energy
which corresponds to process b) in Fig.2 (we recall that only that part
contributes to thermodynamics) is be replaced by $U^{2}(0)+U^{2}(2k_{F})$.
The diagram in Fig. \ref{fig1}b requires more care, but we know from the
analysis of the ``sunrise'' diagram for the self-energy (Fig.4b) that a
non-analytic piece comes from the range where two internal momenta in the
self-energy diagram are near $-\mathbf{k}$, and the third is near $\mathbf{k.%
}$ For diagram Fig. \ref{fig1}b, this implies that the momenta are labeled
as in Fig.~\ref{fig:crossed}.

It is then obvious that the overall factor for the diagram in Fig. \ref{fig1}%
b is $U(0)U(2k_{F})$. Process a) in Fig. 2 determines that part of the
self-energy which is singular on the mass-shell and does not contribute to
thermodynamics. The overall factor for that part is $U\left( 0\right) ^{2}$.
Collecting all contributions, we find that
\begin{eqnarray}
&&\Sigma _{R}^{\prime \prime }(\omega ,T) =\frac{mU^{2}(0)}{16\pi
^{3}v_{F}^{2}}~\left[ \omega ^{2}+(\pi T)^{2}\right] ~\ln \frac{W}{|\omega
-\epsilon _{k}|}  \notag \\
&&+\frac{m(U^{2}(0)+U^{2}(2k_{F})-U(0)U(2k_{F}))}{8\pi ^{3}v_{F}^{2}}\left(
[(\pi T)^{2}+\omega ^{2}]\ln \frac{{\bar{A}}}{T}-\omega ^{2}~f\left( \frac{%
\omega }{\pi T}\right) \right) ,  \label{u1}
\end{eqnarray}
where $A$ and ${\bar{A}}$ are constants, and the scaling function $f(x)$ is
given by (\ref{2.7}). The real part of the self-energy is given by
\begin{equation}
\Sigma _{R}^{\prime }(\omega )=-\frac{m(U^{2}(0)+U^{2}(2k_{F})-U(0)U(2k_{F}))%
}{16\pi ^{2}v_{F}^{2}}\omega |\omega |g\left( \frac{\omega }{T}\right) .
\label{u1.1}
\end{equation}
The limiting forms of the scaling function $g(x)$ are $g(\infty )=1$ and $%
g(x\ll 1)\approx 4 log 2/(x)$.

\subsection{Spin and charge susceptibilities}

\label{sec:chiUofq}

The same consideration holds for the susceptibility--the very fact that all
non-analytic contributions come from the vertices with near zero total
momentum and transferred momentum either near zero or near $2k_{F}$ implies
that for $U=U(q)$, an overall factor of $U^{2}$ is replaced either by $%
U^{2}(0)$ or $U^{2}(2k_{F})$, as in diagrams 1, 3, 6 and 7 in Fig.\ref
{fig:diag}, and by $U(0)U(2k_{F})$, as in diagrams 2, 4 and 5. With this
substitution, we have, finally
\begin{eqnarray}
\delta \chi _{1}(Q,T) &=&K(Q,T)(U^{2}(0)+U^{2}(2k_{F}));\delta \chi
_{2}(Q,T)=-K(Q,T)U(0)U(2k_{F})  \notag \\
\delta \chi _{3}(Q,T) &=&K(Q,T)(U^{2}(2k_{F})-U^{2}(0));\delta \chi
_{4}(Q,T)=K(Q,T)U(0)U(2k_{F});  \notag \\
\delta \chi _{5}(Q,T) &=&0,~\delta \chi
_{6}(Q,T)=-K(Q,T)(U^{2}(0)-U^{2}(2k_{F})); \notag \\
\delta \chi
_{7}(Q,T)&=& -K(Q,T)(U^{2}(0)+U^{2}(2k_{F})),  \label{u2}
\end{eqnarray}
where $K(Q,0)$ and $K(0,T)$ are given by Eqs. (\ref{c5_5}) and (\ref{c20}),
respectively:
\begin{equation}
K(Q,0)=\chi _{0}^{2D}~\frac{2}{3\pi }~\left( \frac{mU}{4\pi }\right) ^{2}~%
\frac{|Q|}{k_{F}};~~K(0,T)=\chi _{0}^{2D}~\left( \frac{mU}{4\pi }\right)
^{2}~\frac{T}{E_{F}}  \label{u3}
\end{equation}
where $\chi _{0}=m/\pi $. When both $Q$ and $T$ are nonzero, $K(Q,T)=K(Q,0){%
\tilde{g}}(v_{F}Q/T)$, where ${\tilde{g}}(x)$ is a scaling function subject
to ${\tilde{g}}(x\gg 1)\propto 1/x$. However, we did not attempt to compute $%
{\tilde{g}}(x)$ at $x$ other than $x=0$ and $x=\infty $.

Collecting all contributions we find for the spin susceptibility
\begin{equation}
\delta \chi _{s}(Q,T)=2K(Q,T)U^{2}(2k_{F}).  \label{u4}
\end{equation}
As for the case $\ U=$const$,$ the charge susceptibility is regular because
all non-analytic corrections from individual diagrams cancel out. \vspace{2mm%
}

Eqs. (\ref{u1}), (\ref{u1.1}), (\ref{u3}) and (\ref{u4}) are the central
results of this paper.

While it is intuitively obvious that the momentum dependence of the
susceptibility should only include $U(0)$ and $U(2k_{F})$, this intuition is
based on the analysis of the self-energy but not the susceptibility itself.
It is therefore worthwhile to demonstrate \textit{explicitly} that
non-analytic terms in the susceptibility do not depend on the
momentum-averaged interaction. This is what we are going to do in the
remainder of this Section.

\begin{figure}[tbp]
\centerline{\epsfxsize=3in
\epsfbox{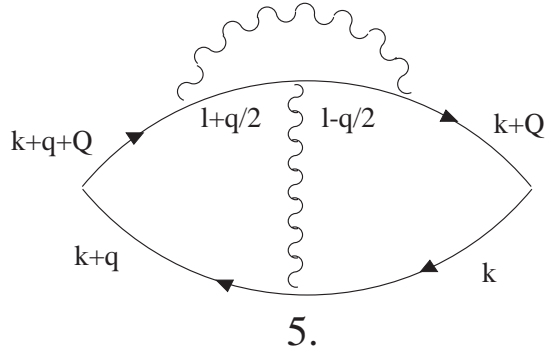}}
\caption{Another way of labeling momenta in $\protect\delta\protect\chi_5$.}
\label{fig:chi5}
\end{figure}

To demonstrate that only $U(2k_{F})$ matters, consider one of the diagrams
for which, as we claim, the non-analytic term scales as $U(0)U(2k_{F})$,
i.e., diagrams 2, 4 and 5. Each of these diagrams has two interaction lines.
Quite obviously, one of momentum transfers should be near zero. The issue is
to prove that the other one is near $2k_{F}$. Consider for definiteness
diagram 5. The net result for this diagram is zero, but this is a result of
the cancelation between two contributions, $\delta \chi _{5}^{a}(Q)$ and $%
\delta \chi _{5}^{b}(Q)$, which differ in the choice of which of the two
interactions carry small momentum. Consider one of the choices. We label the
internal momenta in the diagram as $\mathbf{k}$, $\mathbf{k}+\mathbf{q}$, $%
\mathbf{k}+\mathbf{Q}$, $\mathbf{k}+\mathbf{q}+\mathbf{Q}$, $\mathbf{l}+%
\mathbf{q}/2$, $\mathbf{l}-\mathbf{q}/2$, where $Q$ is the external
momentum, and introduce two angles $\theta _{1}$ and $\theta _{2}$ between $%
\mathbf{q}$ and $\mathbf{l}$ and between $\mathbf{q}$ and $\mathbf{k}$,
respectively (cf. Fig.\ref{fig:chi5}).

The integration over $k$ and the corresponding frequency $\omega $ is
straightforward (see Appendix \ref{app_C}). Introducing then $q=r\cos \phi $
and $\Omega =r\sin \phi $, where $\Omega $ is the frequency associated with $%
q$, we integrate over $r$ and, after redefinition of the variables, obtain
that the non-analytic, linear-in-$Q$ piece of diagram $5$ reduces to
\begin{equation}
\delta \chi _{5}^{a}(Q)=\chi _{0}\frac{m^{2}U(0)}{4\pi ^{5}}\frac{|Q|}{k_{F}}%
~J,  \label{u5}
\end{equation}
where
\begin{equation}
J=\int_{0}^{\infty }dxx^{2}\int_{-\pi }^{\pi }\frac{d\theta _{1}}{x+i\cos
\theta _{1}}~\int_{-\pi }^{\pi }\frac{d\theta _{2}}{(x+i\cos \theta _{2})^{4}%
}~U\left( 2k_{F}\sin ^{2}{\frac{\theta _{1}-\theta _{2}}{2}}\right) .
\label{u6}
\end{equation}
For a constant interaction $U(q)=U$, we can integrate independently over $%
\theta _{1}$ and $\theta _{2}$, and then integrate over $x$, which gives $%
J=\pi ^{2}/6$. The result for $\delta \chi _{5}^{a}(Q)$ then coincides with
one of the two contributions to $\delta \chi _{5}(Q)$, as we discussed in
Sec.\ref{sec:chiQ2D}. A relevant point here is that typical $\cos \theta
_{1,2}$ are of order $x,$ whereas typical $x$ are of order 1. Hence $\theta
_{1}-\theta _{2}\sim 1$, i.e., typical angles between two momenta are
generic. This implies that the argument of $U(2k_{F}\sin ^{2}(\theta
_{1}-\theta _{2})/2)$ is just of the order of $k_{F}$ but not necessarily
close to $2k_{F}.$

We now show that, in fact, only $\theta _{1}-\theta _{2}=\pm \pi $ matter.
To see this, we introduce diagonal variables $a=(\theta _{1}+\theta _{2})/2$
and $b=(\theta _{1}-\theta _{2})/2$ and integrate first over $x$ and then
over $a$. This integration is tedious but straightforward, and carrying it
out we obtain, after some algebra,
\begin{equation}
J=-\int_{0}^{\pi /2}dbU(2k_{F}\sin ^{2}b)\mbox{Re}\left[ S(b)+S(\pi -b)%
\right] ,  \label{u7}
\end{equation}
where
\begin{equation}
S(b)=\left( \frac{4}{3}+\frac{\cos 2b}{\sin ^{4}b}\right) \ln {\cos }%
2b\left( \frac{1}{\sin 2b-i\delta }-\frac{1}{\sin 2b+i\delta }\right) .
\label{n8}
\end{equation}
Then
\begin{equation}
J=i\delta \mbox{Im}\int_{0}^{\pi }dz\ln \cos z\sin ^{2}z{+}\delta {^{2}}%
\left( \frac{4}{3}+\frac{\cos z}{(\sin z/2)^{4}}\right) U(2k_{F}(\sin
z/2)^{2})  \label{n9}
\end{equation}
The integral does not vanish due to divergences near $z=0$ and $z=\pi $. The
divergence near $z=0$ does not contribute to the imaginary part of the
integral, but the one near $z=\pi $ does contribute. Restricting $z$ near $%
\pi ,$ we obtain
\begin{equation}
J=\frac{1}{3}U(2k_{F})\int_{0}^{\infty }\frac{dy\delta }{y^{2}+\delta ^{2}}=%
\frac{\pi ^{2}}{6}U(2k_{F}).  \label{n10}
\end{equation}
This consideration shows that although for a momentum-independent
interaction we could evaluate $\delta \chi _{5}^{a}(Q)$ in a scheme in which
the angular integrals were not restricted to a particular $\theta _{1}$ or $%
\theta _{2}$, the calculation performed in another way demonstrates that the
whole integral comes only from the range where $\theta _{1}-\theta _{2}=\pm
\pi $. For a momentum-dependent interaction, this implies that only $%
U(2k_{F})$ matters, precisely as we anticipated. Similar calculations can be
repeated for other cross diagrams with the result that the overall factor is
always $U(0)U(2k_{F})$.

The above consideration is another indication that the non-analyticities in
the specific heat and spin susceptibility come from the two interaction
vertices in which the transferred momentum is either near $0$ or $2k_{F}$,
and simultaneously the total momentum for both vertices is near $2k_{F}$.

\section{Conclusions}

\label{sec:concl} We now summarize the key results of the paper. We
considered the universal corrections to the Fermi-liquid forms of the
effective mass, specific heat, and spin and charge susceptibilities of the
2D Fermi liquid. We assumed that the Born approximation is valid, i.e., $%
mU(q)/4\pi \ll 1$, and performed calculations to second order in the
interaction potential $U(q)$. We found that the corrections to the mass and
specific heat are non-analytic and linear in $T$, and obtained for the first
time the explicit results for these corrections. We next found that the
corrections to the static spin susceptibility are also non-analytic and
yield the $|Q|$-dependence of $\chi _{s}(Q,T=0)$ and $T$ dependence of $\chi
_{s}(Q=0,T)$. We obtained for the first time the explicit expressions for
the linear-in-$Q$ and linear-in-$T$ terms in the susceptibility. We found
that the corrections to the charge susceptibility are all analytic. We also
performed calculations in 3D and confirmed the results of BKV and others
that the correction to $\chi _{s}(Q,T=0)$ scales as $Q^{2}\ln Q$, but the
correction to $\chi _{s}(Q=0,T)$ scales as $T^{2}$ without a logarithmic
prefactor.

We analyzed in detail the physical origin of the non-analytic corrections to
the Fermi liquid and clarified the discrepancy between earlier papers. We
argued that the non-analyticities in the fermionic self-energy and in $\chi
_{s}(Q,T)$ are due to the non-analyticities in the \emph{dynamical}
particle-hole susceptibility. We argued that the non-analyticities in the
fermionic self-energy and in $\chi _{s}(Q,0)$ are due to the
non-analyticities in the dynamical two-particle response functions. We have
shown that non-analytic terms in the self-energy and the spin susceptibility
come from the processes which involve the scattering amplitude with a small
momentum transfer \emph{and} a small total momentum. We explicitly
demonstrated that the non-analytic terms can be viewed equivalently as
coming from \emph{either} of the two non-analyticities in the dynamical
particle-hole bubble--the one near $q=0$ and the other one near $q=2k_{F}$%
--or from the singularity in the dynamical particle-particle bubble near
zero total momentum. We also demonstrated explicitly that the non-analytic
terms in all diagrams for the susceptibility and the self-energy depend only
on $U(0)$ and $U(2k_{F})$, but not on averaged interactions over the Fermi
surface. Only under this condition, is there a substantial cancelation
between different diagrams for the susceptibility. Due to these
cancelations, the non-analytic correction to the spin susceptibility
depends only on $U(2k_{F})$, but not on $U(0)$, and scales as $U^{2}(2k_{F})$%
. The non-analytic corrections to the effective mass and the spin
susceptibility scale as $U^{2}(0)+U^{2}(2k_{F})-U(0)U(2k_{F})$.

The non-analytic $Q$ behavior of $\chi _{s}(Q,T=0)$ obtained in both 3D and
2D questioned the validity of the Hertz-Millis-Moriya phenomenological
theory of quantum phase transitions. This theory assumes a regular $q^{2}$%
-expansion of the spin susceptibility. Indeed, extending the results for $%
\chi _{s}(Q,T)$ to the critical region one obtains a rather complex quantum
critical behavior \cite{BK}, which is very different from the
Hertz-Millis-Moriya theory. We caution, however, that the non-analytic
behavior of $\chi _{s}(Q,T)$ was obtained within the Born approximation,
when fermions behave as sharp quasiparticles. Near a magnetic transition,
the fermionic self-energy is large, and destroys the coherent Fermi-liquid
behavior beginning at a frequency which vanishes at the quantum critical
point. In this situation, the second-order perturbation theory is
unreliable. The issue whether non-analytic corrections to the static $\chi
_{s}(Q,T)$ survive at criticality is now under consideration and we refrain
from further speculations on this matter.

We acknowledge stimulating discussions with A. Abanov, I. Aleiner, B.
 Altshuler, D. Belitz, S. DasSarma, 
 A. Finkelstein, A. Millis, M. Norman, C. Pepin, A.
 Rosch, J. Rech, M. Reizer, and Q. Si. We also thank M. Mar'enko for
 bringing Ref.\cite{marenko} to our attention and G. Martin for critical
 reading of the manuscript.

The research has been supported by NSF DMR 9979749 (A. V. Ch.) and NSF
DMR-0077825 (DLM).

\appendix

\section{Mass-shell singularity}

\label{app_F}

\label{sec:massshell} 
\begin{figure}[tbp]
\centerline{\epsfxsize=4in \epsfbox{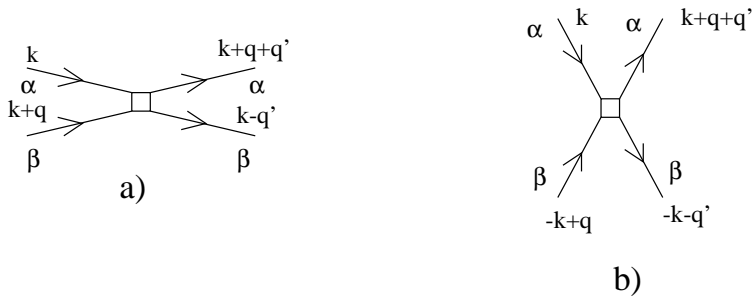}}
\caption{Two processes contributing to the log-singularities in the
self-energy.}
\label{twoprocesses}
\end{figure}

In this Appendix, we take a deeper look into the
origin of the logarithmic divergence of the self-energy on the mass shell.
To better understand where it comes from, we come back to the derivation of (%
\ref{sigmams1}). Re-writing (\ref{sigmams1}) as (\ref{sigma12}) to
logarithmic accuracy, we now argue that the two logarithmic terms in (\ref
{sigma12}) come from two different processes, as shown in Fig.\ref
{twoprocesses}. In the first process (Fig.\ref{twoprocesses}a), all four
momenta are close to each other, and in the second one (Fig.\ref
{twoprocesses}b), the net momentum of the two incoming particles is close to
zero, whereas the momenta of the outgoing particles are close to their
initial values. In terms of the momentum transfers, both processes are of
forward-scattering type.
To see this, we notice that for generic $\omega /\epsilon _{k}$, i.e., not
too close to the mass shell, the logarithmic form of the self-energy is due
to $1/Q$ behavior of the momentum integrand at $v_{F}Q\gg \Omega ,\omega $.
This $1/Q$ form in 2D results from the combination of two facts: (i) the
polarization operator $\Pi (Q,\Omega )$ behaves as $\Omega /v_{F}Q$, and
(ii) the imaginary part of the fermionic propagator, integrated over the
angle $\theta $ between $Q$ and external momentum $k$, behaves as $1/Q$. The
product of the two terms yields $\int QdQ/Q^{2}$ that gives rise to a
logarithm. It is easy to make sure that for $v_{F}Q\gg \Omega $, typical
values of $\theta $ are close to $\pm \pi /2$, the deviation from these
values being of order $\left| \Omega \right| /v_{F}Q$. That means that the
external momentum ($\mathbf{k)}$ and the internal (small) one ($\mathbf{Q)}$
(as labeled in Fig.\ref{fig1}a) are nearly orthogonal to each other. The
same reasoning also works for the polarization bubble. If the two internal
momenta in that bubble are $\mathbf{p}$ and $\mathbf{p}+\mathbf{Q}$, then
typical $\mathbf{p}$ and $\mathbf{Q}$ are also nearly orthogonal. Since both
$\mathbf{k}$ and $\mathbf{p}$ are orthogonal to the same $\mathbf{Q}$, and
both are confined to the near vicinity of the Fermi surface, they are either
near each other, or near the opposite points of the Fermi surface. If $%
\mathbf{p}$ and $\mathbf{k}$ are close to each other, all three internal
fermionic momenta in the second-order diagram are close to external $\mathbf{%
k}$, if $\mathbf{p}$ is close to $-\mathbf{k}$, out of three internal
momenta one is close to $\mathbf{k}$, while the other two are close to $-%
\mathbf{k}$. These two regions of intermediate momenta give rise to two
logarithms in (\ref{sigma12}). The logarithm that diverges on the mass shell
comes from a region where all momenta are close to $\mathbf{k}$. To see
this, we recall that the actual divergence is the consequence of the fact
that both the polarization bubble and the angle-averaged $G^{\prime \prime }(%
\mathbf{k}+\mathbf{Q},\omega +\Omega )$ at the mass shell possess
square-root singularities in the form $1/\sqrt{(v_{F}Q)^{2}-(\Omega )^{2}}$
such that the product of the two gives $(v_{F}Q)^{2}-(\Omega )^{2})^{-1}$,
and the momentum integral diverges. The square-root singularities come from
near parallel $\mathbf{p}$ and $\mathbf{Q}$ and $\mathbf{k}$ and $\mathbf{Q}$%
, respectively. Obviously then, $\mathbf{k}$ and $\mathbf{p}$ are near
parallel, i.e., they are located near the same point at the Fermi surface.
With a little more effort, one can show that as $\omega $ approaches $%
\epsilon _{k}$, typical angles between $\mathbf{p}$ and $\mathbf{Q}$ and
between $\mathbf{k}$ and $\mathbf{Q}$, both move from near $\pi /2$ (or $%
-\pi /2$) to near zero, but in such a way that $\mathbf{k}$ and $\mathbf{p}$
remain parallel. This once again confirms that the divergent logarithm comes
from the process in Fig.\ref{twoprocesses}a (all internal momenta are close
to $k$), while the ``conventional'', non-divergent $\omega ^{2}\ln \omega $%
-term comes from the process in Fig.\ref{twoprocesses}b.

The analysis can be extended to finite $T$, and the (anticipated) result is
that $\Sigma _{1}^{\prime \prime }$ given by (\ref{2.51}) comes from forward
scattering, while $\Sigma _{2}^{\prime \prime }$ given by (\ref{2.52}) comes
from back scattering.
\begin{figure}[tbp]
\centerline{\epsfxsize=4in \epsfbox{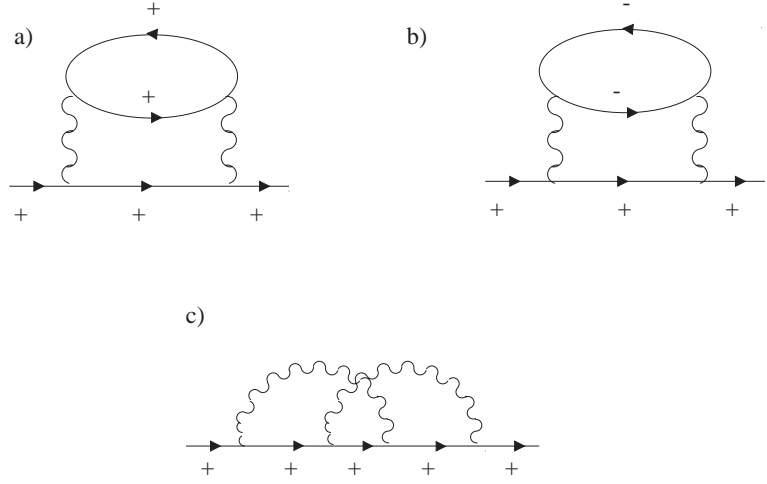}}
\caption{Non-trivial diagrams for the self-energy in 1D. $\pm $ denotes the
propagator of a right/left moving fermion.}
\label{selfenergy1D}
\end{figure}

It is interesting follow the same arguments for $D=1.$ In this case,
processes in Fig.\ref{twoprocesses} acquire even simpler physical meanings:
process a) is forward scattering of fermions of the same chirality, e.g.,
two right-moving fermions scatter into two right-moving ones, whereas
process b) is forward scattering of fermions of opposite chirality, e.g., a
right-moving fermion scatters at a left-moving one so that their respective
chiralities are conserved. In the g-ology notations, vertex a) is $g_{2}$
and vertex b) is $g_{4}$ \cite{emery}. In the Luttinger model, when only
forward scattering is taken into account, the self-energy of, e.g.,
right-moving fermions is represented by the set of diagrams shown in Fig.\ref
{selfenergy1D} \cite{DL}, where $\pm $ denotes propagators of right/left
moving species
\begin{equation}
G_{\pm }\left( k,\omega \right) =\frac{1}{i\omega -\epsilon _{k}^{\pm }},%
\text{ }\epsilon _{k}^{\pm }=v_{F}\left( k\mp k_{F}\right) .
\end{equation}
Diagrams a) and c) contain two vertices of type a) from Fig.\ref
{twoprocesses} whereas diagram b) contain vertices of type b) from Fig.\ref
{twoprocesses}. The imaginary parts of the retarded polarization bubbles for
right- and left-moving fermions for $\left| Q\right| \rightarrow 0$ are
given by
\begin{equation}
\Pi _{R\pm }^{^{\prime \prime }}=\frac{Q}{2}\delta \left( \Omega \mp
v_{F}Q\right) .
\end{equation}
The delta-function form of $\Pi _{R\pm }^{^{\prime \prime }}$ is due to the
fact that in 1D and for $\left| Q\right| \rightarrow 0$ the particle-hole
continuum shrinks to two lines in the $\left( \Omega ,Q\right) $ plane
described by $\Omega =v_{F}\left| Q\right| .$ The combination of the
diagrams a) and c) in Fig.\ref{selfenergy1D} yields for the imaginary part
of the self-energy
\begin{equation}
\left[ \Sigma _{R+}^{\prime \prime }\left( \mathbf{k},\omega \right) \right]
_{a)+c)}=\frac{U^{2}}{8\pi v_{F}^{2}}\omega ^{2}\delta \left( \omega
-\epsilon _{k}^{+}\right) ,  \label{sigma1D}
\end{equation}
We see that $\Sigma _{R+}^{\prime \prime }$ given by (\ref{sigma1D}), which
is a 1D analog of our $\Sigma _{2}^{\prime \prime }$ from(\ref{sigma2}), is
very singular on the mass shell but vanishes outside the mass shell. At the
same time, diagram b) in Fig.\ref{selfenergy1D} yields

\begin{equation}
\left[ \Sigma _{R+}^{\prime \prime }\left( \mathbf{k},\omega \right) \right]
_{b)}=\left\{
\begin{array}{l}
\frac{U^{2}}{2\pi v_{F}^{2}}\left( \left|
\omega \right| -\left| \epsilon _{k}^{+}\right| \right) ,\text{ for }|\omega|
>\left| \epsilon _{k}^{+}\right| ;\nonumber \\
0,\text{ otherwise.}
\end{array}
\right. .
\end{equation}

This self-energy vanishes on the mass shell, but for a generic $\omega
/\epsilon _{k}$ it yields $\left[ \Sigma _{R+}^{\prime \prime }\left(
k,\omega \right) \right] _{b)}\propto \left| \omega \right| $. This $|\omega
|$ dependence obviously implies that Fermi-liquid behavior is in danger.

Which of the two terms is actually relevant? In 1D, the answer is well
known: the summation of infinite series of the diagrams yields the
non-Fermi-liquid behavior, and the resulting state--the Luttinger liquid--is
free of singularities on the mass shell. This implies that the mass shell
singularity of Eq.(\ref{sigma1D}) is completely eliminated by the
re-summation of diagrams to all orders in the interaction. This can be shown
explicitly either via Ward identities or using the bosonization~\cite
{bosonization}. Furthermore, the exact solution of the model with only type
a) scattering (``$g_{4}$ -model'') yields a \emph{free-}Fermi-gas behavior
with a renormalized Fermi velocity, i.e., no mass-shell singularity. This
all implies that the mass shell singularity found in the second-order
self-energy diagram in 1D is an artificial one and is eliminated by higher
order diagrams.

The same elimination of the mass shell singularity holds in 2D, as we now
demonstrate. Indeed, as we mentioned before, the logarithmic divergence in (%
\ref{sigma12}) at $\omega =\epsilon _{k}$ is the consequence of the matching
of the two square-root singularities: one resulting from the angular
integration of the fermionic Green's function, and another one being the $1/%
\sqrt{(v_{F}Q)^{2}-\Omega ^{2}}$ singularity in $\Pi _{R}^{\prime \prime
}\left( Q,\Omega \right) $. Suppose now that the interaction gets
renormalized (screened) by higher-order terms in $U$ \ so that $U\rightarrow
U(Q,\Omega )$. The combination $U^{2}\Pi _{R}^{\prime \prime }\left(
Q,\Omega \right) $ in (\ref{2.2}) is now replaced by $U_{R}^{\prime \prime
}(Q,\Omega ).$ In the RPA approximation (which is not a controllable one for
a short-range interaction),
\begin{eqnarray}
&&U_{R}^{\prime \prime }\left( Q,\Omega \right) =\frac{U^{2}\Pi _{R}^{\prime
\prime }\left( \Omega ,Q\right) }{\left[ 1+U\Pi _{R}^{\prime }\left(
Q,\Omega \right) \right] ^{2}+\left[ U\Pi _{R}^{\prime \prime }\left( \Omega
,Q\right) \right] ^{2}}  \notag \\
&=&\frac{2\pi }{m}\tilde{U}^{2}~\Omega ~\frac{\sqrt{(v_{F}Q)^{2}-\Omega ^{2}}%
\theta \left( v_{F}Q-\left| \Omega \right| \right) }{(1+\tilde{U})^{2}\left[
(v_{F}Q)^{2}-\Omega ^{2}\right] +\tilde{U}^{2}\Omega ^{2}},
\end{eqnarray}
where $\tilde{U}\equiv mU/2\pi .$ Obviously, $U_{R}^{\prime \prime }$ now
vanishes at $Q=\left| \Omega \right| /v_{F}$, and the divergence is
eliminated. At the same time, the logarithmic dependence on $\Omega $ in (%
\ref{2.5}), and hence the $\omega ^{2}\ln \omega $ form of the self-energy,
survive as they come from typical $\Omega \sim v_{F}Q$ for which $\Pi
_{R}^{\prime }(Q,\Omega )$ and $\Pi _{R}^{\prime \prime }(Q,\Omega )$ are of
the same order, and hence the screened interaction is of the order of the
bare one. Note that this reasoning is also valid for the Coulomb
interaction, for which the RPA approximation is asymptotically exact in the
high-density limit.

Another argument that the mass-shell singularity is artificial is that it is
eliminated, already at the second order of interaction, if one takes into
account the curvature of the fermionic dispersion. Indeed, in obtaining (\ref
{sigmams1}), we linearized the fermionic dispersion near the Fermi surface,
i.e., approximated $\epsilon _{k+q}$ by $\epsilon _{k}+v_{F}q\cos \theta $.
Using the full quadratic dispersion, we obtain, instead of (\ref{sigmams2})
\begin{equation}
\Sigma _{R}^{\prime \prime }(\mathbf{k},\omega )=\frac{mU^{2}}{8\pi
^{3}v_{F}^{2}}\int_{0}^{\omega }d\Omega ~\Omega ~\ln \frac{W^{2}}{\epsilon
_{k}-\omega }][2\Omega -\omega +\epsilon _{k}]+\Delta \left( \omega ,\Omega
\right) |~,  \label{sigmams4}
\end{equation}
where
\begin{equation}
\Delta \left( \omega ,\Omega \right) =\frac{\Omega ^{2}}{2E_{F}}~\left(
3\omega -\epsilon _{k}-\Omega \right)
\end{equation}
and where, for the sake of definiteness, we assumed $\omega >0$. On the mass
shell, $\omega =\epsilon _{k}$, the integration over $\Omega $ yields a
finite result
\begin{equation}
\Sigma _{R}^{\prime \prime }(\mathbf{k},\omega )|_{\omega =\epsilon _{k}}=%
\frac{3U^{2}m}{16\pi ^{3}v_{F}^{2}}\omega ^{2}\ln \frac{W}{\left| \omega
\right| }.  \label{sigmams3}
\end{equation}
The crossover between Eqs.(\ref{sigmams1}) and (\ref{sigmams3}) occurs when,
inside the log in Eq.(\ref{sigmams1}), $\Delta \left( \omega ,\Omega \right)
$ becomes comparable to the other term in the denominator, i.e., when
\begin{equation}
\left| \omega -\epsilon _{k}\right| \sim \omega ^{2}/W.
\end{equation}
For $\left| \omega -\epsilon _{k}\right| \gg \omega ^{2}/W,$ the leading
asymptotic behavior of $\Sigma _{R}^{\prime \prime }(\mathbf{k},\omega )$ is
given by (\ref{sigmams1}) and for $\left| \omega -\epsilon _{k}\right| \ll
\omega ^{2}/W$ it is given by (\ref{sigmams3}). A general formula which
interpolates between the two limiting cases might, in principle, be obtained
but we do not dwell on this here. Notice that $\Sigma _{R}^{\prime \prime }(%
\mathbf{k},\omega )$ on the mass shell is by a factor of 3/2 bigger than its
value on the Fermi surface, which means that, for fixed $\omega ,$ the slope
of $\Sigma _{R}^{\prime \prime }(\mathbf{k},\omega )$ as function of $%
\epsilon _{k}$ becomes steeper as the mass shell is approached.

The same result can be also obtained by calculating the quasiparticle
lifetime for $T=0$ which, by definition, is taken directly on the mass
shell. For $D=2,$ the Fermi Golden Rule gives
\begin{equation}
1/\tau \left( \omega \right) =\frac{U^{2}m}{8\pi ^{3}}\int_{0}^{\omega
}d\Omega \int_{-\Omega }^{0}d\omega ^{\prime }\int^{W/v_{F}}dQQ\int d\theta
\int d\theta ^{\prime }\delta \left( \Omega -\varepsilon _{\mathbf{k}%
}+\varepsilon _{\mathbf{k-Q}}\right) \delta \left( \Omega -\varepsilon _{%
\mathbf{p+Q}}+\varepsilon _{\mathbf{p}}\right) ,  \label{tau}
\end{equation}
where $\varepsilon =\epsilon _{\mathbf{k}}$ ,$\omega ^{\prime }=\epsilon _{%
\mathbf{p}}$, \ and $\theta $,$\theta ^{\prime }$ are the angles between $%
\mathbf{k}$ and $\mathbf{q}$ and $\mathbf{p}$ and $\mathbf{q}$\textbf{, }%
respectively, and $W$ is the ultraviolet energy cutoff. For linearized
dispersion the arguments of the first and second delta-functions in (\ref
{tau}) reduce to $\Omega +v_{F}Q\cos \theta ,\Omega -v_{F}Q\cos \theta
^{\prime },$ respectively. Each of the angular integrations yields a factor
of $2/\sqrt{(v_{F}Q)^{2}-\Omega ^{2}},$ and the integral over $Q$%
% reduces to
\begin{equation}
A=\int_{|\Omega |/v_{F}}^{W/v_{F}}dQQ\frac{1}{(v_{F}Q)^{2}-\Omega ^{2}}
\end{equation}
diverges logarithmically at the lower limit. To regularize the singularity,
one must to keep the higher-order terms in $\varepsilon _{\mathbf{k-Q}}$ and
$\varepsilon _{\mathbf{p+Q}}$. On the mass shell,
\begin{subequations}
\begin{eqnarray}
\varepsilon _{\mathbf{k}}-\varepsilon _{\mathbf{k-Q}} &=&v_{F}Q\left( 1+%
\frac{\omega }{2E_{F}}\right) \cos \theta -\frac{Q^{2}}{2m}; \\
\varepsilon _{\mathbf{p}}-\varepsilon _{\mathbf{p+Q}} &=&-v_{F}Q\left( 1+%
\frac{\omega ^{\prime }}{2E_{F}}\right) \cos \theta ^{\prime }-\frac{Q^{2}}{%
2m}.
\end{eqnarray}
Now the integral over $Q$ takes the form
\end{subequations}
\begin{equation}
A=\int^{W/v_{F}}dQQ\frac{1}{\sqrt{\left( v_{F}Q\right) ^{2}-\Omega
^{2}-\delta }}\frac{1}{\sqrt{\left( v_{F}Q\right) ^{2}-\Omega ^{2}+\delta
^{\prime }}},
\end{equation}
where
\begin{subequations}
\begin{eqnarray}
\delta &=&\Omega \left( v_{F}Q\frac{\omega }{E_{F}}+\frac{Q^{2}}{m}\right)
\label{delta} \\
\delta ^{\prime } &=&\Omega \left( v_{F}Q\frac{\omega ^{\prime }}{E_{F}}+%
\frac{Q^{2}}{m}\right)  \label{deltaprime}
\end{eqnarray}
The lower limit in the integral is such that the arguments of the square
roots are positive. The momentum integral is controlled by $Q\sim |\Omega
|/v_{F}$. To logarithmic accuracy, one can then just replace $Q$ by $|\Omega
|/v_{F}$ in (\ref{delta},\ref{deltaprime}). After this replacement, the
momentum integration can be easily performed and gives
\end{subequations}
\begin{equation}
A=\frac{1}{2v_{F}^{2}}\ln \frac{E_{F}^{2}W}{\Omega ^{2}\left( \omega +\omega
^{\prime }\right) +\Omega ^{3}}.
\end{equation}
We next have to perform the frequency integration. It is easy to verify
that, in the two integrals over frequency, the dominant contributions come
from the regions $\Omega \sim \omega ^{\prime }\sim \omega $. To logarithmic
accuracy, one can then simplify $A$ to
\begin{equation}
\frac{3}{2v_{F}^{2}}\ln \frac{W}{\omega }.
\end{equation}
We also used the fact that $E_{F}\sim W$. Substituting this into (\ref{tau}
and performing frequency integrations we obtain finally
\begin{equation}
\frac{1}{\tau \left( \omega \right) }=\frac{3U^{2}m}{8\pi ^{3}v_{F}^{2}}%
\omega ^{2}\ln \frac{W}{\omega }.
\end{equation}
We see that $1/\tau (\omega )$ is finite--the only memory left about the
mass-shell singularity for the linearized spectrum is the enhanced numerical
prefactor. Identifying $1/\tau $ with $2\Sigma _{R}^{\prime \prime }$, we
see that the results for $1/\tau $ and $\Sigma ^{\prime \prime }(\omega
=\epsilon _{k})$ coincide, as indeed they should.

\section{Polarization bubble near $2k_F$}

%\section{Appendix A}
\label{app_A}

In this Appendix, we show that the computation of a non-analytic piece in
the particle-hole bubble at $Q\approx 2k_{F}$ can be always performed in
such a way that the dominant contribution comes from fermions near the Fermi
surface and with nearly antiparallel momenta $\pm \mathbf{Q}/2$. We do this
in two ways. First, we compute $\Pi _{ph}(Q,\Omega _{m})$ explicitly and
check where the non-analyticity comes from. Second, we compute $\Pi
_{ph}(Q,\Omega _{m})$ by linearizing the dispersion of fermions, forming the
polarization bubble, near $\pm \mathbf{Q}/2$ and show that the
non-analyticity in $\Pi _{ph}(Q,\Omega _{m})$ comes from the lower limit of
momentum integration and therefore does not depend on the upper cutoff
imposed by the linearization procedure.

\subsection{Explicit computation}

\label{sec:appA_1} Consider first $T=0$. Labeling the momenta of internal
fermionic lines in the polarization bubble as\textbf{\ }$\mathbf{p}\pm
\mathbf{Q}/2$ and for $T=0$, we obtain in Matsubara frequencies
\begin{equation}
\Pi (Q,\Omega _{m})=-\int \frac{d^{2}pd\omega }{(2\pi )^{3}}G(\mathbf{p}+%
\frac{\mathbf{Q}}{2},\omega _{n}+\Omega _{m})~G(\mathbf{p}-\frac{\mathbf{Q}}{%
2},\omega _{n}).  \label{a1.1}
\end{equation}
For a circular Fermi surface
\begin{equation}
\epsilon _{\mathbf{p}\pm \mathbf{Q}/2}=\frac{p^{2}-k_{F}^{2}}{2m}\pm \frac{%
pQ\cos \theta }{2m}+\frac{Q^{2}}{8m}.  \label{a1.1a}
\end{equation}
Substituting (\ref{a1.1a}) into (\ref{a1.1}) and integrating over frequency
and then over $p,$ we obtain for $Q<2k_{F}$
\begin{equation}
\Pi (Q,\Omega _{m})=\frac{m}{2\pi }\left( 1-2\frac{m\Omega _{m}}{\pi Q^{2}}%
\int_{0}^{\pi /2}\frac{d\theta }{\cos ^{2}\theta }\left[ \arctan \frac{p_{1}%
}{m\Omega _{m}}-\arctan \frac{p_{2}}{m\Omega _{m}}\right] \right) ,
\label{a1.2}
\end{equation}
where
\begin{equation}
p_{1,2}=Q\cos {\theta }\sqrt{k_{F}^{2}-\frac{Q^{2}}{4}\sin ^{2}\theta }\pm
\frac{1}{2}Q^{2}\cos ^{2}\theta .
\end{equation}
For $Q=2k_{F}$, we have $p_{1}=4k_{F}^{2}\cos ^{2}\theta ,~p_{2}=0$, and (%
\ref{a1.2}) reduces to
\begin{equation}
\Pi (Q,\Omega _{m})=\frac{m}{2\pi }\left( 1-2\frac{m\Omega _{m}}{\pi Q^{2}}%
\int_{0}^{\pi /2}\frac{d\theta }{\cos ^{2}\theta }\arctan \frac{%
4k_{F}^{2}\cos ^{2}\theta }{m\Omega _{m}}\right) =\frac{m}{2\pi }\left( 1-%
\frac{1}{2}\left( \frac{|\Omega _{m}|}{E_{F}}\right) ^{1/2}\right) .
\label{a1.3}
\end{equation}
It is easy to see that the integral comes from $\cos ^{2}\theta \sim |\Omega
|/E_{F}$, i.e. typical $\mathbf{p}$ are nearly orthogonal to $\mathbf{Q}$.
Furthermore, in the integral over $p$, typical $p$ were of order $Q\cos
\theta $. Hence typical $p$ are of order $\sqrt{m|\Omega _{m}|}$, i.e. at
vanishing $\Omega $, the integration is indeed confined to internal momenta
which nearly coincide with $\pm \mathbf{Q/2}$.

The same reasoning is valid also for $Q$ in a narrow range near $2k_{F}$.
For $Q\leq 2k_{F}$, Eq. (\ref{a1.3}) can be re-written as
\begin{equation}
\Pi (Q,\Omega _{m})=\frac{m}{2\pi }\left( 1-2\frac{m\Omega _{m}}{\pi Q^{2}}%
\int_{0}^{\pi /2}\frac{d\theta }{\cos ^{2}\theta }\arctan \frac{Q^{2}\cos
^{2}\theta }{(m\Omega _{m})(1-Q^{2}\cos ^{2}\theta \epsilon ^{2})}\right) ,
\label{a1.4}
\end{equation}
where $\epsilon ^{2}=(Q^{2}/4-k_{F}^{2})/m|\Omega _{m}|$. Assuming that the
integral is dominated by $\theta $ near $\pi /2$ and expanding $\theta $ to
linear order near $\pi /2$, we obtain after simple manipulations
\begin{equation}
\Pi (Q,\Omega _{m})=\frac{m}{2\pi }\left( 1-\frac{(m|\Omega _{m}|)^{1/2}}{%
\pi k_{F}}\int_{0}^{\infty }dz\arctan \frac{1}{z^{2}-\epsilon ^{2}}\right).
\label{a1.5}
\end{equation}
We see that the integral is convergent, i.e., the linearization of $\cos
\theta $ near $\pi /2$ does not lead to cutoff-dependent integrals. This
implies that the non-analytic piece in the polarization operator comes from
typically small $\cos \theta $ and hence from typically small internal $%
p\propto \cos \theta $. Evaluating the integral over $z$ in (\ref{a1.5}), we
obtain
\begin{equation}
\Pi (Q,\Omega _{m})=\frac{m}{2\pi }\left( 1-\frac{1}{2}~\left( \frac{|\Omega
_{m}|}{E_{F}}\right) ^{1/2}~\left( \frac{v_{F}{\tilde{Q}}}{|\Omega _{m}|}+%
\sqrt{1+\left( \frac{v_{F}{\tilde{Q}}}{|\Omega _{m}|}\right) ^{2}}\right)
^{1/2}\right) .  \label{a1.6}
\end{equation}
where ${\tilde{Q}}=Q-2k_{F}$. This is the result that we cited in the text
(Eq. (\ref{2.02})).

For $Q>2k_{F}$, i.e., ${\tilde{Q}}>0$, the calculations proceed in the same
way. Integrating over $p$ and over $\omega $ and again expanding to linear
order near $\theta =\pi /2$ we obtain after straightforward manipulations
\begin{equation}
\Pi (Q,\Omega _{m})=\frac{m}{2\pi }\left( 1-\left( \frac{\tilde{Q}}{k_{F}}%
\right) ^{1/2}-\frac{1}{\pi \sqrt{2}}~\left( \frac{|\Omega _{m}|}{E_{F}}%
\right) ^{1/2}\int_{0}^{1/2\epsilon }dz\arctan \frac{(1-4\epsilon
^{2}z^{2})^{1/2}}{z^{2}+\epsilon ^{2}}\right) .  \label{a1.7}
\end{equation}
Evaluating the integral we find that the result reduces to Eq. (\ref{a1.6}).

\subsection{Another way of calculating $\Pi (Q\approx 2k_{F},\Omega _{m})$}

\label{sec:appA_2}

For completeness, we also compute the non-analytic part in $\Pi (Q,\Omega
_{m})$ near $2k_{F}$ by explicitly restricting the integral over $p$ in (\ref
{a1.1}) to small $p$ and assuming that $\mathbf{p}$ is nearly orthogonal to $%
\mathbf{Q}$. This calculation shows in a more direct way that typical values
$p$ are indeed small. To avoid lengthy calculations, we assume that $%
Q=2k_{F} $ and aim at reproducing the $\sqrt{\Omega _{m}}$ non-analyticity.
For $Q=2k_{F}$, the energies on the internal fermionic lines are $\epsilon
_{k_{F}\hat{n}+\mathbf{p}}$ and $\epsilon _{-k_{F}\hat{n}+\mathbf{p}}$.
Introducing $x=v_{F}p$ and $\gamma =1/(2mv_{F}^{2})=1/(4E_{F})$, expanding $%
\cos \theta \approx \tilde{\theta}$, where $\tilde{\theta}=\pi /2-\theta $
and substituting into (\ref{a1.1}), we obtain
\begin{equation}
\Pi (2k_{F},\Omega _{m})=\frac{1}{4\pi ^{3}v_{F}^{2}}\int_{-\infty }^{\infty
}d\tilde{\theta}\int_{-\infty }^{\infty }d\omega \int_{0}^{\infty }dx~\frac{x%
}{(x\tilde{\theta}+\gamma x^{2}-i\omega _{n})~(x\tilde{\theta}-\gamma
x^{2}+i(\omega _{n}+\Omega _{m}))}.  \label{a1.8}
\end{equation}
Introducing $y=x\tilde{\theta}$ and integrating over $y,$ we obtain after
simple manipulations with variables
\begin{equation}
\Pi (2k_{F},\Omega _{m})=\frac{1}{2\pi ^{2}v_{F}^{2}}\int_{0}^{\infty
}dx\int_{|\Omega _{m}|}^{\infty }\frac{zdz}{z^{2}+4\gamma ^{2}x^{4}}.
\label{a1.9}
\end{equation}
The integration is elementary and yields
\begin{equation}
\Pi (2k_{F},\Omega )=\frac{1}{8\pi v_{F}^{2}\sqrt{\gamma }}\int_{|\Omega
_{m}|}^{\infty }\frac{dz}{\sqrt{z}}.  \label{a1.10}
\end{equation}
The divergence of the integral at the upper limit simply reflects that a
constant term in the polarization bubble cannot be reproduced this way.
However, the lower limit of the integral over $z$ yields a universal and
non-analytic contribution to $\Pi (2k_{F},\Omega _{m})$ of the form
\begin{equation}
\Pi _{\text{sing}}(2k_{F},\Omega _{m})=-\frac{1}{4\pi v_{F}^{2}}\left( \frac{%
|\Omega _{m}|}{\gamma }\right) ^{1/2}=-\frac{m}{4\pi }\left( \frac{|\Omega
_{m}|}{E_{F}}\right) ^{1/2}.  \label{a1.11}
\end{equation}
This coincides with Eq. (\ref{a1.6}). It is essential that this result does
not depend on the upper limit, and hence typical internal momenta scale with
external $\Omega $. This obviously implies that typical values of $p$ are
indeed small.

\subsection{Finite temperature}

\label{appA_3}

At finite $T$, a sharp $\sqrt{Q-2k_{F}}$ non-analyticity in the static
polarization operator is softened in qualitatively the same way as it is
softened by a finite $\Omega _{m}$ at $T=0$. In general,
\begin{equation}
\Pi (Q,\Omega _{m},T)=\frac{m}{2\pi }\left( 1-\left( \frac{T}{E_{F}}\right)
^{1/2}\Phi \left( \frac{v_{F}|Q-2k_{F}|}{T},\frac{\Omega _{m}}{T}\right)
\right).  \label{a1.12}
\end{equation}
We could not find a simple analytical expression for the scaling function $%
\Phi (x,y)$ at arbitrary values of its arguments. At $Q=2k_{F}$ and $\Omega
=0$, $\Phi (0,0)\approx 0.339$.

\section{{Equivalence of $q=0$ and $q=2k_{F}$ contributions to the
self-energy} \label{app_B}}

In this appendix, we explicitly compute the contribution to the self-energy
from the $2k_{F}$ non-analyticity in the particle-hole bubble, and show that
it is equal to the backscattering part of the self-energy from the $q=0$
non-analyticity. We will also show that the non-analytic self-energy can be
equally viewed as coming from the singularity in the particle-particle
channel at zero total momentum and frequency.

\subsection{$2k_{F}$ part of the self-energy from the particle-hole channel}

\label{sec:appB_1} Since our goal is to verify a general reasoning that $q=0
$ and $2k_{F}$ contributions to $\Sigma (k,\omega )$ are equal, we focus on
the case $T=\epsilon _{k}=0$, compute the $2k_{F}$ part of the self-energy
in Matsubara frequencies and compare the prefactor for $\omega _{n}\ln
|\omega _{n}|$ term with $1/2$ of that in Eq. (\ref{2.50111}).

For a contact interaction, the second-order self-energy is
\begin{equation}
\Sigma (k,\omega _{n})=-U^{2}\int \int \frac{d^{2}qd\Omega _{m}}{(2\pi )^{3}}%
~G_{0}(\mathbf{k}+\mathbf{q},\omega _{n}+\Omega _{m})~\Pi _{ph}(q,\Omega
_{m}).  \label{a1.1_1}
\end{equation}
Assuming $q=2k_{F}+\tilde{q}$, where $\tilde{q}$ is small, we expand $%
\epsilon _{k+q}$ as $\epsilon _{\mathbf{k+q}}=-\epsilon _{k}+v_{F}\tilde{q}%
+2v_{F}k_{F}(1+\cos \theta )$, where $\theta $ is the angle between $\mathbf{%
k}$ and $\mathbf{q}$. As we already discussed in Appendix \ref{app_A}, only $%
\theta $ near $\theta =\pi $ matter (i.e., typical $\mathbf{q}$ is nearly
antiparallel to $\mathbf{k}$), hence we can further approximate $\epsilon _{%
\mathbf{k+q}}$ as
\begin{equation}
\epsilon _{\mathbf{k+q}}\approx -\epsilon _{k}+v_{F}\tilde{q}+v_{F}k_{F}%
\tilde{\theta}^{2},  \label{a1.2_1}
\end{equation}
where ${\tilde{\theta}}=\pi -\theta $. Substituting (\ref{a1.2_1}) into (\ref
{a1.1}), we obtain, setting $\epsilon _{k}=0$,
\begin{equation}
\Sigma (\omega _{n})=\frac{2U^{2}k_{F}}{(2\pi )^{3}}~\int_{-\infty }^{\infty
}d\tilde{q}~d\Omega _{m}~\int_{0}^{\infty }d{\tilde{\theta}}~\frac{1}{v_{F}%
\tilde{q}+v_{F}k_{F}{\tilde{\theta}}^{2}-i(\omega _{n}+\Omega _{m})}~\Pi
_{ph}(\tilde{q},\Omega _{m}),  \label{a1.3_1}
\end{equation}
where $\Pi _{ph}(\tilde{q},\Omega _{m})$ is given by (\ref{2.02}).

As an exercise, consider first a model case where $\Pi _{ph}(\tilde{q}%
,\Omega _{m})$ is static. To ensure convergence, we assume that the static
behavior holds for $\Omega _{m}\ll \Omega _{0},$ where $\Omega _{0}$ is some
ultraviolet cutoff (of order bandwidth), and for larger $\Omega _{m}$, $\Pi
_{ph}({\tilde{q}},\Omega _{m})$ rapidly falls off. The angular integration
in (\ref{a1.1_1}) reduces the range of integration over $\Omega _{m}$ to $%
-\omega _{n}\leq \Omega _{m}\leq \omega _{n},$ hence at the smallest $\omega
_{n}$, $\Sigma \propto \omega _{n}$. This accounts for the conventional mass
renormalization. We now show that there are no non-analytic corrections to $%
\Sigma $ in this model. A static $\Pi _{ph}(\tilde{q},0)$ is non-analytic
only for $\tilde{q}>0$, where $\Pi _{ph}(\tilde{q},\Omega _{m})=(m/2\pi )(1-(%
\tilde{q}/k_{F})^{1/2})$. Substituting non-analytic part of the polarization
bubble into (\ref{a1.3}) introducing $\tilde{\theta}=\sqrt{r/v_{F}k_{F}}\cos
\phi ,~\sqrt{\tilde{q}}=\sqrt{r/v_{F}}\sin \phi $, and integrating over $%
\phi $, we obtain for a potentially non-analytic part of the self-energy
\begin{equation}
\Sigma (\omega )=-\frac{mU^{2}}{32\pi ^{3}v_{F}^{2}}~\int_{-\infty }^{\infty
}d\Omega _{m}~\int_{0}^{\infty }\frac{rdr}{r-i(\omega _{n}+\Omega _{m})}.
\label{a1.4_1}
\end{equation}
One can easily make sure that this integral yields a regular $\omega $ term
(determined by high-energy states), but no universal $\omega ^{2}\ln \omega $%
-term. This implies, as we mentioned several times in the text, that static $%
\Pi _{ph}(\tilde{q},0)$ does not give rise to a non-analyticity in the
fermionic self-energy.

\begin{figure}[tbp]
\centerline{\epsfxsize=2in
 \epsfbox{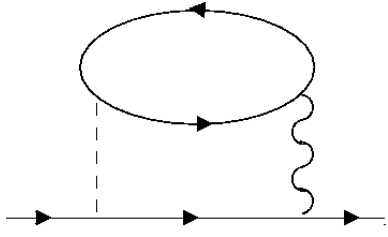}}
\caption{Hartree contribution to the self-energy for scattering at a Friedel
oscillation.}
\label{fig:impurity}
\end{figure}

It is instructive to distinguish this case from the impurity problem. If one
of the interaction lines in Fig.1a is replaced by an impurity line, as shown
in Fig.~\ref{fig:impurity}, the diagram in Fig. 1a transforms into the
Hartree diagram describing the scattering of fermions by Friedel
oscillations produced by impurities. In the ballistic limit, $\left| \omega
_{n}\right| \tau \gg 1,$ it suffices to keep only a single impurity line
connecting $G$ and $\Pi _{ph}$ and also neglect disorder in G. For
delta-correlated disorder with amplitude $V$, the analytic expression for
the diagram in Fig. \ref{fig:impurity} takes the form
\begin{equation}
\Sigma (k,\omega _{n})=-2UV\int \frac{d^{2}q}{(2\pi )^{3}}~G_{0}(\mathbf{k}+%
\mathbf{q},\omega _{n})~\Pi _{ph}(q,0).
\end{equation}
The particle-hole bubble is still static, but in distinction to \ref{a1.1_1}
we no longer have to perform a summation over frequencies. The non-analytic
piece in $\Sigma (\omega )$ is then given by, instead of Eq.(\ref{a1.4}),
\begin{equation}
\Sigma (\omega _{n})=-\frac{mUV}{4\pi ^{3}v_{F}^{2}}~~\int_{0}^{\infty }%
\frac{rdr}{r-i\omega _{n}}.
\end{equation}
$\allowbreak $Due to the absence of the integral over $\Omega _{m},$ (\ref
{a1.4_1}) does yield a universal contribution $\Sigma (\omega _{n})\propto
-i\omega _{n}\ln \left( -i\omega _{n}\right) $ which comes from the lower
limit of the integral over $r.$ Upon analytic continuation, one obtains $%
\Sigma _{R}^{\prime }\propto \omega \ln \left| \omega \right| $ and $\Sigma
_{R}^{^{\prime \prime }}\left( \omega \right) \propto \left| \omega \right|
. $ The linear in $\omega $ form of $\Sigma _{R}^{^{\prime \prime }}\left(
\omega \right) $ is related to the Hartree part of the linear-in-$T$ term in
the conductivity at finite $T$ \cite{zna} .

We now come back to the electron-electron interaction, when a
non-analytic-in-$\omega _{n}$ behavior of $\Sigma \left( \omega _{n}\right) $
can be obtained if the $\Omega _{m}-$ dependence is retained in $\Pi _{ph}(%
\tilde{q},\Omega _{m})$. As with any logarithmic singularity, typical $%
\tilde{q}$ should well exceed $\omega _{n}/v_{F}$. We will see that typical $%
\Omega _{m}$ are of order $\omega _{n}$. Typical values of $v_{F}\tilde{q}$
well exceed then typical values of $\Omega _{m}$, and one can expand $\Pi
_{ph}(\tilde{q},\Omega _{m})$ in powers of $\Omega _{m}/v_{F}\tilde{q}$. \
For $\tilde{q}>0$, the frequency expansion of $\Pi _{ph}(\tilde{q},\Omega
_{m})$ starts at a constant and holds in even powers of $\Omega _{m}/v_{F}%
\tilde{q}$. We have already verified that the constant term does not give
rise to an $\omega ^{2}\ln \omega $-piece in $\Sigma (\omega )$. At $\tilde{q%
}<0$, however, the leading expansion term has the same $|\Omega _{m}|$
non-analyticity as the polarization operator near $q=0$. The non-analytic
behavior in frequency is crucial as it prevents one from eliminating a
low-energy non-analyticity by closing the integration contour in the
integral over $\Omega _{m}$ over a distant semi-circle in a half-plane where
the denominator in (\ref{a1.3}) has no poles.

Expanding $\Pi _{ph}(\tilde{q},\Omega _{m})$ at $\tilde{q}<0$ and $\Omega
_{m}\ll v_{F}|{\tilde{q}}|$ , we find
\begin{equation}
\Pi _{ph}({\tilde{q}},\Omega _{m})=\frac{m}{2\pi }\left( 1-\frac{|\Omega
_{m}|}{2v_{F}(k_{F}|\tilde{q}|)^{1/2}}\right) .  \label{a1.5_1}
\end{equation}
Substituting this result into Eq.(\ref{a1.3}) and keeping only potentially
non-analytic piece, we obtain
\begin{equation}
\Sigma (\omega )=-\frac{2mU^{2}k_{F}}{(2\pi )^{4}}~\int_{-\infty }^{\infty
}d\Omega _{m}~\int_{-\infty }^{0}d\tilde{q}~\int_{0}^{\infty }d\tilde{\theta}%
~\frac{1}{v_{F}\tilde{q}+v_{F}k_{F}\tilde{\theta}^{2}-i(\omega _{n}+\Omega
_{m})}~\frac{|\Omega _{m}|}{v_{F}(k_{F}|\tilde{q}|)^{1/2}}.  \label{a1.6_1}
\end{equation}
Introducing $x^{2}=-v_{F}\tilde{q}~$and $y^{2}=v_{F}k_{F}\tilde{\theta}^{2}$
, we obtain from (\ref{a1.6_1})
\begin{equation}
\Sigma (\omega )=-\frac{mU^{2}}{4\pi ^{4}v_{F}^{2}}~\int_{-\infty }^{\infty
}d\Omega _{m}~|\Omega _{m}|~\int_{0}^{\infty }\int_{0}^{\infty }\frac{dx~dy}{%
y^{2}-x^{2}-i(\omega _{n}+\Omega _{m})}.  \label{a1.7_1}
\end{equation}
Introducing further $y=\sqrt{r}\cos \phi /2,~x=\sqrt{r}\sin \phi /2$ and
integrating over $\phi $ first, we obtain
\begin{eqnarray}
\Sigma (\omega ) &=&-\frac{mU^{2}}{8\pi ^{4}v_{F}^{2}}~\int_{-\infty
}^{\infty }d\Omega _{m}~|\Omega _{m}|~\int_{0}^{W}\frac{dr}{r}\int_{0}^{\pi }%
\frac{d\phi }{\cos \phi -i(\omega +\Omega )/r}  \notag \\
&=&-i\frac{mU^{2}}{8\pi ^{3}v_{F}^{2}}~\int_{-\infty }^{\infty }d\Omega
_{m}~|\Omega _{m}|~\text{sgn}(\omega _{n}+\Omega _{m})~\int_{0}^{W^{2}}\frac{%
dr}{(r^{2}+(\omega _{n}+\Omega _{m})^{2})^{1/2}}.  \label{a1.8_1}
\end{eqnarray}
Evaluating the integral over $r$ with logarithmic accuracy and integrating
finally over $\Omega _{m},$ we obtain
\begin{equation}
\Sigma (\omega )=-i\frac{mU^{2}}{16\pi ^{3}v_{F}^{2}}~\omega _{n}^{2}\ln
\frac{W^{2}}{\omega _{n}^{2}}.  \label{a1.9_1}
\end{equation}
This coincides with the half of Eq.(\ref{2.50111}) for $\epsilon _{k}=0.$

To further clarify this issue, we redo the calculation in a different way.
Namely, we use the fact that for $\mathbf{Q}=-2\mathbf{k}+\mathbf{Q}^{\prime
}$, and $Q^{\prime }$ small, the non-analytic part of the bubble $\Pi
_{ph}(Q^{\prime },\Omega _{m})$ comes from the region of small $Q^{\prime
\prime }$ in the following integral:
\begin{equation}
\Pi _{ph}(Q^{\prime },\Omega _{m})=-\int \int \frac{d^{2}Q^{\prime \prime
}d\omega _{n}}{(2\pi )^{3}}~G_{\mathbf{k+Q}^{\prime \prime },\omega
_{n}}~G_{-\mathbf{k+Q}^{\prime }\mathbf{+Q}^{^{\prime \prime }},\omega
_{n}+\Omega _{m}}.  \label{a1.10_1}
\end{equation}
Now, we want to re-express the $2k_{F}$ contribution as an effective $Q=0$
contribution. To do this, we substitute (\ref{a1.10_1}) into (\ref{a1.1})
and change the order of the integrations over $Q^{\prime }$ and $Q^{^{\prime
\prime }}$. The non-analytic ``$2k_{F}$'' piece in the self-energy then
becomes
\begin{equation}
\Sigma (k,\omega _{n})=-U^{2}\int \int \frac{d^{2}Q^{\prime \prime }d\Omega
_{m}}{(2\pi )^{3}}~G_{\mathbf{k+Q}^{\prime },\omega _{n}+\Omega _{m}}{\tilde{%
\Pi}}(Q^{\prime },\Omega _{m}),  \label{a1.11_1}
\end{equation}
where the effective particle hole-bubble
\begin{equation}
{\tilde{\Pi}}(Q^{\prime \prime },\Omega _{m})=-\int \int \frac{%
d^{2}Q^{^{\prime \prime }}d\omega _{n}}{(2\pi )^{3}}~G_{-\mathbf{k+Q}%
^{\prime \prime },\omega _{n}}~G_{-\mathbf{k+Q}^{\prime }\mathbf{+Q}%
^{^{\prime \prime }},\omega _{n}+\Omega _{m}}.  \label{a1.12_1}
\end{equation}
This $\tilde{\Pi}$ is a part of the particle-hole polarization bubble at
small momentum transfer, which comes from the integration over \emph{small} $%
Q^{^{\prime \prime }}$. We now show that for $\Omega _{m}\ll
v_{F}Q^{^{\prime \prime }}$, i.e., in the momentum/frequency range which
yields the logarithm in the self-energy, the non-analytic part of ${\tilde{%
\Pi}}(Q,\Omega _{m})$ is a half of that in $\Pi (Q,\Omega _{m})$. This would
again imply that the $2k_{F}$ contribution to the self-energy coincides with
the (non-divergent) $\Sigma _{2}$ part of ``$q=0$'' contribution.

The calculation proceeds as follows. We set $\epsilon _{k}=0$ and write $%
\epsilon _{-\mathbf{k+Q}^{\prime }}=-x\cos \theta _{1}+\gamma x^{2}$ where $%
x=v_{F}Q^{\prime }$, $\gamma =(2mv_{F}^{2})^{-1}$, and $\theta _{1}$ is the
angle between $\mathbf{k}$ and $\mathbf{Q}^{\prime }$. Similarly, $\epsilon
_{-\mathbf{k+Q}^{\prime }}=-x\cos \theta _{1}-y\cos \theta _{2}+\gamma
(x^{2}+y^{2}+2xy\cos (\theta _{1}-\theta _{2}))$, where $y=v_{F}Q$, and $%
\theta _{2}$ is the angle between $\mathbf{k}$ and $\mathbf{Q}$. As we said,
we need to evaluate ${\tilde{\Pi}}$ for $\theta _{2}$ close to $\pm \pi /2$,
and small $y$. We therefore neglect $y^{2}$ terms and set $\theta
_{2}\approx \pi /2$ for definiteness. We assume and then verify that $\Omega
/v_{F}Q$ term in the polarization operator comes from $\theta _{1}$ near $%
\pm \pi _{/}2$ and linearize $\cos \theta _{1}$ near these points. The
integration over $\theta _{1}$ is then straightforward, and performing it we
obtain that the integration over $\omega _{n}$ is confined to $-\Omega
_{m}<\omega _{n}<0$ (for definiteness we assumed that $\Omega _{m}>0$). The
result is
\begin{equation}
{\tilde{\Pi}}(Q,\Omega _{m})=\frac{i\Omega _{m}}{4\pi ^{2}v_{F}^{2}\gamma y}%
\int_{0}^{\infty }dp\left( \frac{1}{\cos \theta _{2}-2p-i\Omega _{m}}+\frac{1%
}{\cos \theta _{2}+2p-i\Omega _{m}}\right) ,  \label{a1.13}
\end{equation}
where we introduced $p=\gamma x$. The integration over $p$ is
straightforward, and for small $\Omega _{m}$ and $\cos \theta _{2}$ the
integral over $dp$ yields $i\pi /2$. Substituting this into (\ref{a1.13}) we
obtain
\begin{equation}
{\tilde{\Pi}}(Q,\Omega _{m})=\frac{1}{2}~\frac{m}{2\pi }\frac{\Omega _{m}}{%
v_{F}Q}.  \label{a1.14}
\end{equation}
It is essential that the momentum integral is confined to small $p=Q^{\prime
}/k_{F}$ (typical $p\sim \cos \theta _{2}\sim \Omega _{m}/v_{F}Q$), and
hence we are really restricting our momentum integral to small $Q^{\prime }$%
. Comparing (\ref{a1.14}) and (\ref{2.01}) we see that, as we expected, (\ref
{a1.14}) is a half of a non-analytic part of $\Pi (Q,\Omega _{m})$ at $%
\Omega _{m}\ll v_{F}Q$. Another half obviously comes from the region of
large $Q^{\prime }$, which cannot be re-expressed as a ``$2k_{F}$
contribution.''

\subsection{An alternative computation of the self-energy, via $\Pi_{pp} (q,
\Omega)$}

\label{sec:appB_2} We discussed in the text that the second-order
self-energy can be equivalently presented as a convolution of the fermionic
Green's function and the particle-particle bubble
\begin{equation}
\Sigma (\omega _{n})=-U^{2}\int \int \frac{d^{2}qd\Omega _{m}}{(2\pi )^{3}}%
~G_{0}(-k_{F}\hat{n}+\mathbf{q},-\omega _{n}+\Omega _{m})~\Pi _{pp}(q,\Omega
_{m}),  \label{a2_2}
\end{equation}
where $\Pi _{pp}(q,\Omega _{m})=(m/2\pi )\ln \left[ B/\left( |\Omega _{m}|+%
\sqrt{\Omega _{m}+(v_{F}q)^{2}}\right) \right] $. Substituting this $\Pi
_{pp}$ into the self-energy and expanding $\epsilon _{-k_{F}\hat{n}+\mathbf{q%
}}$ as $-v_{F}q\cos \theta $, we obtain for $\epsilon _{k}=0$
\begin{equation}
\Sigma (\omega _{n})=-\frac{mU^{2}}{8\pi ^{4}v_{F}^{2}}~\int_{-\infty
}^{\infty }d\Omega _{m}\int_{0}^{\pi }d\theta \int_{0}^{W}dx\frac{x}{x\cos
\theta +i(\Omega _{m}-\omega _{n})}~\ln \frac{B}{|\Omega _{m}|+\sqrt{\Omega
_{m}+x^{2}}}.  \label{a2_3}
\end{equation}
Assuming, as before, that typical $\Omega _{m}$ are of order $\omega _{n}$,
while typical $x=v_{F}q$ are much larger, we can further expand under the
logarithm and obtain
\begin{equation}
\Sigma (\omega _{n})=\frac{mU^{2}}{8\pi ^{4}v_{F}^{2}}~\int_{-\infty
}^{\infty }d\Omega _{m}|\Omega _{m}|~\int_{0}^{\pi }d\theta \int_{0}^{W}dx%
\frac{1}{x\cos \theta +i(\Omega _{m}-\omega _{n})}.  \label{a2_4}
\end{equation}
The integration over $\theta $ yields
\begin{equation}
\Sigma (\omega _{n})=i\frac{mU^{2}}{8\pi ^{3}v_{F}^{2}}~\int_{-\infty
}^{\infty }d\Omega _{m}|\Omega _{m}|~\text{sgn}(\Omega _{m}-\omega
_{n})~\int_{0}^{W}dx\frac{1}{\sqrt{x^{2}+(\Omega _{m}-\omega _{n})^{2}}}.
\label{a2_5}
\end{equation}
Evaluating the integral over $x$ to logarithmic accuracy, we finally
obtain
\begin{equation}
\Sigma (\omega _{n})=-i\frac{mU^{2}}{16\pi ^{3}v_{F}^{2}}\omega _{n}\ln
\frac{W^{2}}{\omega _{n}}.  \label{a2.6}
\end{equation}
This  coincides precisely with Eq.(\ref{a1.9}).

\section{Evaluation of $\Sigma _{R}^{\prime }(\protect\omega ,\protect%
\epsilon _{k})$ on the mass shell}

\label{app_se}

In this Appendix, we present the calculation of the real part of the
fermionic self-energy on the mass shell. We will be only interested in the
non-analytic piece of the self-energy. The non-analytic part of $\Sigma
_{R}^{\prime }(\omega )$ is simply twice of $\Sigma _{2}^{\prime }(\omega ),$
where, according to Eq. (\ref{2.712}), $\Sigma _{2}^{\prime }(\omega )$ can
be written as
\begin{equation}
\Sigma _{2}^{\prime }(\omega )=-\frac{mU^{2}}{16\pi ^{4}v_{F}^{2}}~\omega
~Z(\omega ,T),  \label{p.1}
\end{equation}
where
\begin{eqnarray}
Z(\omega ,T) &=&\int_{-\infty }^{\infty }d\Omega \Omega \text{ }\mathcal{P}%
\int_{0}^{\infty }\frac{dE}{E^{2}-\omega ^{2}}~\left( \coth \frac{\Omega }{2T%
}-\tanh \frac{\Omega +E}{2T}\right)  \notag \\
&&~\left( \frac{E}{\omega }\ln \left| \frac{2\Omega +E-\omega }{2\Omega
+E+\omega }\right| +\ln \frac{|(2\Omega +E)^{2}-\omega ^{2}|}{W^{2}}\right) .
\label{p.2}
\end{eqnarray}

We first find $Z(\omega )$ at $T=0$. The term with $\coth $ and $\tanh $
functions restricts the integration over $\Omega $ to the interval $-E\leq
\Omega \leq 0$. Introducing the rescaled variables $E=\omega z$ and $\Omega
=-\omega zx$ and assuming for definiteness that $\omega >0$ (and thus $z>0$%
), we obtain
\begin{equation}
Z(\omega )=2\omega ~\int_{0}^{\infty }\frac{dz}{z^{2}-1}~\int_{0}^{1}xdx%
\left[ z\ln \left| \frac{z(2x-1)+1}{z(2x-1)-1}\right| +\ln
[z^{2}(2x-1)^{2}-1]\right].\label{p.3}
\end{equation}
Introducing a new variable via $y=2x-1$ and eliminating terms that vanish by
parity we obtain, instead of Eq.(\ref{p.3}),
\begin{equation}
Z(\omega )=\omega ~\int_{0}^{\infty }\frac{dz}{z^{2}-1}~\int_{0}^{1}dy\left[
zy\ln \left| \frac{zy+1}{zy-1}\right| +\ln (z^{2}y^{2}-1)\right] .
\label{p.4}
\end{equation}
The integration over $y$ is now straightforward, and performing it we obtain
\begin{equation}
Z(\omega )=\omega \left[ \int_{0}^{\infty }\frac{dz}{z^{2}-1}\left( \frac{1}{%
z}\ln {\left| \frac{z+1}{z-1}\right| }+\ln {z^{2}-1}\right)
+\int_{0}^{\infty }\frac{dz}{2z}\ln \left| \frac{z+1}{z-1}\right| \right] .
\label{p.5}
\end{equation}
Finally, we use the values of the following integrals
\begin{equation}
\int_{0}^{\infty }dz~\frac{\ln {z^{2}-1}}{z^{2}-1}=\frac{\pi ^{2}}{2}%
;~~\int_{0}^{\infty }\frac{dz}{z^{2}-1}~\frac{1}{z}\ln \left| \frac{z+1}{z-1}%
\right| =-\frac{\pi ^{2}}{4};~~\int_{0}^{\infty }\frac{dz}{2z}\ln \left|
\frac{z+1}{z-1}\right| =\frac{\pi ^{2}}{4}.  \label{p.6}
\end{equation}
Substituting these results into Eq.(\ref{p.5}) we obtain
\begin{equation}
Z(\omega )=\omega \frac{\pi ^{2}}{2}.  \label{p.61}
\end{equation}
Substituting this further into Eq.(\ref{p.1}) we reproduce Eq.(\ref{2.713}).

We next consider finite $T$. As a first step, we show that one can safely
replace $\coth \Omega /(2T)$ by $\tanh \Omega /(2T)$ in (\ref{p.2}). Indeed,
this replacement changes $Z(\omega )$ by
\begin{eqnarray}
&&Z_{\text{extra}}(\omega ,T)=2\int_{-\infty }^{\infty }d\Omega \frac{\Omega
}{\sinh {\frac{\Omega }{T}}}\mathcal{P}\int_{0}^{\infty }\frac{dE}{%
E^{2}-\omega ^{2}}~  \notag \\
&&~\left( \frac{E}{\omega }\ln \left| \frac{2\Omega +E-\omega }{2\Omega
+E+\omega }\right| +\ln \frac{|(2\Omega +E)^{2}-\omega ^{2}|}{W^{2}}\right) .
\label{p.7}
\end{eqnarray}
The integration over $E$ in $Z_{\text{extra}}(\omega ,T)$ is straightforward
and performing it we obtain
\begin{equation}
Z_{\text{extra}}(\omega ,T)=2\int_{-\infty }^{\infty }d\Omega \frac{\Omega }{%
\sinh \frac{\Omega }{T}}\frac{\ln ^{2}|2\Omega +\omega |-\ln ^{2}|2\Omega
-\omega |}{\omega }.  \label{p.8}
\end{equation}
This integral obviously vanishes as the integrand is odd in $\Omega $.

Next, one can readily check that in the expression for $Z,$ obtained by
replacing $\coth \Omega /(2T)$ $\rightarrow \tanh \Omega /(2T),$ i.e., in
\begin{eqnarray}
Z(\omega ,T) &=&\int_{-\infty }^{\infty }d\Omega \Omega \text{ }\mathcal{P}%
\int_{0}^{\infty }\frac{dE}{E^{2}-\omega ^{2}}~\left( \tanh {\frac{\Omega }{%
2T}}-\tanh {\frac{\Omega +E}{2T}}\right)  \notag \\
&&~\left( \frac{E}{\omega }\ln \left| {\frac{2\Omega +E-\omega }{2\Omega
+E+\omega }}\right| +\ln \frac{|(2\Omega +E)^{2}-\omega ^{2}|}{W^{2}}\right)
,  \label{p.9}
\end{eqnarray}
the integrand vanishes at large $|\Omega |$, $E$. Hence the integration can
be performed in the infinite limits and Eq.(\ref{p.9}) can be rewritten as a
difference of two terms with the same argument of $\tanh $, upon changing in
the second term to a new variable $\Omega +E$. Carrying out this procedure,
introducing new variables, and converting the $\Omega $ integration to the
integral over positive $\Omega $, we obtain
\begin{equation}
Z(\omega ,T)=\int_{0}^{\infty }d\Omega \tanh \frac{\Omega }{2T}~\Psi \left(
\frac{2\Omega }{|\omega |}\right) ,  \label{p.10}
\end{equation}
where
\begin{equation}
\Psi (a)=\mathcal{P}\int_{0}^{\infty }\frac{dxx}{x^{2}-1}\left[ a\ln \left|
\frac{a^{2}-(x-1)^{2}}{a^{2}-(x+1)^{2}}\right| +x\ln \left| \frac{%
(a-1)^{2}-x^{2}}{(a+1)^{2}-x^{2}}\right| +\ln \left| \frac{(a-x)^{2}-1}{%
(a+x)^{2}-1}\right| \right]  \label{p.11}
\end{equation}
The integration over $x$ is tedious but straightforward, and yields
\begin{equation}
\Psi (x)=\left\{
\begin{array}{l}
-\pi ^{2}\frac{a}{2},\text{ for }a<2;\nonumber \\
-\pi ^{2},\text{ for }a>2.
\end{array}
\right. .
\end{equation}
Substituting this into Eq.(\ref{p.10}) and integrating over $\Omega $ we
obtain
\begin{equation}
Z(\omega )=A+\frac{\pi ^{2}|\omega |}{2}g\left( \frac{\omega }{T}\right) ,
\label{p.12}
\end{equation}
where $A<0$ is a (formally infinite) constant which is irrelevant to us as
it accounts for the high energy contribution to a linear in $\omega $ term
in $\Sigma _{2}^{\prime }(\omega )$, $g(x)$ is universal scaling
function
\begin{equation}
g(x)=1+\frac{4}{x^{2}}\left[ \frac{\pi ^{2}}{12}+\text{Li}_{2}\left(
-e^{-x}\right) \right],  \label{p.14}
\end{equation}
and Li$_{2}(x)$ is a polylogarithmic function. This is the result we cited
in Eq.(\ref{2.111}).

At $x=\infty $, \emph{i.e}., at $T=0$, we have $g(\infty )=1$ and thus $%
Z(\omega )=(\pi ^{2}/2)|\omega |$. This coincides with Eq.(\ref{p.61}). In
the opposite limit of $|\omega |\ll T$, we use property
\begin{equation}
\text{Li}_{2}\left( -e^{-x}\right) =\sum_{k=1}^{\infty }\frac{\left(
-e^{-x}\right) ^{k}}{k^{2}}\approx -\frac{\pi ^{2}}{12}+x\ln 2+\mathcal{O}%
(x^{2}).  \label{p.15}
\end{equation}
Substituting this into Eqs.(\ref{p.14}) and (\ref{p.12}) we obtain that up
to a constant,
\begin{equation}
Z(\omega \ll T)\approx 2\pi ^{2}\ln 2~T  \label{p.151}
\end{equation}
Substituting this further into Eq.(\ref{p.1}) we obtain
\begin{equation}
\Sigma _{2}^{\prime }(\omega )=-\frac{mU^{2}\ln 2}{8\pi ^{2}v_{F}^{2}}\omega
T.  \label{p.16}
\end{equation}
This is the result we cited in Eq.(\ref{2.714}).

As an independent verification, we reproduced (\ref{p.16}) by computing the
temperature derivative of $Z(\omega )$ in the limit $\omega \rightarrow 0$.
[It is essential to take the limit, not just set $\omega =0$]. Evaluating
the derivative, setting $\omega \rightarrow 0$, introducing dimensionless
variables, and eliminating the terms which vanish by parity, we obtain
\begin{equation}
\frac{\partial Z(\omega ,T)}{\partial T}=4\int_{0}^{\infty }\frac{dxx}{\cosh
^{2}x}\text{ }\mathcal{P}\int_{0}^{\infty }\frac{dy}{y}\ln \left| \frac{y+1}{%
y-1}\right| .  \label{p.17}
\end{equation}
The integral over $x$ gives $\ln 2$, whereas that over $y$ yields, upon
integrating by parts,
\begin{equation}
\text{ }\mathcal{P}\int_{0}^{\infty }\frac{dy\ln y}{y^{2}-1}=\frac{\pi ^{2}}{%
4}.  \label{p.18}
\end{equation}
Combining the two terms we obtain $\frac{\partial Z(\omega ,T)}{\partial T}%
=2\pi ^{2}\ln 2$, i.e., up to a constant $Z(\omega \ll T)=2\pi ^{2}\ln 2~T$.
This coincides with Eq.(\ref{p.151}).

\section{$2k_F$ contributions to diagrams 1 and 3 in Fig.5}

\label{app_C}

In this Appendix we present explicit calculations of the $2k_{F}-$
contributions to diagrams 1 and 3 in Fig. \ref{fig:diag}.

\subsection{$2k_{F}$ part of diagram 1}

\label{appC_1} We first verify that the non-analytic $O(|Q|)$ term that
results from the $2k_{F}$ non-analyticity in the particle-hole bubble is
indeed the same as the contribution from the $q=0$ non-analyticity. For $%
\delta \chi _{1}^{q=0}(Q,0)$ we obtained in (\ref{c5_5})
\begin{equation}
\delta \chi _{1}^{q=0}(Q,0)=\chi _{0}~\frac{2}{3\pi }~\left( \frac{mU}{4\pi }%
\right) ^{2}~\frac{|Q|}{k_{F}},  \label{c1.1}
\end{equation}
Now we explicitly evaluate $\delta \chi _{1}^{2k_{F}}(Q,0)$. The general
expression for the diagram 1 is
\begin{equation}
\delta \chi _{1}(Q,0)=-8U^{2}\int \frac{d^{2}k~d^{2}q~d\omega d\Omega }{%
(2\pi )^{6}}G_{0}^{2}(\mathbf{k},\omega )G_{0}(\mathbf{k}+\mathbf{Q},\omega
)G_{0}(\mathbf{k}+\mathbf{q},\omega +\Omega )\Pi (q,\Omega ).  \label{c1.2}
\end{equation}
For $q\approx 2k_{F}$ the quasiparticle energies can be approximated by
\begin{equation}
\epsilon _{k}=v_{F}(k-k_{F}),~~\epsilon _{k+Q}=\epsilon _{k}+v_{F}Q\cos
\theta _{1},~~\ \epsilon _{k+q}=-\epsilon _{k}+v_{F}\tilde{q}%
+2v_{F}k_{F}(1+\cos \theta _{2}),  \label{c1.11}
\end{equation}
where $\tilde{q}=q-2k_{F}$, and $\theta _{1}$ and $\theta _{2}$ are the
angles between $\mathbf{k}$ and $\mathbf{Q}$ and between $\mathbf{k}$ and $%
\mathbf{q}$, respectively. As we have said several times before, the $2k_{F}$
non-analyticity comes from internal fermionic momenta in the particle-hole
bubble that nearly coincide with the external one. In our notations, this
implies that $\theta _{2}$ is close to $\pi $. We can then expand in $\cos
\theta _{2}$ upon which $\epsilon _{k+q}$ reduces to $\epsilon
_{k+q}=-\epsilon _{k}+v_{F}\tilde{q}+v_{F}k_{F}(\pi -\theta _{2})^{2}$.
Substituting this expansion into (\ref{c1}), integrating over $\epsilon _{k}$
and then over $\omega $ (this requires more care than for the $q=0$ case),
and introducing dimensionless variables $\bar{q}={\tilde{q}}/|Q|,~{\bar{%
\omega}}=\Omega /(v_{F}|Q|)$, $k_{F}(\pi -\theta _{2})^{2}=|Q|{\bar{\theta}}%
^{2}$ and polar coordinates as $\bar{q}=r\cos \phi ,{\bar{\omega}}=r\sin
\phi $, we obtain from (\ref{c1})
\begin{eqnarray}
\delta \chi _{1}^{2k_{F}}(Q,0) &=&\frac{4mU^{2}(k_{F}|Q|)^{1/2}}{\pi
^{4}v_{F}}~\int_{0}^{\pi }d\phi \Pi (\phi )~\mbox{Re}\int rdr\int_{0}^{\pi }%
\frac{d\theta _{1}}{\cos ^{2}\theta _{1}}~ \\
&&\int_{0}^{\infty }d{\tilde{\theta}_{2}}~\left[ \frac{\cos \theta _{1}}{{%
\tilde{\theta}}^{2}+re^{i\phi }}-\ln \frac{{\tilde{\theta}}^{2}+re^{i\phi
}+\cos \theta _{1}}{{\tilde{\theta}}^{2}+re^{i\phi }}\right] .  \label{c8}
\end{eqnarray}
The polarization operator is now given by (\ref{2.02}) which in the new
variables takes the form
\begin{equation}
\Pi (\phi )=\frac{m}{2\pi }\left( 1-\left( \frac{r|Q|}{k_{F}}\right)
^{1/2}~\cos \frac{\phi }{2}\right).
\end{equation}
Performing the integration over $r$ and keeping only the contribution which
comes from low energies, we again find that only the non-analytic piece in $%
\Pi (\phi )$ contributes to order $|Q|$, and this universal contribution is
\begin{eqnarray}
\delta \chi _{1}^{2k_{F}}(Q,0) &=&\frac{2m^{2}U^{2}|Q|}{3\pi ^{5}v_{F}}%
~\int_{0}^{\pi }d\phi \cos \frac{\phi }{2}~\mbox{Re}~~\int_{0}^{\infty }%
\frac{d{\tilde{\theta}_{2}}}{({\tilde{\theta}}^{2}+e^{i\phi })^{3}}~
\label{c9} \\
&&\times \int_{0}^{\pi }d\theta _{1}~\cos \theta _{1}~\ln \frac{{\tilde{%
\theta}}^{2}+e^{i\phi }}{\cos \theta _{1}}.
\end{eqnarray}
The integral over $\theta _{1}$ yields $i\pi $. Evaluating then the integral
over ${\tilde{\theta}_{2}}$, we obtain
\begin{eqnarray}
\delta \chi _{1}^{2k_{F}}(Q,0) &=&\frac{m^{2}U^{2}|Q|}{8\pi ^{4}v_{F}}%
~\int_{0}^{\pi }d\phi \cos \frac{\phi }{2}~\sin \frac{5\phi }{2}  \notag \\
&=&\chi _{0}~\frac{2}{3\pi }~\left( \frac{mU}{4\pi }\right) ^{2}~\frac{|Q|}{%
k_{F}};~~~\chi _{0}\equiv \frac{m}{\pi }.  \label{c10}
\end{eqnarray}
Comparing this result with Eq. (\ref{c5_5}), we see that the two expressions
are indeed equal. We emphasize again that in order to obtain this result,
one has to include the frequency dependence of $\Pi (q,\omega )$ near $%
q=2k_{F}$. Had we replaced $\Pi (q,\omega )$ by its static value ($\Pi
(q,0)) $, we would have not obtained Eq. (\ref{c10}).

\subsection{$2k_{F}$ part of the diagram 3}

\label{appC_2} In an explicit form,
\begin{equation}
\delta \chi _{3}(Q,0)=-4U^{2}\int \frac{d^{2}k~d^{2}q~d\omega d\Omega }{%
(2\pi )^{6}}G_{0}(\mathbf{k},\omega )G_{0}(\mathbf{k}+\mathbf{Q},\omega
)G_{0}(\mathbf{k}+\mathbf{q},\omega +\Omega )G_{0}(\mathbf{k}+\mathbf{q}+%
\mathbf{Q},\omega +\Omega )\Pi _{ph}(q,\Omega ),  \label{c2.1}
\end{equation}
Assuming that $q$ is close to $2k_{F}$ and expanding quasiparticle energies
as in Eq.(\ref{c1.11}) we obtain after rescaling the variables and
restricting with only the non-analytic part
\begin{eqnarray}
&&\delta \chi _{3}^{2k_{F}}(Q,0)=\chi _{0}\frac{m^{2}U^{2}}{4\pi ^{6}}\frac{%
|Q|}{k_{F}}\int_{-\infty }^{\infty }dx\int_{0}^{\infty }d\Omega \left( \sqrt{%
x+i\Omega }+\sqrt{x-i\Omega }\right)  \notag \\
&&\int_{0}^{\pi }d\theta \int_{0}^{\infty }dy\int_{-\infty }^{\infty
}dz\int_{-\infty }^{\infty }d\omega \frac{1}{(z-i\omega )(z+\cos \theta
-i\omega )}  \label{c2.2} \\
&&\times \frac{1}{~(z-x-y^{2}+i(\omega +\Omega ))~(z-x-y^{2}+\cos \theta
+i(\omega +\Omega ))},
\end{eqnarray}
where $\chi _{0}=m/\pi $. Performing the integration over $z$ first we
obtain after straightforward manipulations
\begin{eqnarray}
&&\delta \chi _{3}^{2k_{F}}(Q,0)=-\chi _{0}\frac{m^{2}U^{2}}{\pi ^{5}}\frac{%
|Q|}{k_{F}}\int_{-\infty }^{\infty }dx\int_{0}^{\infty }d\Omega \left( \sqrt{%
x+i\Omega }+\sqrt{x-i\Omega }\right)  \notag \\
&&\int_{0}^{\pi }d\theta \int_{0}^{\infty }dy\mbox{Im}\left[
\int_{0}^{\infty }d\omega \frac{1}{(x+y^{2}-i(\omega +\Omega
))~((x+y^{2}-i(\omega +\Omega ))^{2}-\cos ^{2}\theta )}\right].  \label{c2.3}
\end{eqnarray}
Introducing then $q=r\cos \phi $ and $\Omega =r\sin \pi $ such that $\left(
\sqrt{x+i\Omega }+\sqrt{x-i\Omega }\right) =2\sqrt{r}\cos \phi _{2}$ and
rescaling $\omega \rightarrow r\omega $, $y\rightarrow \sqrt{r}y$, we obtain
\begin{eqnarray}
&&\delta \chi _{3}^{2k_{F}}(Q,0)=-2\chi _{0}\frac{m^{2}U^{2}}{\pi ^{5}}\frac{%
|Q|}{k_{F}}\int_{0}^{\pi }d\phi \cos {\phi /2} \times  \notag \\
&&\mbox{Im}\left[ \int_{0}^{\infty }dy\int_{0}^{\infty }d\omega
\int_{0}^{\infty }r^{2}dr\int_{0}^{\pi }d\theta \frac{1}{(e^{-i\phi
}+y^{2}-i\omega )~(r^{2}(e^{-i\phi }+y^{2}-i\omega )^{2}-\cos ^{2}\theta )}%
\right].  \label{c2.4}
\end{eqnarray}
Introducing then $p=r(e^{-i\phi }+y^{2}-i\omega )$, replacing the
integration over $r$ by the integration over $p$, and restricting with the
universal contribution from the lower limit of the $p-$integral, we obtain,
after integrating over $p$ and then over $\theta $
\begin{equation}
\delta \chi _{3}^{2k_{F}}(Q,0)=-2\chi _{0}\frac{m^{2}U^{2}}{\pi ^{4}}\frac{%
|Q|}{k_{F}}\int_{0}^{\pi }d\phi \cos {\phi /2}\int_{0}^{\infty
}dy\int_{0}^{\infty }d\omega Re\left[ \frac{1}{(\omega +i(y^{2}+e^{-i\phi
}))^{4}}\right].  \label{c2.5}
\end{equation}
The integration over $\omega $ is now straightforward. Performing it and
then evaluating the integral over $y$ we finally obtain
\begin{equation}
\delta \chi _{3}^{2k_{F}}(Q,0)=\chi _{0}\frac{m^{2}U^{2}}{8\pi ^{3}}\frac{|Q|%
}{k_{F}}\int_{0}^{\pi }d\phi \cos {\frac{\phi }{2}}\sin {\frac{5\phi }{2}}%
=\chi _{0}\frac{2}{3\pi }~\left( \frac{mU}{4\pi }\right) ^{2}~\frac{|Q|}{%
k_{F}}.  \label{c2.6}
\end{equation}
This is the result that we cited in the text.

\section{$2k_{F}$ contribution to $\protect\chi _{s}(Q=0,T)$ for a static
Lindhard function}

\label{app_d}

In this appendix we show that the thermal smearing of the static Lindhard
function by itself does give rise to a linear-in-$T$ term in the uniform
spin susceptibility, but does not account for the full linear-in-$T$
dependence of $\chi _{s}(0,T)$--the latter also contains a contribution from
finite frequencies.

\begin{figure}[tbp]
\centerline{\epsfxsize=3in
 \epsfbox{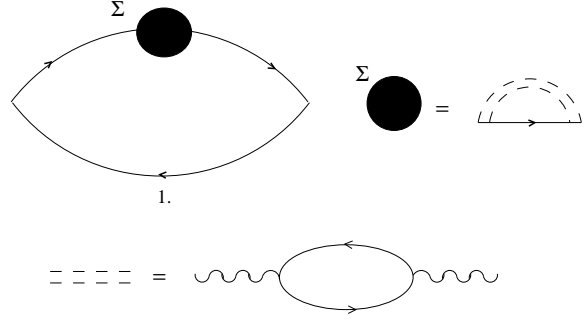}}
\caption{Diagram 1 of Fig.3 as the first-order self-energy insertion.}
\label{fig:chi1_1}
\end{figure}

The computation proceeds as follows. Because a static polarization operator
can be viewed as an effective interaction, diagram 1 can be re-expressed as
the first-order self-energy insertion (see Fig.\ref{fig:chi1_1})
\begin{equation}
\delta \chi _{1,static}=-4T\sum_{\omega _{n}}\int \frac{d^{2}k}{\left( 2\pi
\right) ^{2}}\left[ G\left( k,\omega _{n}\right) \right] ^{3}\Sigma _{\text{%
eff}}(\epsilon _{k}),
\end{equation}
where the effective self-energy is given by
\begin{equation*}
\Sigma _{\text{eff}}(\epsilon _{k})=2U^{2}~T\sum_{n}\int \frac{d^{2}q}{%
\left( 2\pi \right) ^{2}}\Pi (q,0,T)~G_{0}(\mathbf{k+q},\omega
_{n})=2U^{2}\int \frac{d^{2}q}{\left( 2\pi \right) ^{2}}~\Pi
(q,0,T)n_{F}(\epsilon _{\mathbf{k+q}})
\end{equation*}
This self-energy is obviously independent of $\omega _{n}.$ Although the
static polarization operator $\Pi (q,0,T)$ is not known exactly, it can be
cast into an intergal form~\cite{millis} convenient for further
calculations. We have
\begin{equation}
\Pi (q,0,T)=\frac{m}{2\pi }\left[ 1-\frac{k_{F}^{2}}{8mT}\int_{-1}^{(\frac{q%
}{2k_{F}})^{2}-1}\frac{dz}{\cosh ^{2}\frac{k_{F}^{2}z}{4mT}}~\left( 1-\frac{%
1+z}{(q/2k_{F})^{2}}\right) ^{1/2}\right] .  \label{bo1}
\end{equation}
Re-writing $\left[ G\right] ^{3}=\left( 1/2\right) \partial ^{2}G/\partial
\epsilon _{k}^{2},$ summing over $\omega _{n}$ with the help of an identity $%
T\sum_{\omega _{n}}G\left( k,\omega _{n}\right) =n_{F}\left( \epsilon
_{k}\right) -1/2$ where $n_{F}\left( z\right) =\left( e^{z/T}+1\right) ^{-1}$
is the Fermi distribution function, and integrating by parts twice, we
obtain
\begin{equation}
\delta \chi _{1,static}=-\chi _{0}^{2D}\int_{-\infty }^{\infty }d\epsilon
_{k}n_{F}\left( \epsilon _{k}\right) \frac{d^{2}\Sigma _{\text{eff}%
}(\epsilon _{k})}{d\epsilon _{k}^{2}}.  \label{c23_3}
\end{equation}
where $\chi _{0}^{2D}=m/\pi $. A non-analytic temperature dependence of $%
\delta \chi _{1,static}$ is due the region of $q$ near $2k_{F}$, where $\Pi
(q,0,T)$ is singular. Expanding, as before, $\epsilon _{\mathbf{k+q}}$ near $%
q=2k_{F}$ and along the direction of $\mathbf{q}$ nearly antiparallel to $%
\mathbf{k}$ because only these $\mathbf{q}$ contribute to the
non-analyticity, we obtain
\begin{equation}
\epsilon _{\mathbf{k+q}}=-\epsilon _{k}+v_{F}\left( q-2k_{F}\right)
+v_{F}k_{F}\left( \pi -\theta \right) ^{2},  \label{c23_4}
\end{equation}
where $\theta $ is the angle between $\mathbf{q}$ and $\mathbf{k.}$
Substituting $\epsilon _{\mathbf{k+q}}$ into (\ref{bo1}) and rescaling
variables, we obtain for the effective self-energy
\begin{equation}
\Sigma _{\text{eff}}(\epsilon _{k})=-\frac{mU^{2}k_{F}^{2}}{2\pi ^{3}}%
~\left( \frac{2T}{E_{F}}\right) ^{2}~\int_{-\infty }^{\infty
}dx\int_{0}^{\infty }dy\int_{-\infty }^{x}dz~\frac{(x-z)^{1/2}}{\cosh ^{2}z}%
n_{F}(-\epsilon _{k}+4T(x+y^{2})).  \label{bo2}
\end{equation}
Substituting this self-energy into (\ref{c23_3}), evaluating the derivative
and further rescaling variables we obtain
\begin{equation}
\delta \chi _{1,static}=-\chi _{0}^{2D}~\left( \frac{mU}{4\pi }\right) ^{2}~%
\frac{T}{E_{F}}~Z,  \label{bo3}
\end{equation}
where
\begin{equation}
Z=\frac{4}{\pi }\int_{0}^{\infty }da\int_{-\infty }^{\infty
}db\int_{0}^{\infty }\frac{dc}{\sqrt{c}}~\frac{\sinh (b-c)}{\cosh ^{3}(b-c)}%
~J(a,b),  \label{bo4}
\end{equation}
and
\begin{equation}
J(a,b)=\int_{-\infty }^{\infty }dx\frac{1}{e^{x}+1}~\frac{1}{%
e^{4(b+a^{2}-x)}+1}.  \label{bo5}
\end{equation}
The last integral can be easily evaluated and yields
\begin{equation}
J(a,b)=\frac{4(b+a^{2})}{e^{4(b+a^{2})}-1}.  \label{bo6}
\end{equation}
Substituting this result into (\ref{bo4}), introducing ${\bar{c}}=\sqrt{c}$,
and ${\bar{b}}=a^{2}+b$ and integrating over ${\bar{c}}$ and $a$ using polar
coordinates, we obtain after straightforward calculations
\begin{equation}
Z=-4\int_{-\infty }^{\infty }\frac{d{\bar{b}}{\bar{b}}}{e^{4{\bar{b}}}-1}~%
\frac{1}{\cosh ^{2}{\bar{b}}}  \label{bo7}
\end{equation}
Carrying out the last integration, we finally obtain $Z=-(1+\pi ^{2}/4)$ and
\begin{equation}
\delta \chi _{1,static}=\chi _{0}^{2D}~\left( \frac{mU}{4\pi }\right) ^{2}~%
\frac{T}{\epsilon _{F}}~\left( 1+\frac{\pi ^{2}}{4}\right)  \label{bo8}
\end{equation}
Comparing this result with our $\delta \chi _{1}^{q=2k_{F}}=\delta \chi
_{1}^{q=0}=(1/2)\delta \chi _{s}(0,T)$, given by Eq. (\ref{c20}), we see
that they differ in that $Z\neq 1$. This discrepancy shows that the
frequency dependence of the polarization bubble does contribute to the
non-analytic piece in the thermal static uniform susceptibility.

\end{document}